\documentclass[a4,useAMS,usenatbib,usegraphicx]{mn2e}
\usepackage{subfig}
\usepackage{lscape}
\usepackage{natbib}
\usepackage{amssymb,amsmath}
\usepackage{fixltx2e}
\usepackage{longtable}
\title[The Arecibo Galaxy Environment Survey V : The Virgo Cluster (I)]{The Arecibo Galaxy Environment Survey V : The Virgo Cluster (I)}
\author[R. Taylor, J. I. Davies, R. Auld, R. F. Minchin]{R. Taylor$^{1}$\thanks{Email: rhyst@naic.edu}, J. I. Davies$^2$, R. Auld$^2$, R. F. Minchin$^1$\\
$^1$Arecibo Observatory, HC03 Box 53995, Arecibo, Puerto Rico 00612\\
$^2$School of Physics \& Astronomy, Cardiff University, Queens Buildings, The Parade, Cardiff CF24 3AA, U.K.}
\begin{document}

\date{Submitted Janurary 2012}

\pagerange{\pageref{firstpage}--\pageref{lastpage}} \pubyear{2011}

\maketitle

\label{firstpage}

\begin{abstract}
We present 21 cm observations of a 10 $\times$ 2 degree region in the Virgo cluster, obtained as part of the Arecibo Galaxy Environment Survey. 289 sources are detected over the full redshift range (-2,000 $<$ $v$$_{hel}$ $<$ + 20,000 km/s) with 95 belonging to the cluster ($v$$_{hel}$ $<$ 3,000 km/s). We combine our observations with data from the optically selected Virgo Cluster Catalogue (VCC) and the Sloan Digital Sky Survey (SDSS). Most of our detections can be clearly associated with a unique optical counterpart, and 30\% of the cluster detections are new objects fainter than the VCC optical completeness limit. 7 detections may have no optical counterpart and we discuss the possible origins of these objects. 7 detections appear associated with early-type galaxies. We perform HI stacking on the HI-undetected galaxies listed in the VCC in this region and show that they must have significantly less gas than those actually detected in HI. Galaxies undetected in HI in the cluster appear to be really devoid of gas, in contrast to a sample of field galaxies from ALFALFA.
\end{abstract}


\section{Introduction}

As one of the nearest rich galaxy clusters, the Virgo Cluster represents an important region for studies of galaxy evolution. While the optically selected Virgo Cluster Catalogue (VCC) of \citealt{bst} is complete to an apparent photographic magnitude of +18.0 and is now over 25 years old, it is only much more recently that correspondingly sensitive studies of the HI gas have become possible. The HI Parkes All-Sky Survey (\citealt{s00}) was only sensitive to an HI mass of $\sim$2$\times$10$^{8}$ $M_{\odot}$ at the distance of the cluster (generally assumed to be 17.0 Mpc following \citealt{gv99}), so can offer few insights into dwarf galaxies except at very close distances. The Arecibo Legacy Fast ALFA (ALFALFA) survey (\citealt{g05}, \citealt{haynesaa}) offers a great improvement, with a sensitivity of $\sim$3$\times$10$^{7}$ $M_{\odot}$ at the distance of Virgo.

Very deep HI surveys, with the capability of detecting dwarfs and very gas-poor objects (with gas masses $\lesssim$ $\times$10$^{7}$ $M_{\odot}$), have until now been restricted largely to pointed surveys. These have been limited to small numbers of objects (e.g. \citealt{gv05}) and suffer an inherent optical bias. Conversely, previous and ongoing blind HI surveys, while not subject to an optical bias, have been more limited in depth (e.g. ALFALFA and the Jodrell survey of \citealt{d04}, which was sensitive to about $\sim$6$\times$10$^{7}$ $M_{\odot}$).

The Arecibo Galaxy Environment Survey, AGES, is a complimentary survey to ALFALFA with significantly greater sensitivity ($\sim$8$\times$10$^{6}$ $M_{\odot}$ at the Virgo distance) at the expense of the area of coverage. Of the survey's target of 200 square degrees, 25 square degrees are in the Virgo Cluster region. Many different processes are known to act within clusters to remove gas - ram pressure stripping (\citealt{v01}) and harassment (\citealt{m06}), for example - but their relative importance is still not well understood. By studying the properties of the objects with the lowest gas masses - the most sensitive tracers of environmental influences - we hope to better understand how the gas is affected by its environment. The other targeted regions of AGES cover a wide range of galaxy densities from the local void and isolated galaxies, to galaxy pairs, groups and other clusters (for a more detailed summary see \citealt{a06}).

Although it does not normally dominate even the baryonic mass component of large galaxies, as the fuel for future star formation neutral hydrogen is nonetheless important in understanding galaxy evolution. It is also easily affected by environment, occasionally drawn into spectacular streams on the 100 kpc scale (e.g \citealt{k08} and \citealt{scott}), and relatively easy to detect compared to other tracers of cold gas. Furthermore, it may also offer a way to detect optically dark galaxies, proposed as a solution to the missing satellites problem (\citealt{d06}). In this context the importance of a deep, fully-sampled HI survey in a region of high galaxy density becomes obvious.

The Virgo Cluster is known to have a complex structure, and is thought to be comprised of at least 3 distinct clouds. The dominant component is the so-called sub-cluster A, with the giant elliptical M87 at its center. It is also where the main bulk of the X-ray gas is centered (\citealt{boh}) with other concentrations around M49 and M86, suggesting that these are the centers of other, smaller sub-clusters.

\citealt{gv99} describe clouds at 17, 23 and 32 Mpc distances, with numerous other separate regions of the cluster whose distances are less obviously distinct. Their findings were based on studies of the Tully-Fisher and fundamental plane relations. These velocity-independent distance estimators indicate that the high velocity dispersion of the cluster (a total range of over 4,000 km/s) means that a galaxy's redshift gives little distance information - apart from verifying its membership of the cluster, it has little or no bearing on whether the object is likely to belong to one particular cloud or another.

More recently, \citealt{Mei} studied the cluster using the method of surface brightness fluctuations. Although this survey was geared towards understanding the main body of the cluster at 17 Mpc distance, it also confirmed the presence of an infalling cloud at 23 Mpc. The main cloud, or sub-cluster A, was found to have a depth of 2.4 $\pm$ 0.4 Mpc, with a velocity dispersion of around 590 km/s. The key point is that there is no straightforward way to assign distances. Here we use the distances assignments given in the Galaxy On-Line Milano Network (GOLDMine) database, which are based on measurements of \citealt{gv99} - see section \ref{sec:distribution} for more details. 

The remainder of this paper is organised as follows. In section \ref{sec:Obs} we describe the observations and data reduction procedures. In section \ref{sec:Analysis} we describe the analysis of the data while in section \ref{sec:Results} we describe the results. Finally in section \ref{sec:Conclusions} we suggest some possible implications of these results. We assume a value of $H_{0}$ of 71 km/s/Mpc throughout. Our analysis here is largely restricted to cluster members; we discuss some of the background objects in more detail in Paper II.

\section{Observations and data deduction}
\label{sec:Obs}

\begin{figure}
\includegraphics[width=84mm]{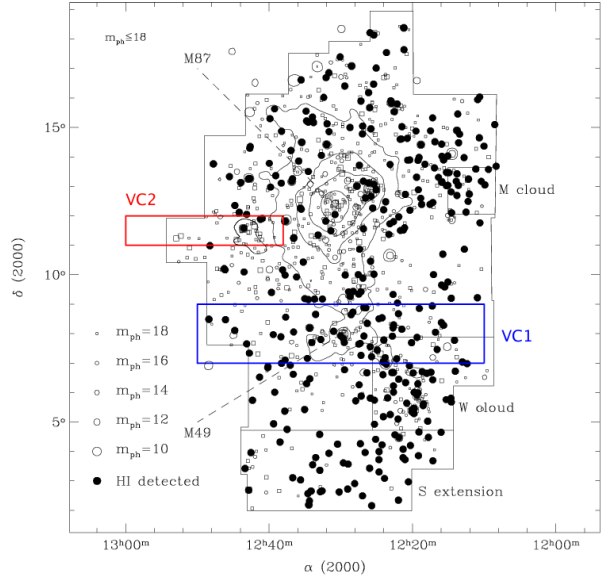}
\caption[Virgo Cluster map showing the AGES fields]{The Virgo Cluster region, highlighting the AGES and VCC areas and detections. Squares are early-type while circles are late-type galaxies. Filled circles are those with known HI detections from previous pointed observations. The contours show X-ray detection from the ROSAT satellite. Physically, the area described in this paper spans 3.0$\times$0.60 Mpc at 17 Mpc distance.}
\label{VMap}
\end{figure}

Two areas of the Virgo Cluster have been selected for study with AGES, VC1 and VC2, shown in figure \ref{VMap}. This paper examines the VC1 area while paper II will consider VC2 and compare the results of the two areas. The areas were selected for study as they span very different regions of the cluster. VC1 spans the interior of the cluster and is centered on M49, with galaxies believed to be in 3 different populations at different distances (\citealt{gv99}). Previous pointed HI observations (\citealt{gv05}, \citealt{gv03})  indicate that this region is denser than VC2, with VC1 and VC2 respectively containing 4.2 and 2.2 detections per square degree. VC2 extends from the cluster interior to beyond the VCC survey region, though cluster members in that region are believed to belong to a single population at a uniform distance (\citealt{gv99}). In this study the focus is on cluster members, though background objects are used where appropriate as a comparison sample.

The observations, data reduction and analysis procedures have been extensively described in \citealt{a06}, as well as in \citealt{c08}, \citealt{m10} and \citealt{d11}. They are here only summarised. Observations were taken in January-June 2008, February-June 2009 January-June 2010 and January 2011, using the ALFA instrument on the Arecibo telescope in spectral line mode. The field observed to full depth spans 10 degrees of R.A. by 2 degrees of declination (in contrast VC2 spans only 5 $\times$ 1 degree) centered on M49. Due to ALFA's hexagonal beam arrangement, a small area outside this range is also included. The full spatial range of the data considered here is from 12:08:36 to 12:49:36 in R.A, and from +06:52:55 to +09:06:55 in declination.

Observations are performed in drift scan mode, where the telescope is driven to the start of the scan position and the sky allowed to drift overhead. Each drift lasts for 20 minutes (covering 5 degrees of R.A.), so two sets of drifts are needed to cover the full R.A. range. Shorter drifts ensure less loss of data in the event of a malfunction, and allow us to reach full depth over successive areas well before the observation campaign is complete. The ALFA instrument is rotated before the start of each scan to correct for the change in parallactic angle. At the start of each scan a signal from a high-temperature noise diode is injected for 5 seconds to provide flux calibration.

At Arecibo a point takes 12 seconds to cross the beam. To ensure Nyquist sampling scans are staggered by half the beam width ($\sim$ 1\arcmin). With the 7 beams of ALFA each point is scanned 25 times, giving a total integration time of 300 seconds per point. Each beam records data from 2 polarisations every second over 4096 channels, which span a velocity range from approximately -2,000 km/s to + 20,000 km/s. The velocity resolution is equivalent to $\sim$ 5 km/s, or 10 km/s after Hanning smoothing to remove Gibbs ringing from the Milky Way and other bright extended sources such as RFI (see below).

The data are reduced using the AIPS++ packages \textsc{livedata} and \textsc{gridzilla}, described in \citealt{b01}, developed by the Australia Telescope National Facility (ATNF). \textsc{livedata} is used to estimate and remove the bandpass and calibrate the resultant spectra; \textsc{gridzilla} co-adds the data to produce the final 3D data cubes. The bandpass across the scan is estimated by the median of the data values, which is robust to bright point sources but can give incorrect values for bright extended sources. The data are analysed using the \textsc{miriad} task $mbspect$. From measurements of individual sources, the median noise level - rms - reached is approximately 0.6 mJy/beam (after Hanning smoothing).

As a sensitivity limit, for an rms of 0.6 mJy a 4$\sigma$ detection with a top-hat profile, 50 km/s velocity width, would have an HI mass at 17 Mpc distance of 8.2$\times$10$^{6}$ M$_{\odot}$. However this is only approximate. Galaxies of any given mass will be detected with lower S/N levels if their velocity width is greater (or equally, if they have the same intrinsic velocity width but higher inclination angles) since their flux is spread out over more channels.

Follow-up observations were performed using the L-wide receiver. This uses the position-switching method, with the ON and OFF source times each of 5 minutes. The data were quickly reduced and additional observations taken if necessary. We use the WAPPs (Wideband Arecibo Pulsar Processors) with 9-level sampling and 1 polarization per board, giving us 4096 channels. We use two bandwidth settings, 50 MHz (giving a velocity resolution of 1.25 km/s) and 25 MHz (0.63 km/s resolution). The criteria for performing follow-up observations are described in section \ref{sec:sourcex}.

\subsection{Data contamination : the Milky Way and Radio Frequency Interference}

Although the bandwidth observed spans over 20,000 km/s of redshift, in certain regions the survey is effectively blinded due to the presence of strong contaminating sources. The signal from the Milky Way is extremely strong and fills the whole sky, and \textsc{livedata} is unable to correctly process the signal using our standard techniques. Consequently we are unable to obtain meaningful data over the velocity range -50 $<$ v $<$ +50 km/s. Similarly the Punta del Este Federal Aviation Administration radar, operating at 1350 MHz, causes strong false signals over the approximate velocity range +15,400 $<$ v $<$ +16,000 km/s, as well as the intermod at approximately 7,000 km/s. Harmonics of the radar are present throughout the cube but are much weaker and do not significantly detract from the available bandwidth. Some scans were also affected by the signal from the L3 GPS satellite, generating false signals at approximately 8,500 km/s. The survey is not blinded within these regions affected by RFI, but sensitivity is greatly reduced.

\subsection{Source extraction}
\label{sec:sourcex}

There is no universally accepted technique for HI source extraction from 3-dimensional data cubes. We therefore employ a variety of methods in order to recover as many real sources as possible, whilst attempting to minimise the number of spurious detections. Firstly we inspect the cube by eye, visually scanning the cube in different projections and recording any possible sources. The cube is scanned at least three times in this way and candidate detections are masked using the \textsc{miriad} task $immask$ to avoid confusion. A source only detected once is rejected; sources detected two or three times are subject to further inspection.

The second technique is an automatic extractor which we developed specifically for AGES called \textsc{glados} (Galaxy Line Analysis for Detection Of Sources). It is  partly based on the existing programs \textsc{polyfind} (described in \citealt{poly}) and \textsc{duchamp} (see \citealt{duch}). \textsc{polyfind} examines spectra and searches for peaks above a S/N threshold, whereas \textsc{duchamp} examines images and searches for a specified number of contiguous pixels above a fixed flux threshold. Both approaches have merits and problems. \textsc{polyfind} only requires a single input threshold, the signal-to-noise ratio (S/N), which does not need to be altered if the actual rms varies. However, since there are many thousands of individual pixels which happen to be above (say) a peak S/N of 4, \textsc{polyfind}'s generated catalogues have a reliability estimated at 3\% (at the 4$\sigma$ level), and are correspondingly laborious to search.

\textsc{duchamp}'s approach, in contrast, can generate catalogues with extremely high levels of reliability, but with poor completeness. There are many problems with using a fixed flux threshold for AGES data. Firstly the noise can vary across the cube, especially at the edges (both spatially and in velocity). To use \textsc{duchamp} to search for faint sources requires these regions to be truncated, otherwise many thousands of spurious detections result. Secondly it is impossible to determine a sensible flux threshold without having searched the cube previously, negating the main benefit of an automatic extractor.

To overcome these problems \textsc{glados} employs a combination of many different techniques. First, both of the individual polarisations are gridded separately, in addition to the averaged cube which is used for the final measurements. \textsc{glados} initially searches spectra along each point in the averaged cube, searching for peaks above a specified S/N threshold (we use 4$\sigma$) and minimum velocity width (we use 25 km/s). The use of a minimum velocity width criteria helps reduce the thousands of noise spikes which plague \textsc{polyfind}. If a candidate signal is found, the individual polarisations are checked at the same point for a corresponding detection.

This step of examining the separate polarisations helps reduce spurious detections, since the noise in each is uncorrelated. The penalty for this is that their noise level is higher, and so they must be smoothed so that weaker detections are not missed. Since this reduces velocity resolution, the end result is that it can fail to detect the narrowest sources. However, since the number of sources found even by a visual search below 25 km/s velocity width is extremely small, this does not seem to be a major sacrifice.

Finally, many of the close bright sources are resolved by AGES and so span many pixels. \textsc{glados} uses the \textsc{starlink} `tmatch' package to combine these repeat detections. As this can result in merging close sources, the original output is retained. Using \textsc{topcat} a map of the original \textsc{glados} detections are plotted, and the post-merging detections are overlaid. An observer then checks by eye for any cases where separate sources may have been unfairly merged.

Our third source extraction technique is to inspect spectra extracted with $mbspect$ at the positions of known galaxies, listed in VCC, which were not detected by the other two methods. The idea here is to search for very weak sources that the other techniques might not be sensitive enough to pick up without allowing many additional spurious detections. Although 217 spectra were examined, this only resulted in 3 additional detections (these are very similar to ALFALFA `code 2' detections - see \citealt{aavirgo}). This demonstrates that our extraction techniques are generally detecting even the weakest sources.

Once a source is accepted as a candidate it is first inspected with \textsc{miriad} and/or a visual inspection of the standard averaged cube and both individual polarisations. These additional checks ensure that any polarised signals, which could be recorded in a visual search (since it is impractical to search the entirety of all 3 cubes by eye), and any unusual artifacts such as may result from RFI, are not accepted as real objects.The remaining sources are then subject to follow-up observations where required. This is performed if a source was only detected by 1 method, has no clear optical counterpart (see section \ref{sec:Analysis}), or is in any way anomalous (such as being unclear if it is polarised emission or not, or if its measurements do not agree with catalogue data).

\section{Source catalogues}
\label{sec:Analysis}

The combination of source extraction methods described above produced a combined candidate list of 322 sources for the entire volume of the cube. 75 of these were considered uncertain, either because they were detected only by 1 method, had no obvious optical counterpart, or are of low S/N. Follow-up observations were performed on all of these, of which 33 were not detected, reducing our final sample to 289 objects.

The data are analysed with $mbspect$, as described in \citealt{c08}. To summarise, the position of the source is found from a Gaussian fit to a moment 0 map within the galaxy's velocity range. The resultant extracted spectrum is a weighted averaged, with the weighting depending on the distance from the determined source position. The HI mass is computed using the equation :
\begin{equation}M_{\rm{HI}} = 2.36\times10^{5}\times d^{2}\times F_{\rm{HI}}\label{HIMass}\end{equation}
Where $M_{\rm{HI}}$ is the HI mass in solar units, $d$ is the distance in Mpc, and $F_{\rm{HI}}$ is the integrated HI flux in Jy km/s. The distance determination is explained in section \ref{sec:distribution}.

Tables \ref{VirgoHIT} and \ref{BackHIT} give the measured parameters for sources within the cluster (cz $<$ 3,000 km/s, \citealt{aavirgo}) and behind it (cz $>$ 3,000 km/s), respectively. The columns in each table are identical and are as follows : (1) Source number in this catalogue (2) Right ascension J2000, error in seconds of time (3) Declination J2000, error in seconds of arc (4) Heliocentric velocity km/s (5) Maximum velocity width at 50$\%$ and (6) 20$\%$ of the peak flux (7) Total flux Jy (8) Distance Mpc (9) Estimated HI mass, $\log$ $M_{\odot}$ (10) Peak signal to noise (11) R.m.s. in mJy.

\subsection{Identification of optical counterparts}
\label{sec:ocs}
\begin{figure}
\begin{center}
\includegraphics[width=84mm]{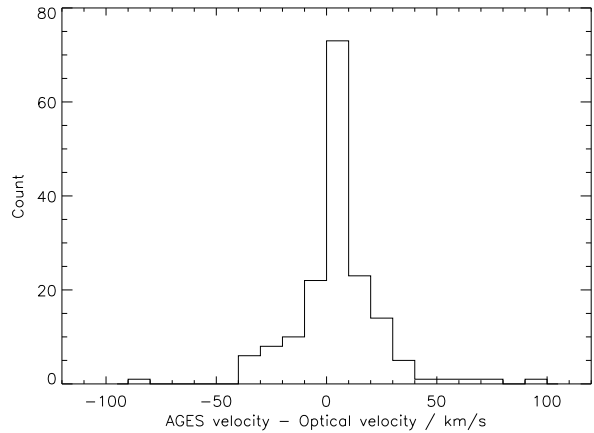}
\caption[Differences between measured HI and optical recessional velocities]{Distribution of the difference between measured HI and optical recessional velocity with a bin size of 10 km/s. The mean difference is +4 km/s.}
\label{RedComp}
\end{center}
\end{figure}

This region has been fully observed with the Sloan Digital Sky Survey (SDSS). We visually search the SDSS images (DR8) within a 3.5\arcmin radius of the HI coordinates as determined by $mbspect$. When a candidate is found, we check its measured redshift from the SDSS, NED and GOLDMine (\citealt{gv03}) - having at least two optical redshift measurements is preferable, since there is a chance that a single measurement might be erroneous and cause problems in identifying the true optical counterpart. If the candidate's optical redshift(s) agrees with the HI redshift to within 200 km/s, the object is defined to have a `sure' optical counterpart. If the optical and HI redshifts differ by more than this then the optical candidate is rejected. The distribution of velocity offsets is shown in figure \ref{RedComp}.

\begin{figure}
\begin{center}
\includegraphics[width=84mm]{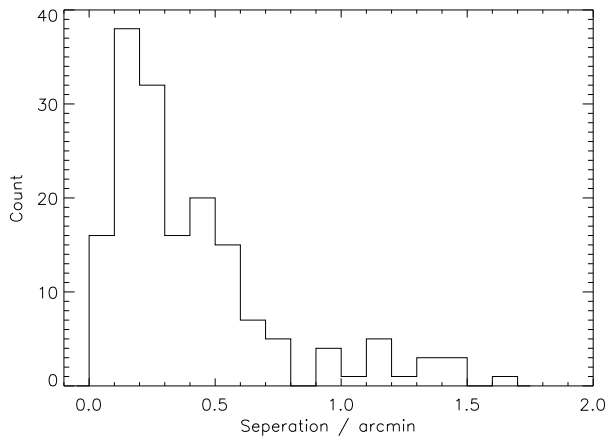}
\caption[Differences between measured HI and optical spatial coordinates]{Distribution of the difference between measured HI and optical spatial coordinates with a bin size of 0.1\arcmin. The mean difference is 24".}
\label{PosComp}
\end{center}
\end{figure}

In many cases there is a lack of optical redshift data available. For those candidates, we accept them as associated with the HI if they are the only unique candidate within 3.5\arcmin (in most cases their optical position differs from the HI by much less than this, as shown in figure \ref{PosComp}). We deem them to be `probable' optical counterparts. Although we treat them as being associated with the HI throughout the analysis, their optical association is flagged as uncertain in the tables. 

The optical and HI measurements of position and redshift are generally in very good agreement, with a mean difference of just 4 km/s for the velocities and 24" for the positions. The few outliers which are present are low S/N HI sources for which it is difficult to precisely define the location of the HI (both in space and velocity).

\subsection{Comparison with other HI surveys}

67 galaxies in this region have also been detected by ALFALFA. We will comment in more detail on the detection rates of the two surveys in a future paper. Here we restrict the analysis to a comparison of our measurements of HI masses and velocity widths. We use the W50 estimate since that is the quantity given in the ALFALFA catalogue. We do not use the HI mass estimate as given in the ALFALFA catalogue, since that assumes a different distance estimate (see \citealt{gv05}); instead we use their measured total flux and convert to an HI mass assuming the distance as described in section \ref{sec:distribution}.

\begin{figure}
\begin{center}
\includegraphics[width=84mm]{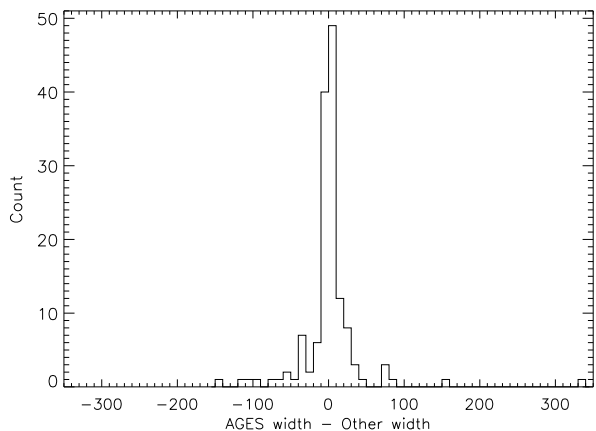}
\caption[Differences between AGES velocity widths and previous observations]{Differences between AGES velocity widths and previous observations, bin size of 10 km/s. The mean difference is 1 km/s.}
\label{VComp}
\end{center}
\end{figure}

\begin{figure}
\begin{center}
\includegraphics[width=84mm]{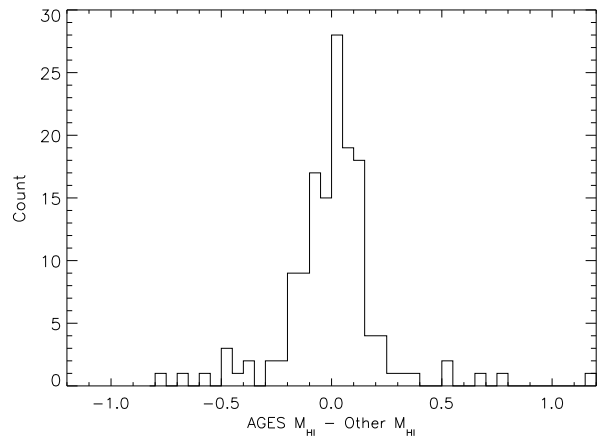}
\caption[Differences between AGES $M_{\rm{HI}}$ and previous observations]{Differences between AGES $M_{\rm{HI}}$ and previous observations, bin size 0.05 (units are logarithmic solar masses). The mean difference is 0.04.}
\label{MComp}
\end{center}
\end{figure}

76 galaxies are listed as having an HI detection in GOLDMine (after discounting those listed as unreliable). For these objects we use the HI mass listed in GOLDMine directly, since the distance assumption is the same. However, GOLDMine only lists the average of the W50 and W20, so for the velocity comparison we use the average of these quantities as determined for the AGES measurements. The velocity width and HI mass estimates are compared in figures \ref{VComp} and \ref{MComp} respectively.

As with the comparisons with the optical data, AGES measurements are generally in good agreement with existing observations, though several outliers are seen. These are low S/N sources, where less sensitive surveys may not detect HI over as wide a velocity range, or may confuse part of the noise with the real detection. We use the AGES measurements in all cases.

\subsection{Analysis of optical data}

Optical data available for this region includes the SDSS and the Isaac Newton Telescope Wide Field Survey (INT WFS). However the INT WFS data is too variable in quality to be of much use in this region. Instead, optical photometry is performed on the SDSS data which is uniformly available in this region, using the aperture photometry tools in the \textsc{ds9} package $funtools$. We restrict photometric measurements to those objects where there is a single sure or probable counterpart, where the HI detection is not in doubt, and where the galaxy's optical emission is uncorrupted by any foreground stars. For instance, in the case of the two bright galaxies VCC 905 and VCC 939, the two objects are so close that the AGES observations cannot resolve the seperate objects, so we cannot accurately determine how much HI each galaxy contains (the HI profile, source 250, is nonetheless listed in table \ref{VirgoHIT}).

This gives a sample size of 233 objects, of which 79 are members of the Virgo Cluster (i.e. at a redshift of $\leq$ 3,000 km/s). For completeness, the HI parameters of all 269 sure detections are shown in tables \ref{VirgoHIT} and \ref{BackHIT}, which detail the Virgo Cluster and background objects respectively. The optical properties of all detections are given in tables \ref{OptVHIT} and \ref{OptBHIT}. Note that if photometry is not relevant (i.e. when studying the velocity distribution of HI detections) then we use all sure HI detections, not just those with good optical photometry data.

\section{Results}
\label{sec:Results}
\subsection{Overall statistics}

A total of 95 sure HI detections were made within the Virgo Cluster. Of these, 26 are not members of the VCC. The VCC is considered to be complete to a photographic magnitude of 18. Taking this to be approximately equal the $g$ band magnitude, 14 of these detections are actually brighter than the completeness limit (4 of these are marginal, with $g$ $>$ 17.5). This is, however, less than 5\% of the total of 282 VCC objects thought to be cluster members (see below) in this region, so the completeness magnitude of the VCC seems reasonable.

The VCC lists 342 objects within this region. 60 of these are excluded as cluster members by previous redshift measurements. The remaining 282 are classed as members, 95 of which are spectroscopically confirmed. AGES provides redshift measurements for 10 objects for which none was previously available, and finds that 3 are cluster members and 7 are background objects.

Though AGES does detect a population of objects missed by the VCC, this population is small in comparison with the number of VCC cluster members. Some basic assumptions show that this is at least partly due to the sensitivity effects of each survey. As described earlier, the sensitivity limit of AGES is approximately 8$\times$10$^{6}$ M$_{\odot}$ at a distance of 17 Mpc. The completeness limit of the VCC is a photographic magnitude of 18.0, but it does contain some objects (in the VC1 region) as faint as 20.0, or 19.6 in the $g$ band (an absolute magnitude of -11.6). Using the relation between HI and stellar mass of \citealt{gv08}, AGES HI mass sensitivity is equivalent to an absolute magnitude of -9.8 (or to put it another way, the mass sensitivities of the two surveys are within a factor of 2 of each other).

The relatively low number of detections from AGES that are not part of the VCC immediately implies that the HI is usually associated with optical sources. This is indeed the case, but there are some exceptions which we discuss in sections \ref{sec:weird} and \ref{sec:Conclusions}.

\subsection{Spatial and Velocity Distribution}
\label{sec:distribution}

This study is focused on the HI detections that are clearly cluster members. For objects outside the cluster, the distance used is determined simply from the Hubble velocity, assuming a Hubble constant of 71 km/s/Mpc. For objects within Virgo, the distance used is that given in the GOLDMine database, which, as mentioned earlier, is based on upon Tully-Fisher and fundamental plane determinations. However this was only possible for a limited sample of 134 objects, as described in \citealt{gv99}. They then defined spatial regions within which all galaxies are assigned to the same distance, based on their sample with actual distance estimates.

\begin{figure}
\begin{center}
\includegraphics[width=84mm]{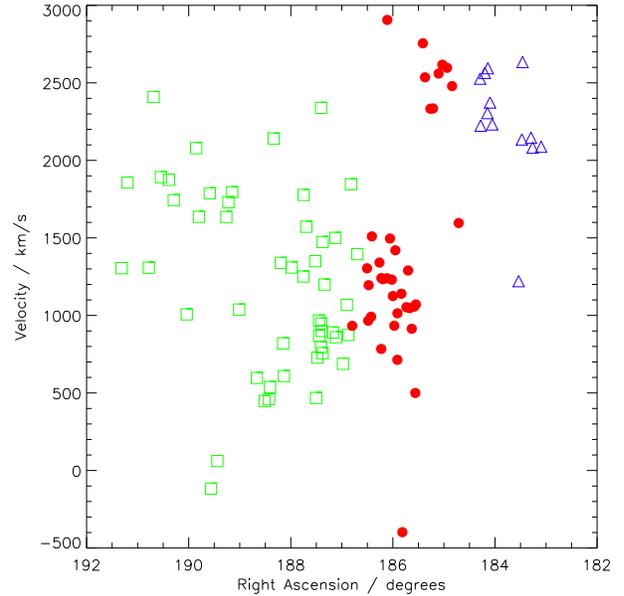}
\caption[P-V diagram of VCC galaxies in VC1]{Position-velocity diagram for the 99 VCC galaxies listed in GOLDMine with redshift measurements. Green squares are those objects thought to be at 17 Mpc, red circles at 23 Mpc and blue triangles are at 32 Mpc distance.}
\label{PV1}
\end{center}
\end{figure}

\begin{figure}
\begin{center}
\includegraphics[width=84mm]{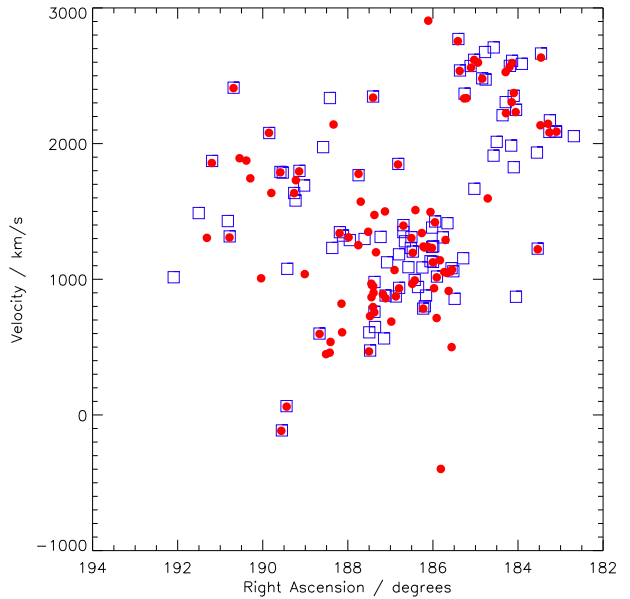}
\caption[P-V diagram of VCC galaxies and AGES detections in VC1]{Position-velocity diagram for the 99 VCC galaxies listed in GOLDMine with redshift measurements (red circles) compared with 95 AGES detections in the cluster (blue squares).}
\label{PV2}
\end{center}
\end{figure}

We apply this same procedure to all our detections within the cluster, including those not listed in the VCC. We note, however, that there remains considerable distance uncertainty, as evidenced by figure \ref{PV1} (we show our detections in figure \ref{PV2}). Although it appears that galaxy populations are quite distinctly separate in position-velocity space, the boundaries chosen by \citealt{gv99} do not respect this. The effects of this are, fortunately, not dramatic. HI mass scales as distance squared, so a change from 17 to 23 Mpc, or from 23 to 32 Mpc, increases the HI mass by a factor 1.8 and 1.9 respectively. Equivalently, this would result in a decrease of absolute magnitude by 0.66 and 0.72 magnitudes respectively.

\begin{figure*}
\begin{center}
\includegraphics[width=180mm]{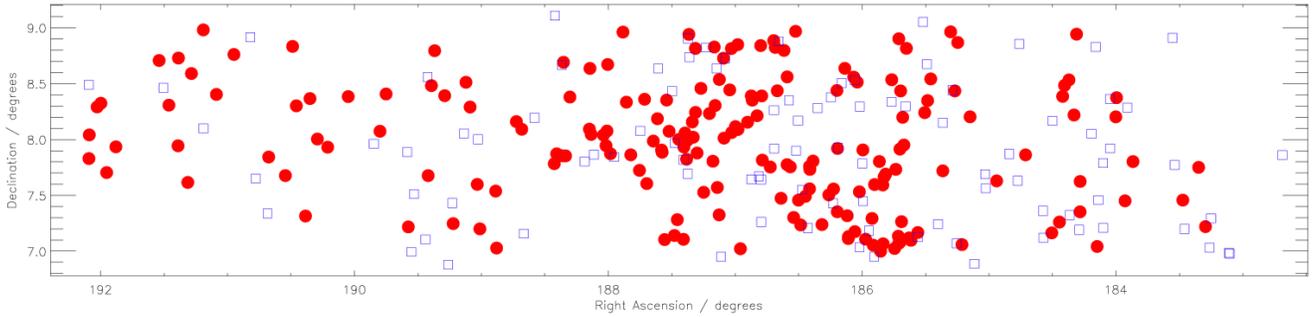}
\caption[Spatial distribution of VC1 HI detections and non-detections]{Spatial distribution of the 95 AGES detections (blue squares) with $cz$ $<$ 3,000 km/s and the 195 VCC non-detected galaxies (red circles) identified as cluster members.}
\label{SpatialVC1}
\end{center}
\end{figure*}

The spatial distribution of the AGES detections and non-detected VCC galaxies is shown in figure \ref{SpatialVC1}. The distributions are broadly similar, with the western region being richer in both HI-detected and non-detected galaxies. The western half contains approximately 22 non-detected and 9 detected objects per square degree, the eastern half just 10 non-detected and 3 detected objects per square degree. Qualitatively, the structures defined by the objects are very similar. Gas-rich galaxies appear to trace roughly the same population of objects as the non-detected galaxies.

More quantitatively, however, 30\% of the HI detections in the cluster are found at velocities greater than 2,000 km/s. In contrast only 10\% of the non-detections have similar velocities. A Kolmogorov-Smirnov test reveals that the distributions are different to a 99.9\% significance level. The influence of the cluster on galaxy gas content is discussed further in sections \ref{sec:gasratio} and \ref{sec:def}.

\begin{figure}
\begin{center}
\includegraphics[scale=0.4]{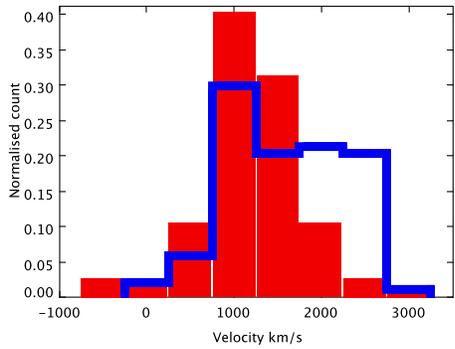}
\caption[Velocity distribution of Virgo Cluster galaxies in VC1]{Velocity distribution of AGES detections (blue) and non-detected VCC galaxies (red) in the VC1 region, bin width of 500 km/s.}
\label{VC1VelHist}
\end{center}
\end{figure}

\subsection{Morphology}
\subsubsection{Late-type galaxies}
\label{sec:LTGs}
\begin{figure}
\begin{center}  
  \subfloat[VCC HI non-detections]{\includegraphics[width=84mm]{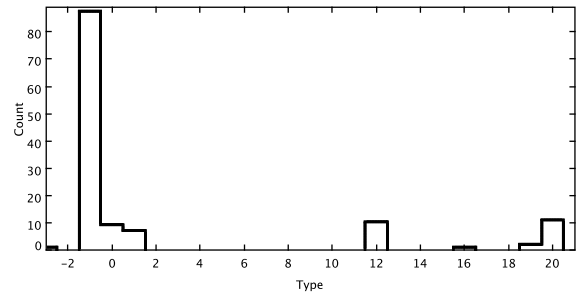}}\\
  \subfloat[AGES detections]{\includegraphics[width=84mm]{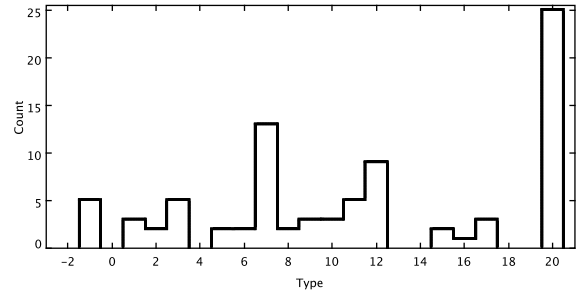}} 
\caption[Morphology distribution of Virgo Cluster galaxies in VC1]{Morphology distribution of VCC non-detections and AGES HI detections.}
\label{MorphDist}
\end{center}
\end{figure}

For morphological classification, we use the scheme adopted by the GOLDMine database, where types -3 to +1 are early-type galaxies (hereafter ETGs), all others are late-type galaxies (LTGs) with +2 to +10 being spirals, +11 to +19 being irregulars and Blue Compact Dwarfs (BCDs), whilst those of type 20 have not been classified. The distribution of the morphologies of HI detections and non-detections is shown in figure \ref{MorphDist}.

At first glance figure \ref{MorphDist} suggests that there is a significant population of undetected LTGs as well as a population of HI detected early-types. We discuss the latter in section \ref{sec:etgs}, however the former can be dismissed. None of the undetected type 20 (unclassified) objects resemble late-types - they are all small, faint and probably dwarf ellipticals. Of the other supposed late-types, those classified as spirals and irregulars, only 2 appear to be genuine LTGs, and those are too close to other HI detections to be resolved as separate sources by AGES. The others appear to be outright misclassifications, and visually do not at all resemble their late-type assignments.

Conversely, the unclassified objects detected in HI are definitely late-type objects, most of them probably dwarf irregulars or BCDs (although 3 appear to be flat, edge-on discs). None of the unclassified objects are likely to be early-type objects but for more discussion on this see the next section. Those undetected unclassified objects all appear to be ETGs, mostly very faint dwarf ellipticals. The type 12 objects undetected by AGES are either very faint (and so possibly misclassified) or have poor quality claimed HI detections which may be spurious. In short it appears that virtually all LTGs are detected in HI in this region by AGES. Note that throughout the rest of the analysis we use the classifications given in GOLDMine, except where these appear to be obviously in error.

\subsubsection{Early-type galaxies}
\label{sec:etgs}

Given the above definition of early-type morphology, 7 HI detections are associated with ETGs; 5 of these are new HI detections. The HI association with VCC 1394 is ambiguous in that the HI coordinates are 2.2\arcmin from the optical position (no optical redshift data exists). It is the only obvious galaxy visible within the Arecibo beam and classed as a cluster member in the VCC, so is by far the most likely candidate.

Additionally, there have been several previous claims of HI detected in the bright S0 VCC 1535, initially by \citealt{bott}, as well as in GOLDMine (which records an $M_{\rm{HI}}$ of 2$\times$10$^{9}$ M$_{\odot}$, though the data used is unpublished). More recently ALFALFA (\citealt{haynesaa}) find a much more modest $M_{\rm{HI}}$ of 1.4$\times$10$^{7}$ $M_{\odot}$, with a S/N of  3.2. However within the reported velocity range of the source AGES finds no structure that could be considered as a detection. We disregard this galaxy in the remainder of the analysis.

It is important to confirm that those objects really are ETGs and not misclassified objects. \citealt{aaETG} argue that several of their detected ETGs show peculiar morphological characteristics and are only borderline ETGs. This does not appear to be the case for the AGES detections, which, with the exception of VCC 758 with its clear dust lane/disc, appear smooth and featureless at optical wavelengths.

\citealt{L06} (hereafter L06) describe two methods employed to enhance any possible structures, which we employ here. The first is to create a residual optical image. A symmetrical model of each galaxy is created using the \textsc{iraf} tasks \textit{ellipse} and \textit{bmodel}, which is subtracted using \textit{imarit} to reveal any asymmetrical features. The position angle and ellipticity are fixed, otherwise, as shown in L06, there is a risk the model image will also incorporate asymmetrical features. This technique is only possible for 3 of the early-type objects detected by AGES. 3 more are too small and too faint for \textit{ellipse} to create a successful model. For the remaining galaxy, VCC 758, there seems little point attempting such a procedure given the presence of a prominent central dust lane or disc.

\begin{figure}
\begin{center}  
  \subfloat[VCC 190 $g$ band]{\includegraphics[scale=0.29]{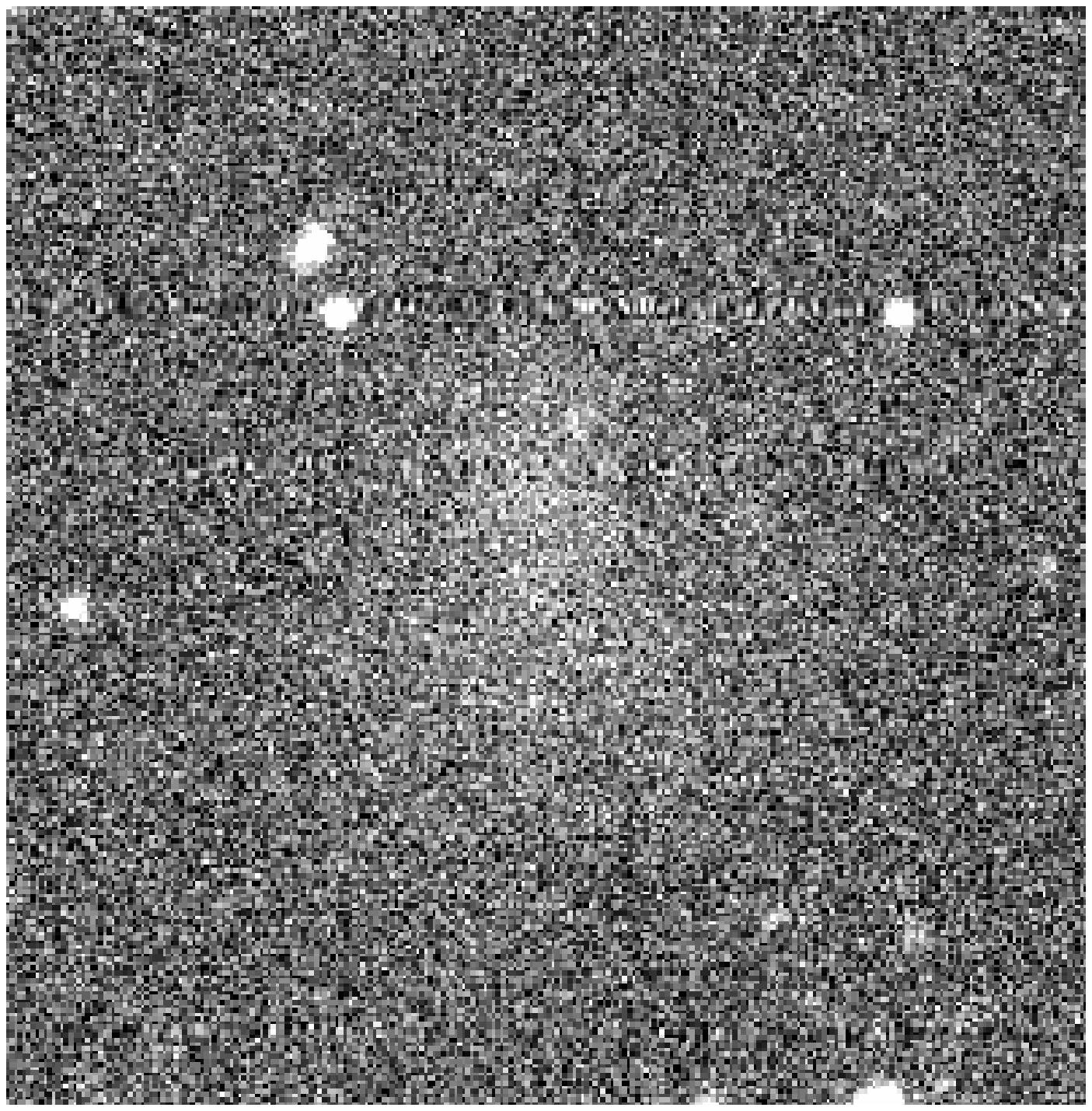}}
  \subfloat[VCC 190 unsharp mask]{\includegraphics[scale=0.29]{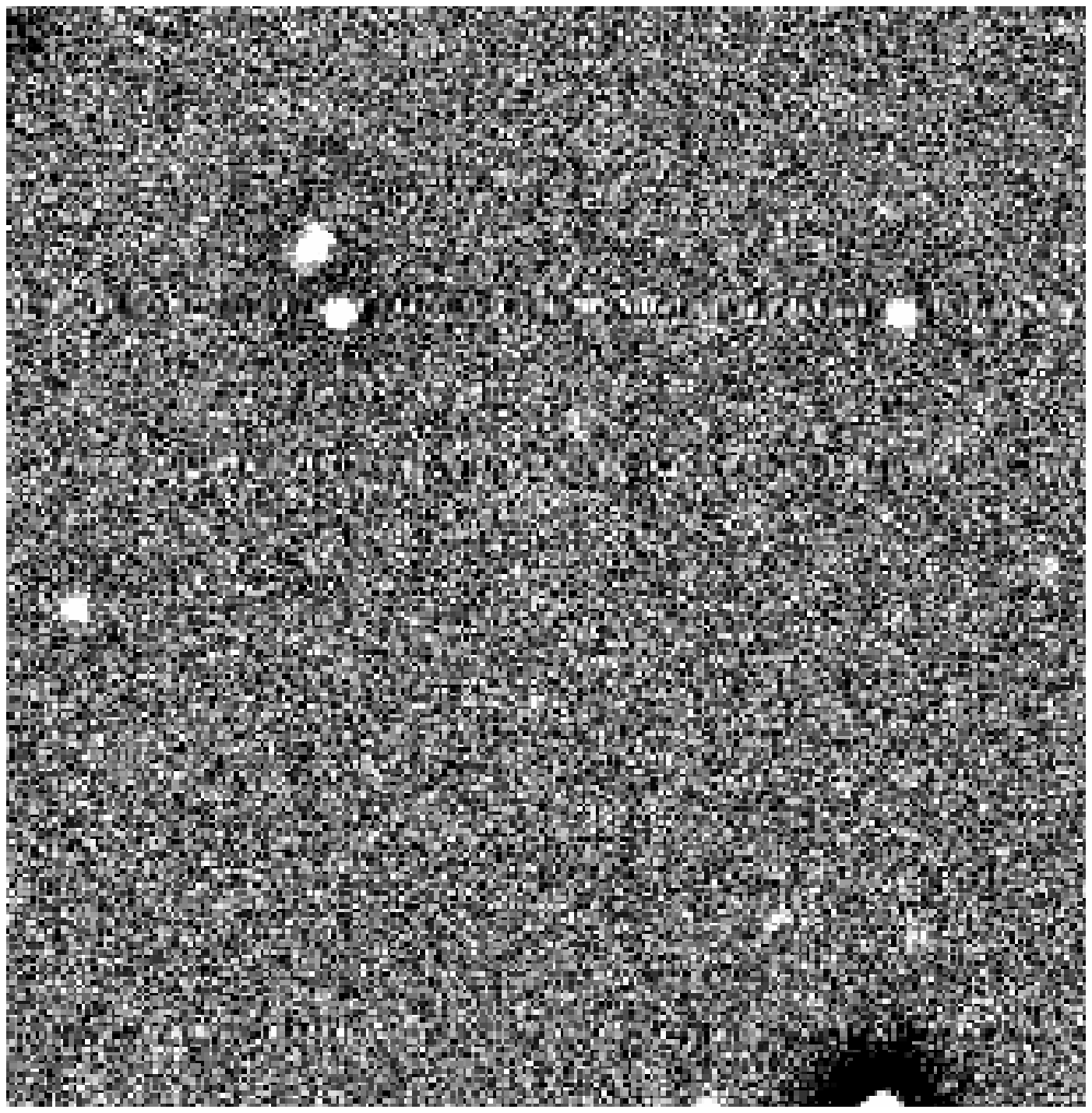}}\\  
  \subfloat[VCC 1394 $g$ band]{\includegraphics[scale=0.29]{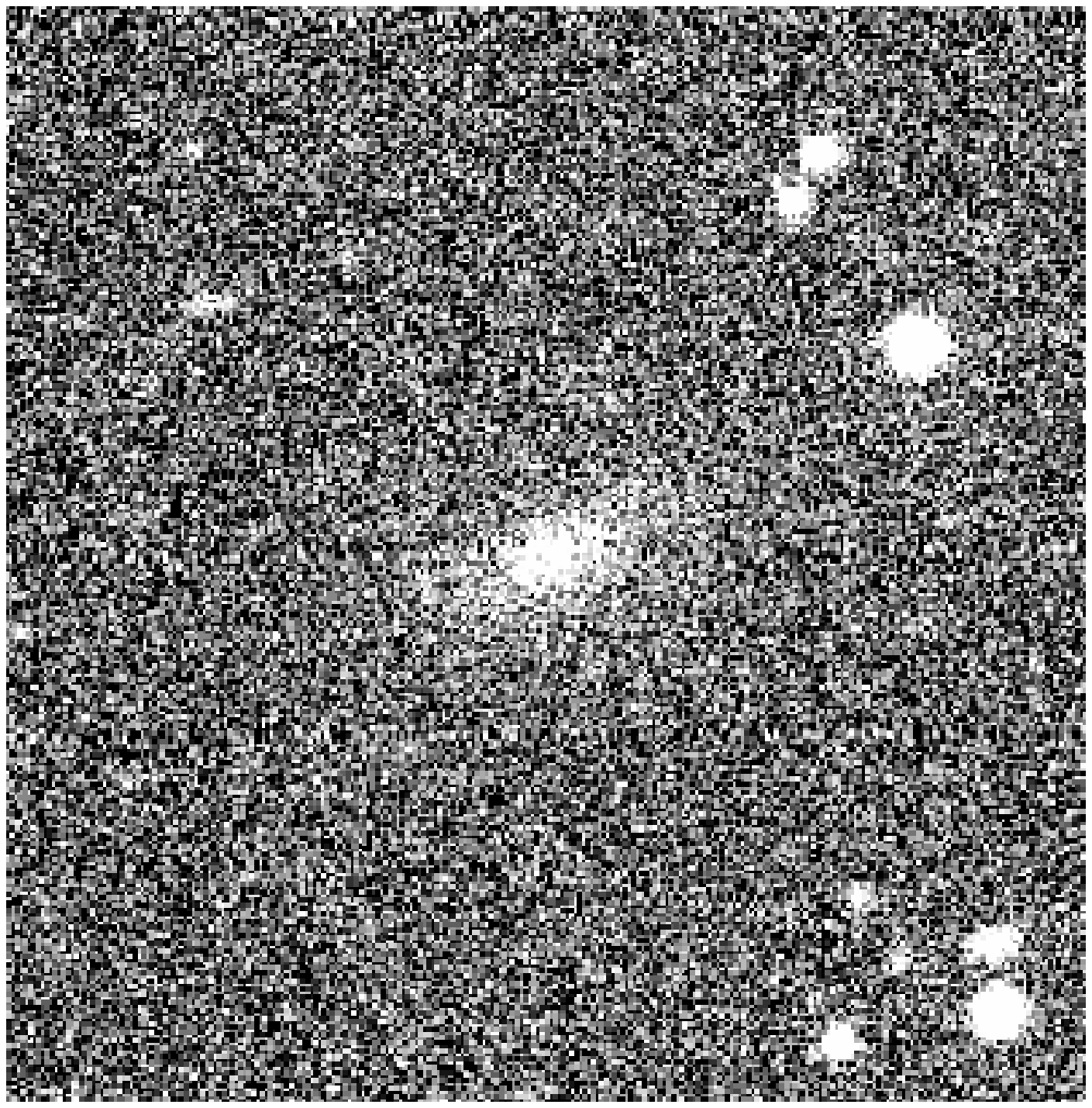}}
  \subfloat[VCC 1394 unsharp mask]{\includegraphics[scale=0.29]{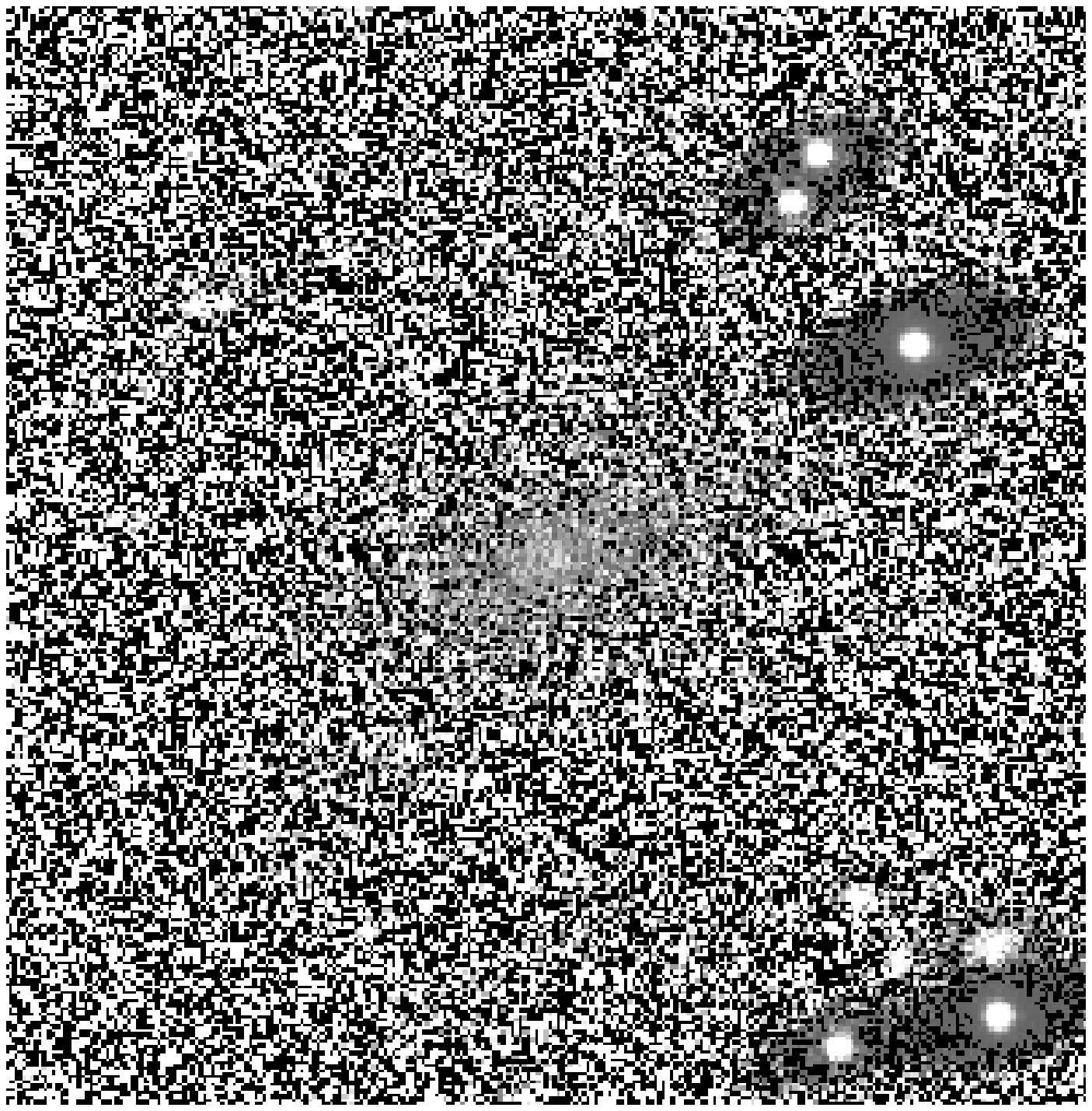}}\\
  \subfloat[VCC 1964 $g$ band]{\includegraphics[width=41mm]{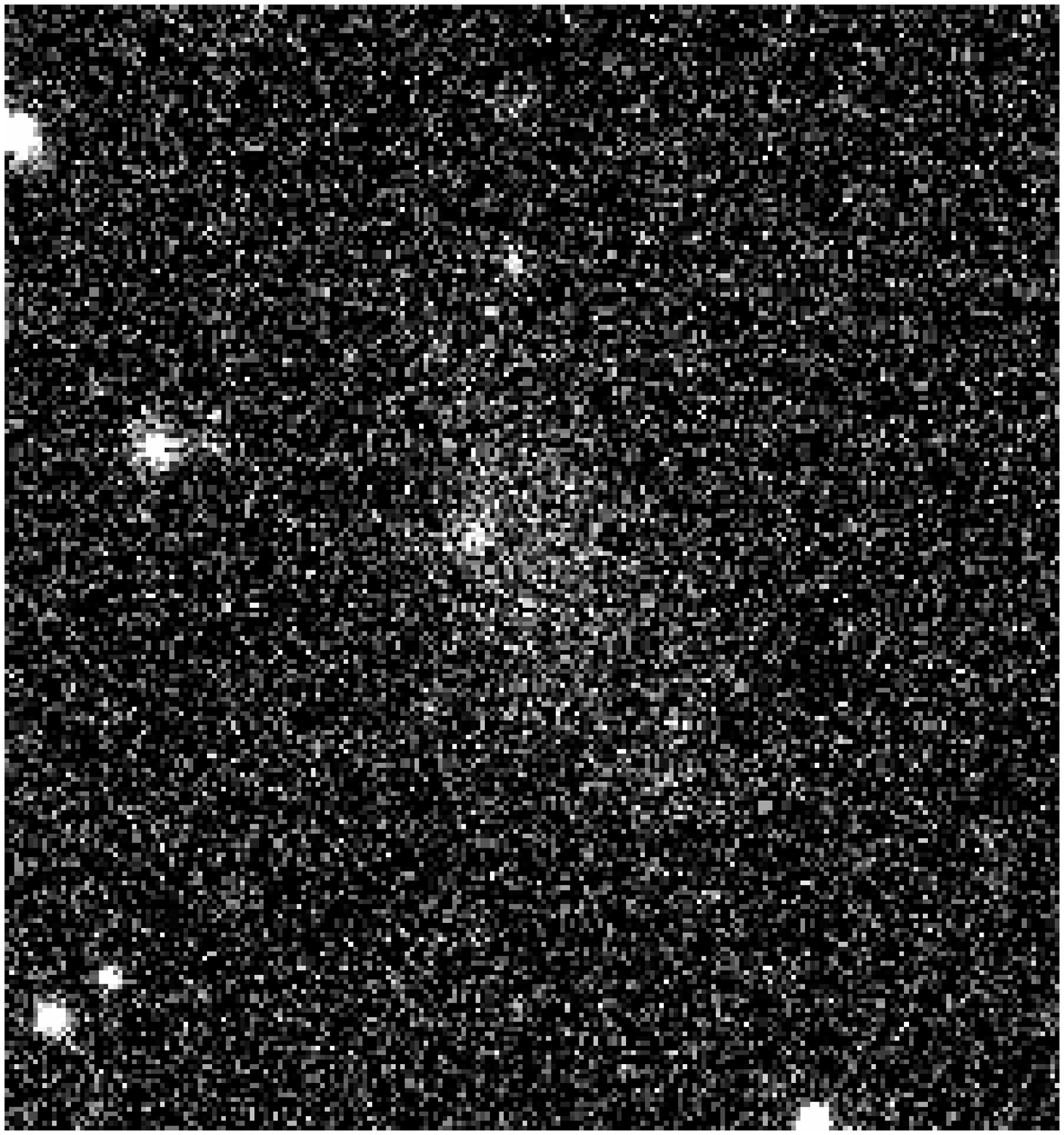}}
  \subfloat[VCC 1964 unsharp mask]{\includegraphics[width=41mm]{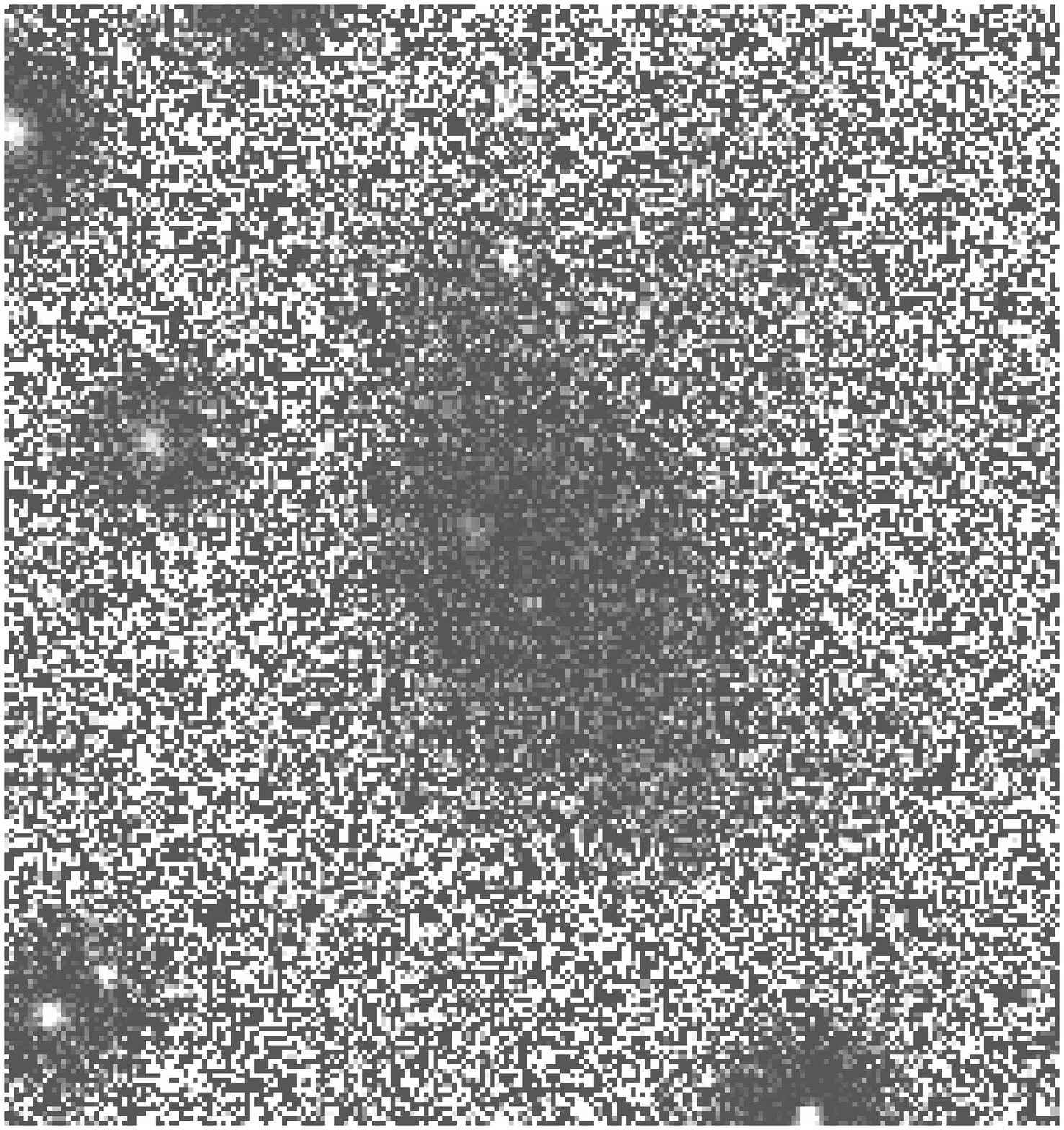}}\\
\caption[SDSS $g$-band and unsharp masks for HI-detected dEs]{SDSS $g$-band and unsharp masks for the faint dwarf ellipticals detected in HI, using circular profiles with kernel sizes of 10 pixels. Images are 1.7\arcmin across.}
\label{Unsharps}
\end{center}
\end{figure}

\begin{figure*}
\begin{center}  
  \subfloat[VCC 450 $g$ band]{\includegraphics[height=35mm]{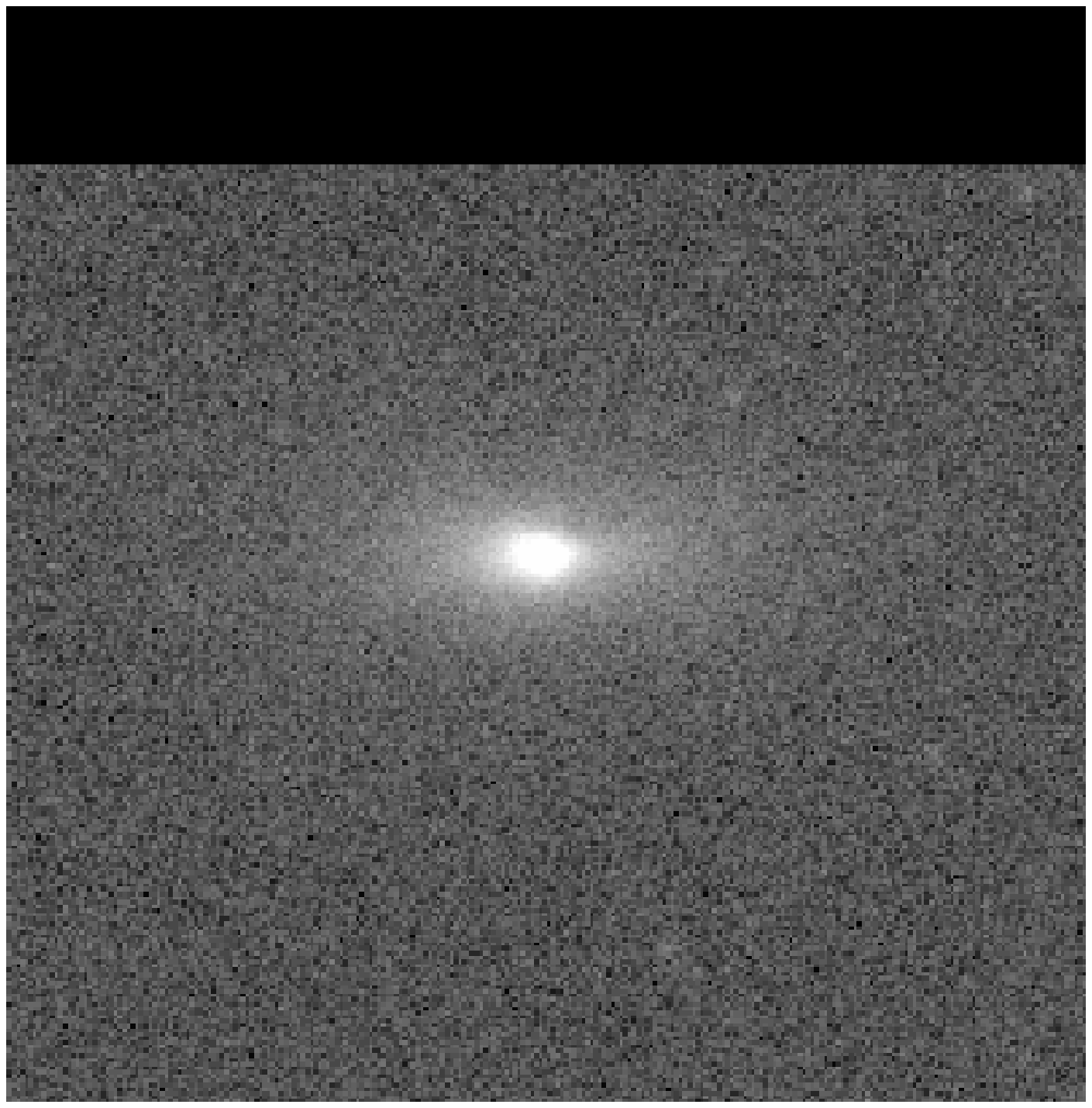}}
  \subfloat[VCC 450 residual]{\includegraphics[height=35mm]{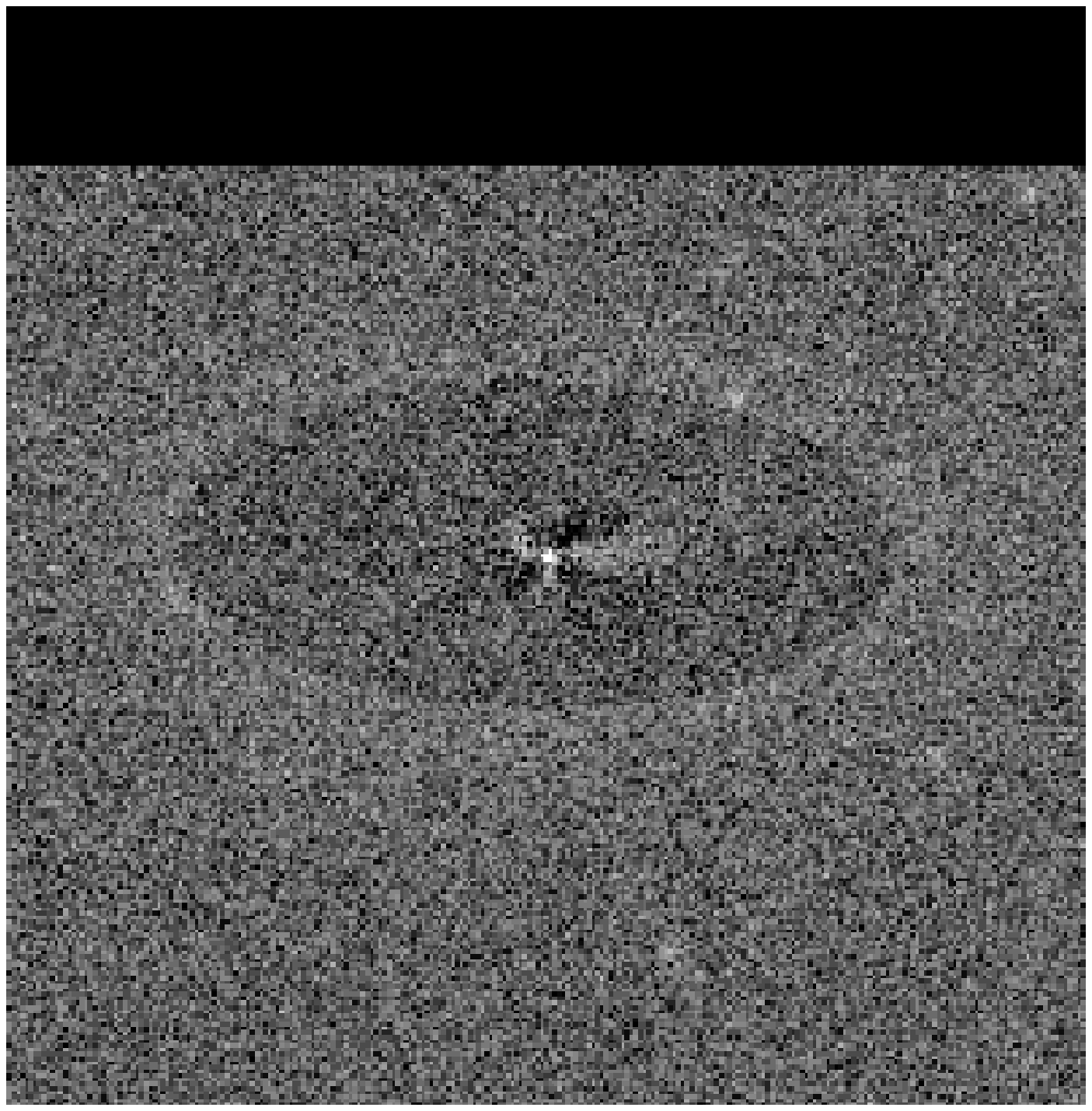}}
  \subfloat[VCC 450 unsharp]{\includegraphics[height=35mm]{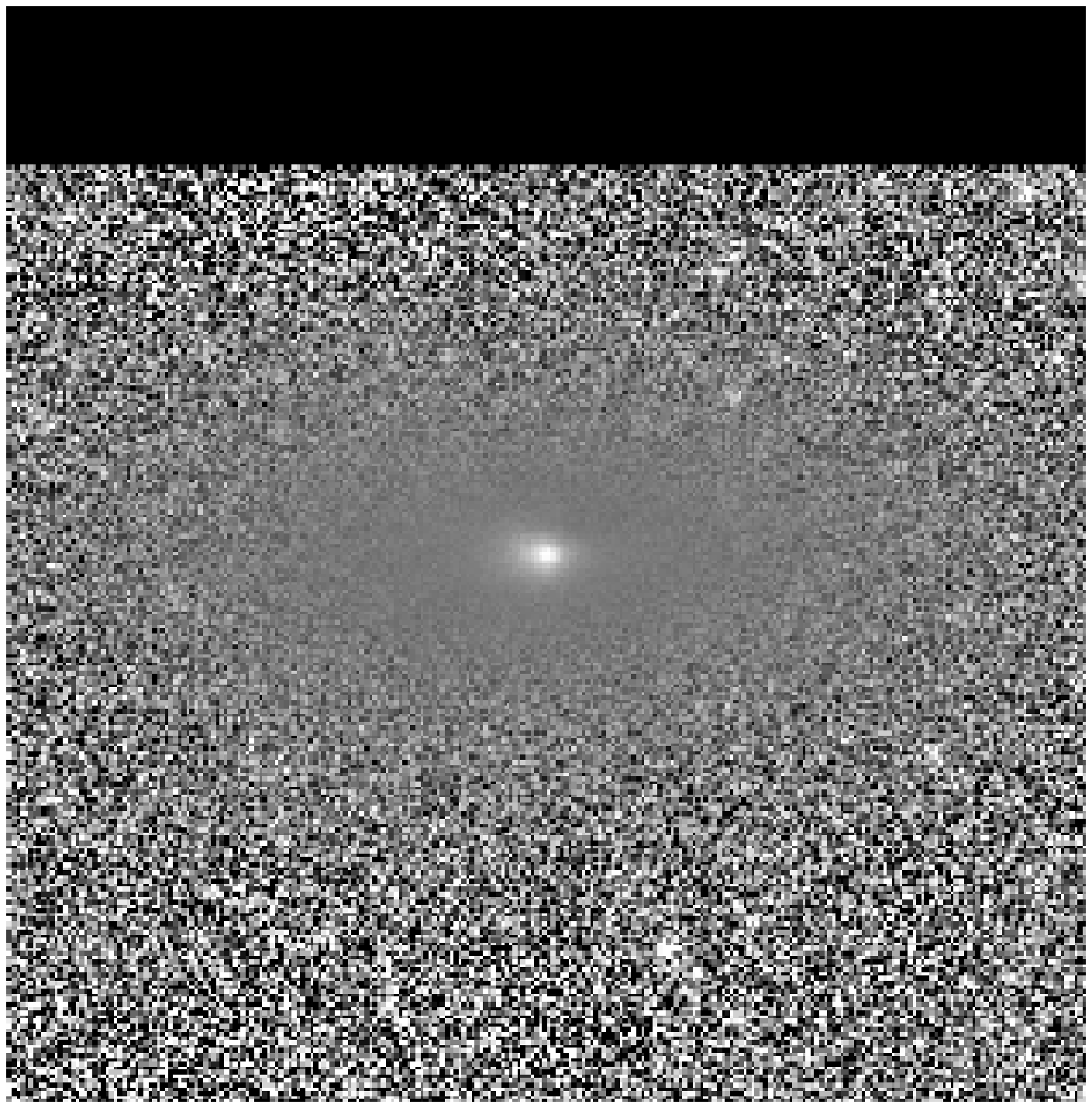}}\\   
  \subfloat[VCC 180 $g$ band]{\includegraphics[height=35mm]{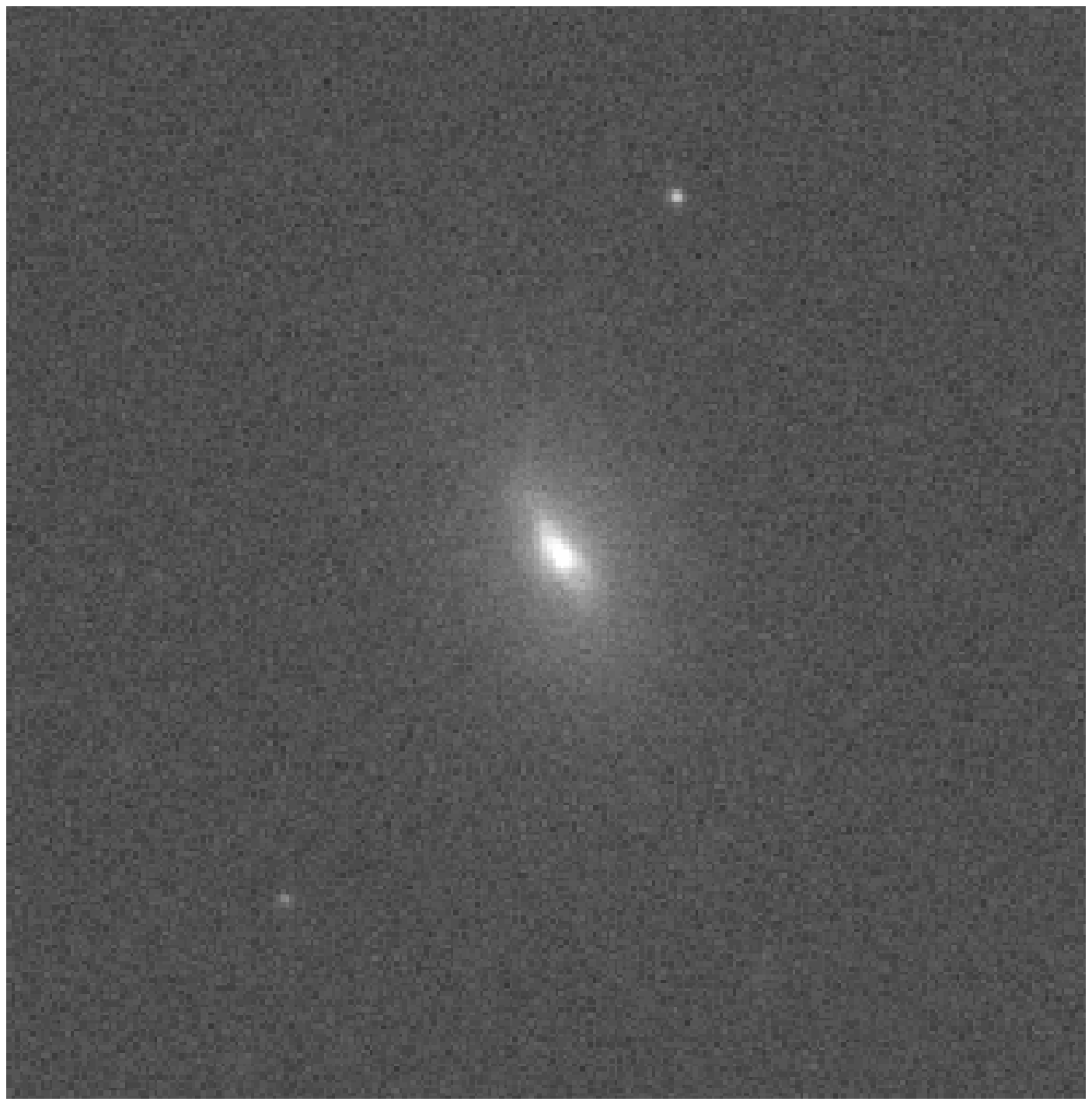}}
  \subfloat[VCC 180 residual]{\includegraphics[height=35mm]{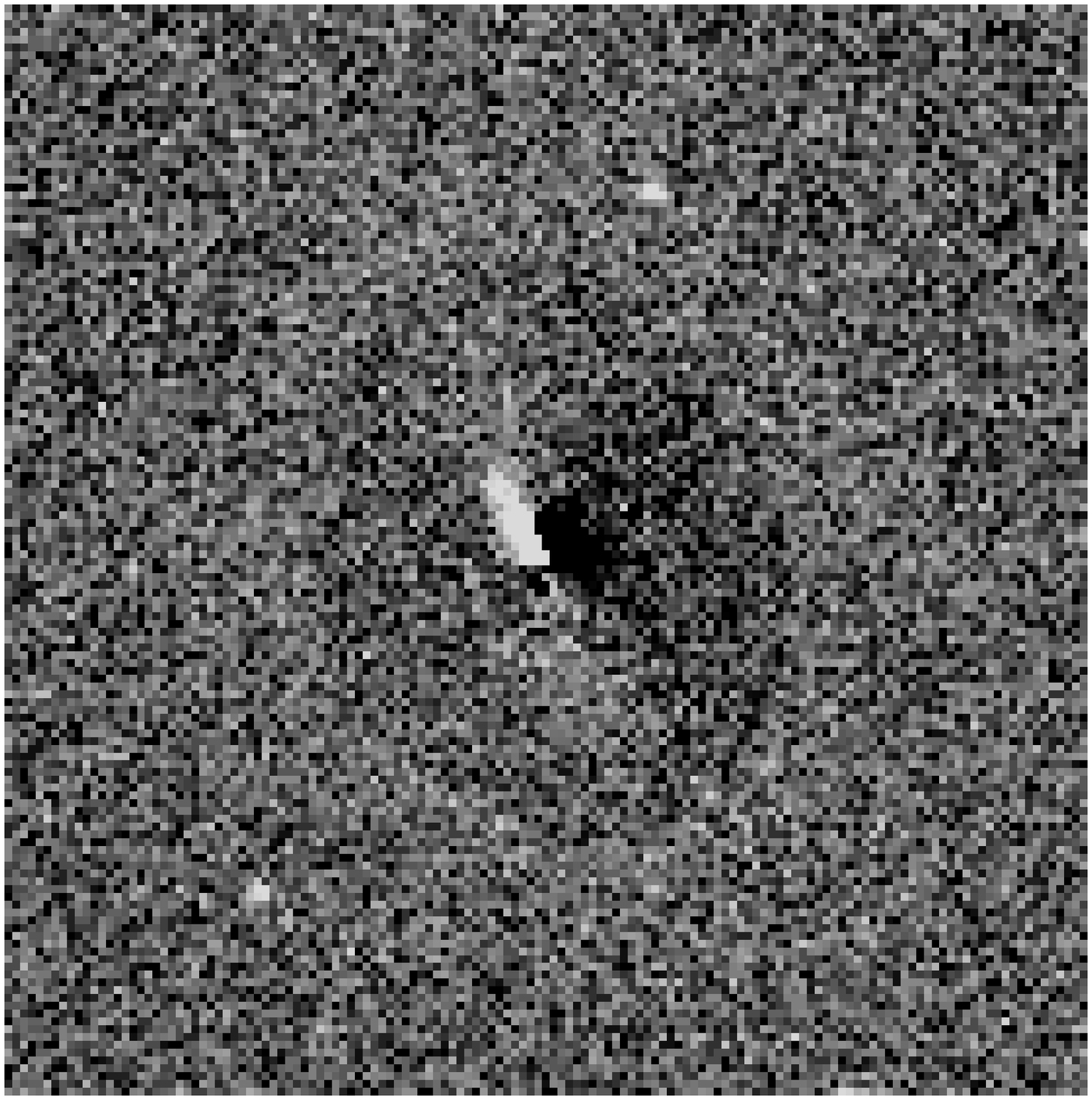}}
  \subfloat[VCC 180 unsharp]{\includegraphics[height=35mm]{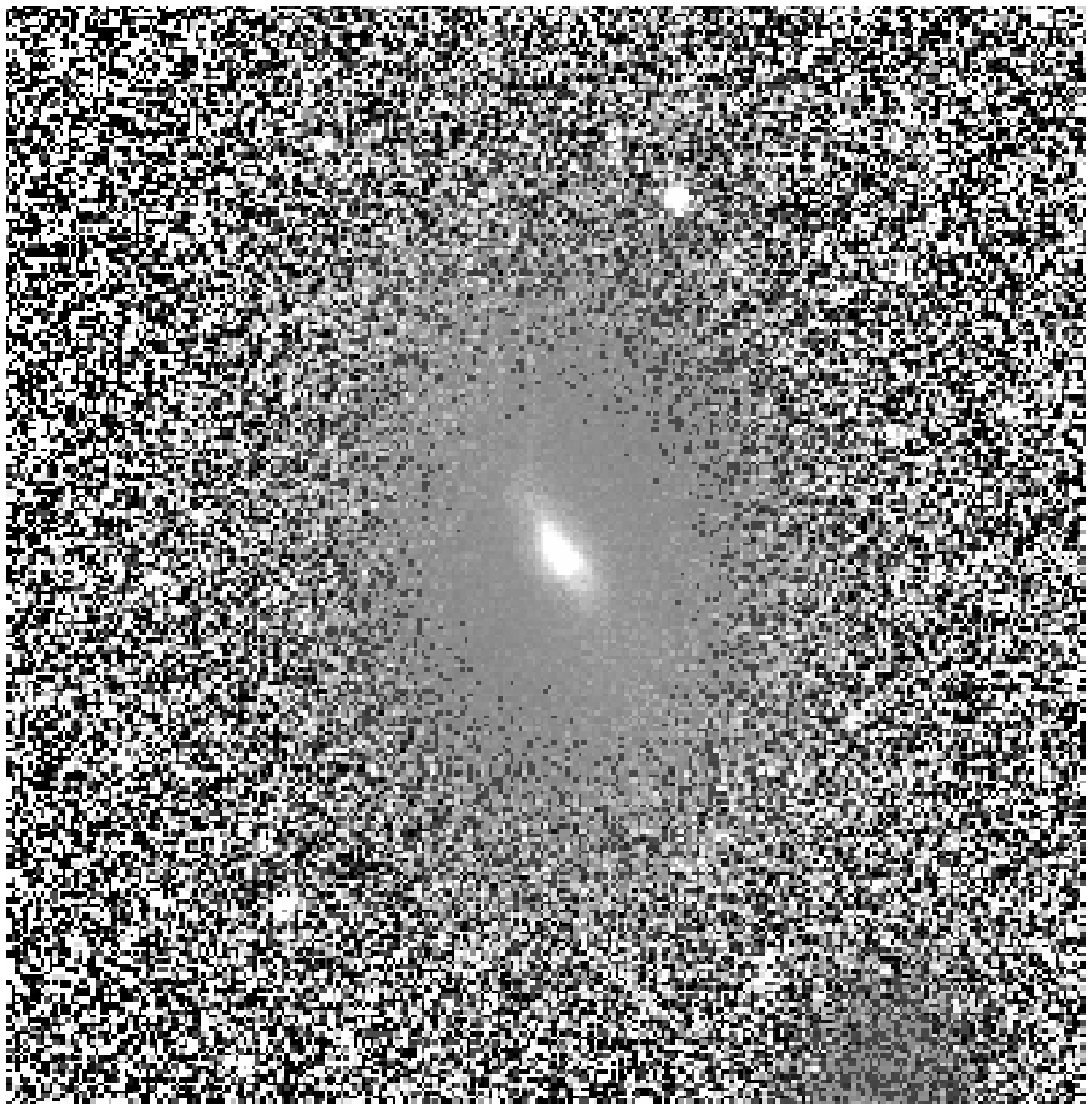}} \\ 
  \subfloat[VCC 611 $g$ band]{\includegraphics[height=37mm]{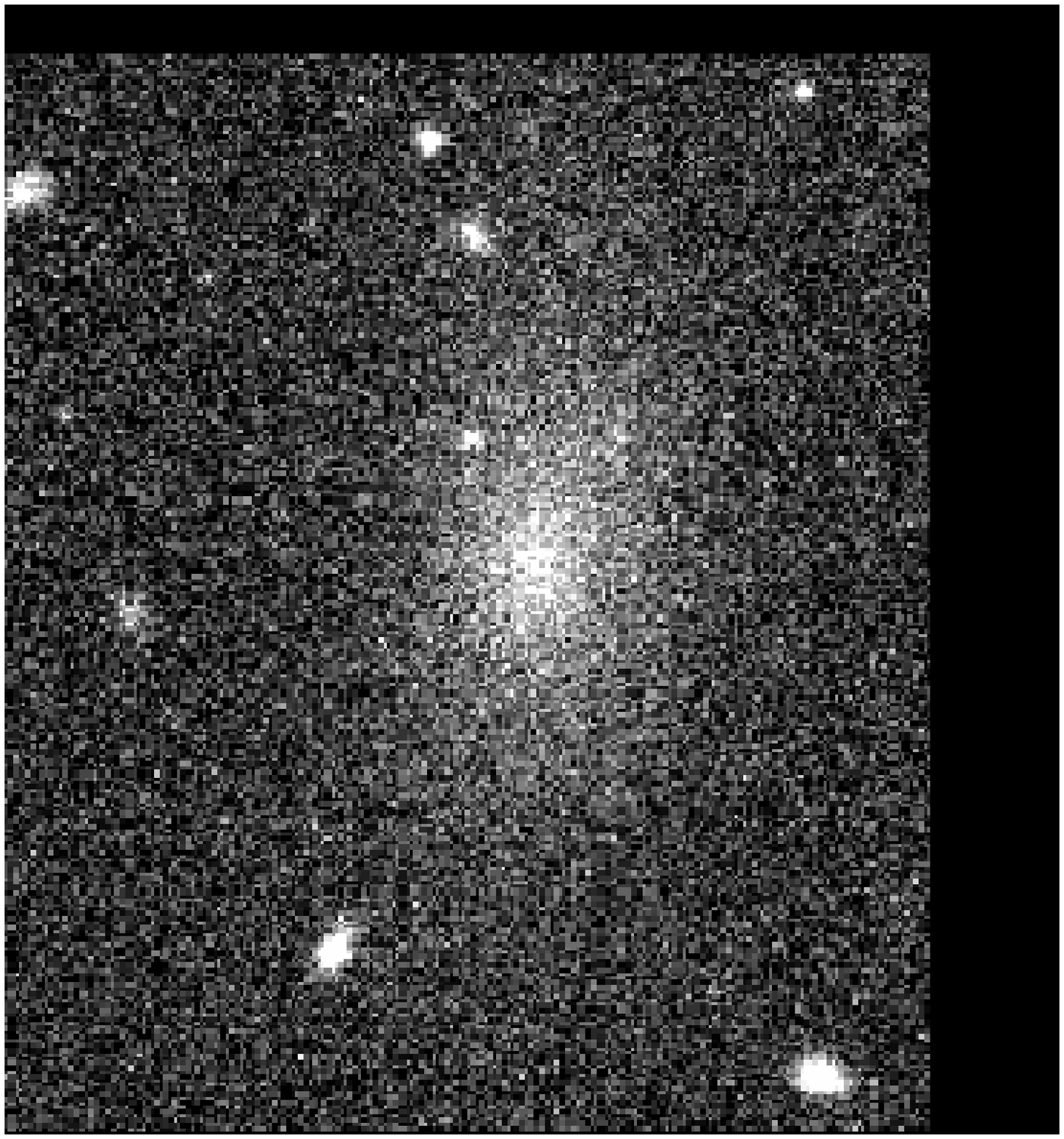}}
  \subfloat[VCC 611 residual]{\includegraphics[height=37mm]{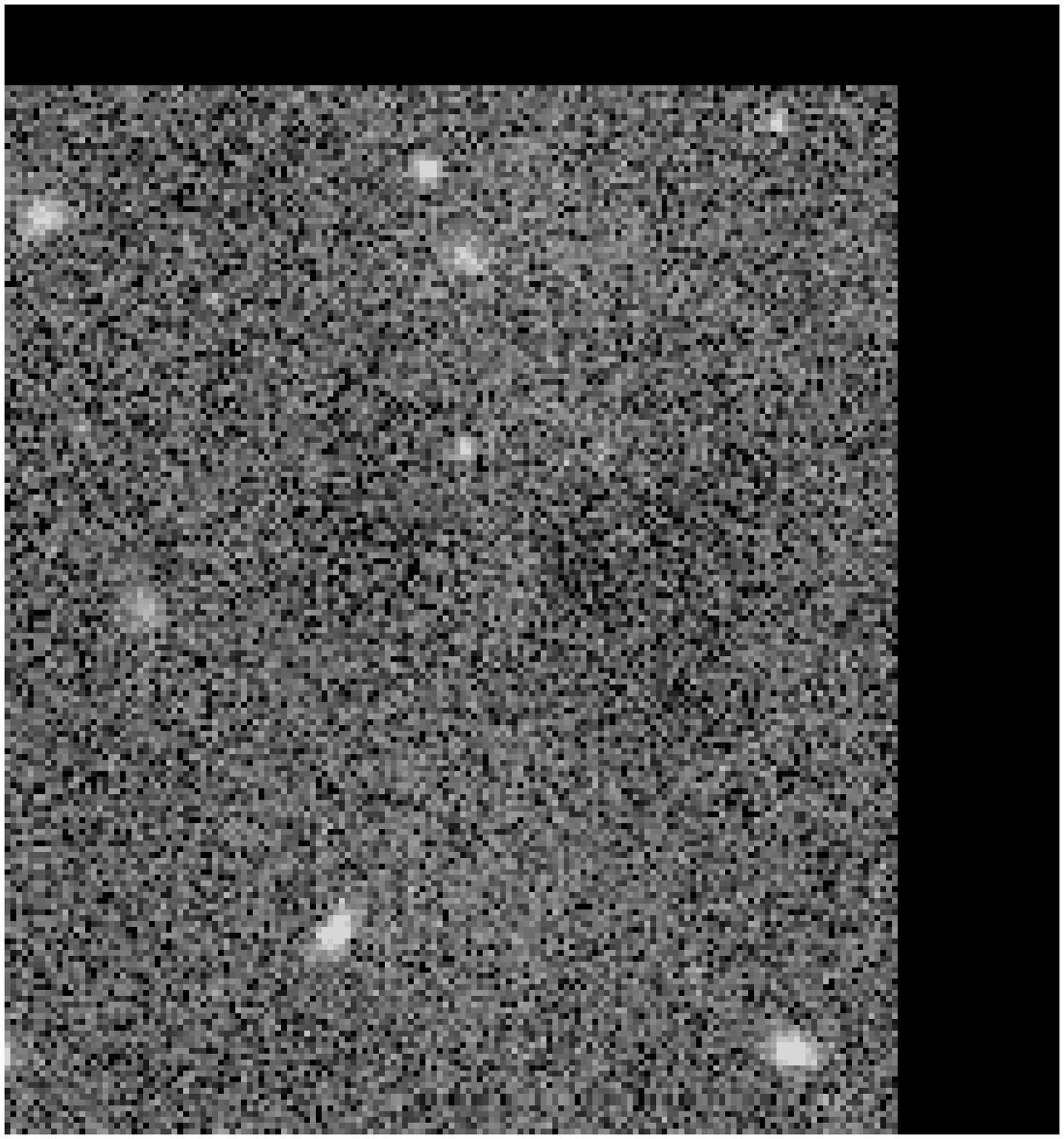}}
  \subfloat[VCC 611 unsharp]{\includegraphics[height=37mm]{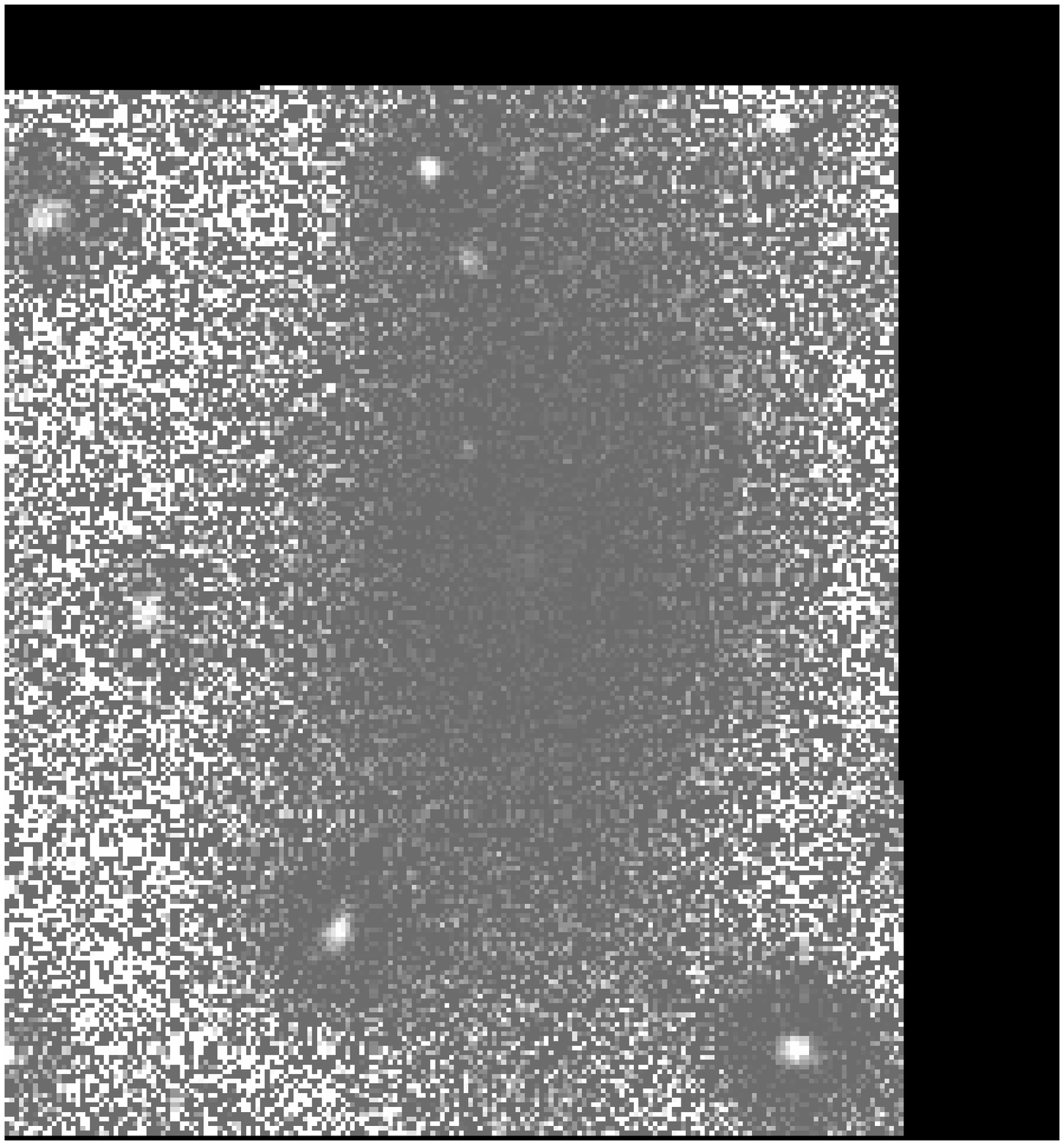}} \\    
  \subfloat[VCC 94 $g$ band]{\includegraphics[height=35mm]{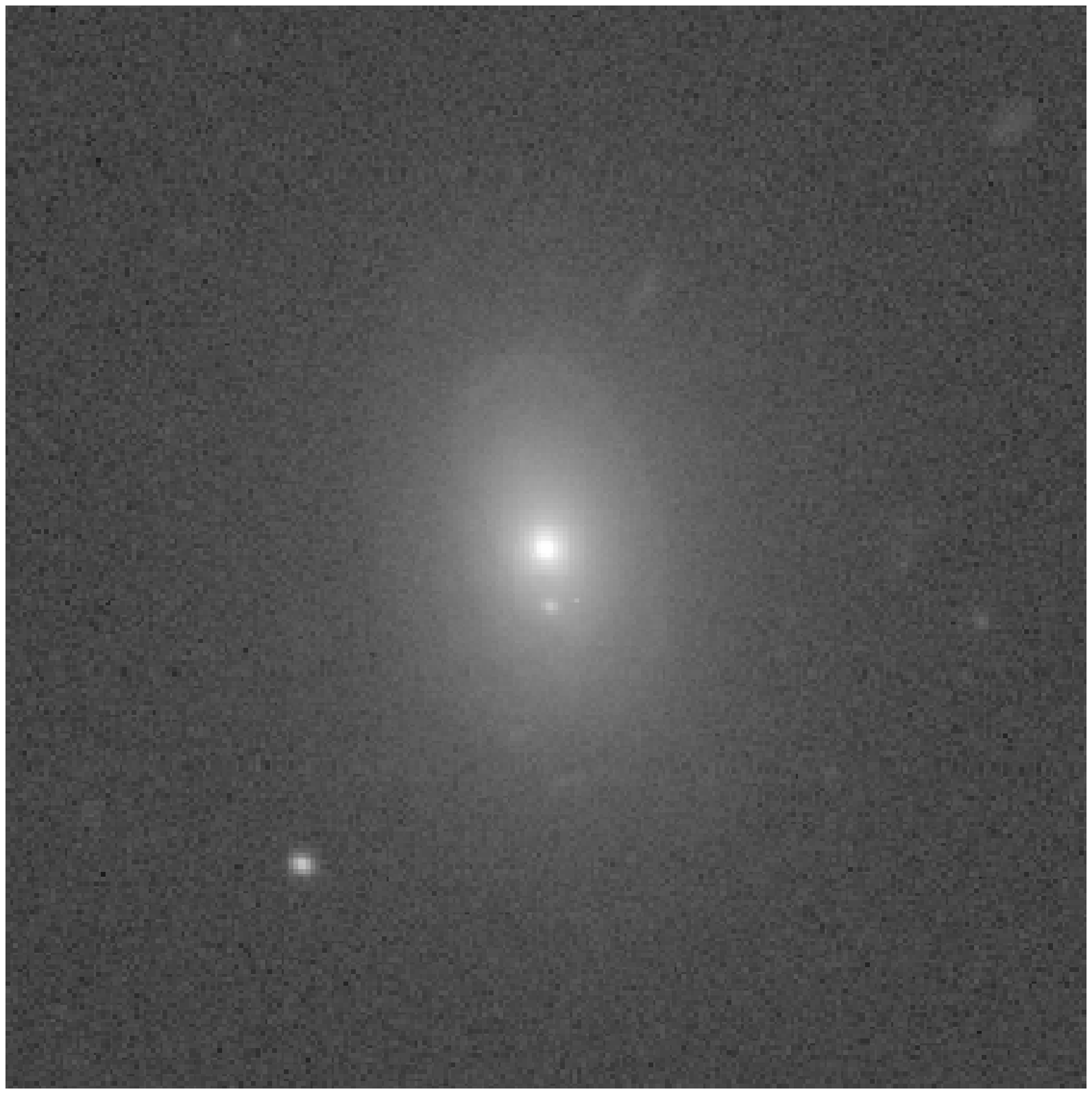}}
  \subfloat[VCC 94 residual]{\includegraphics[height=35mm]{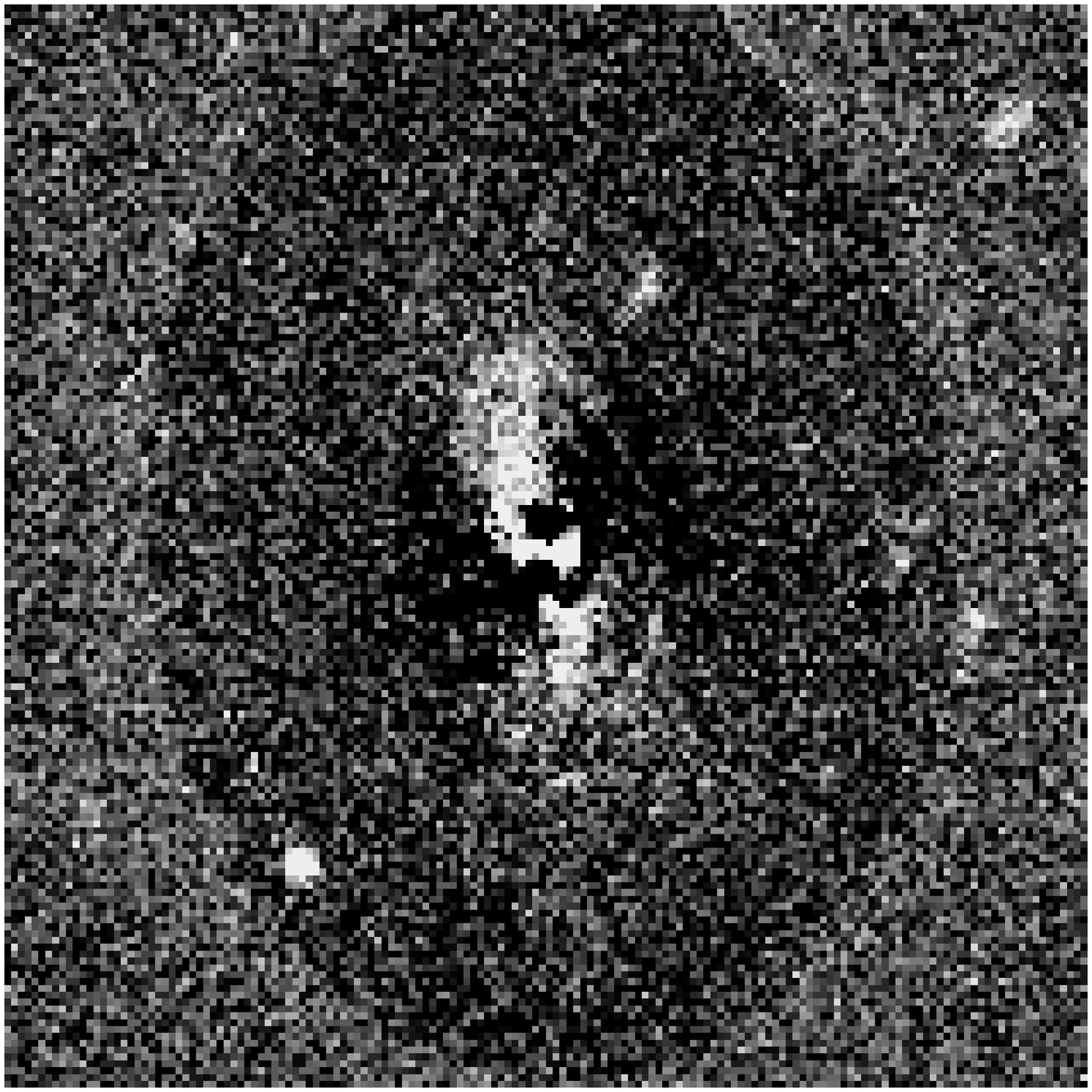}}
  \subfloat[VCC 94 unsharp]{\includegraphics[height=35mm]{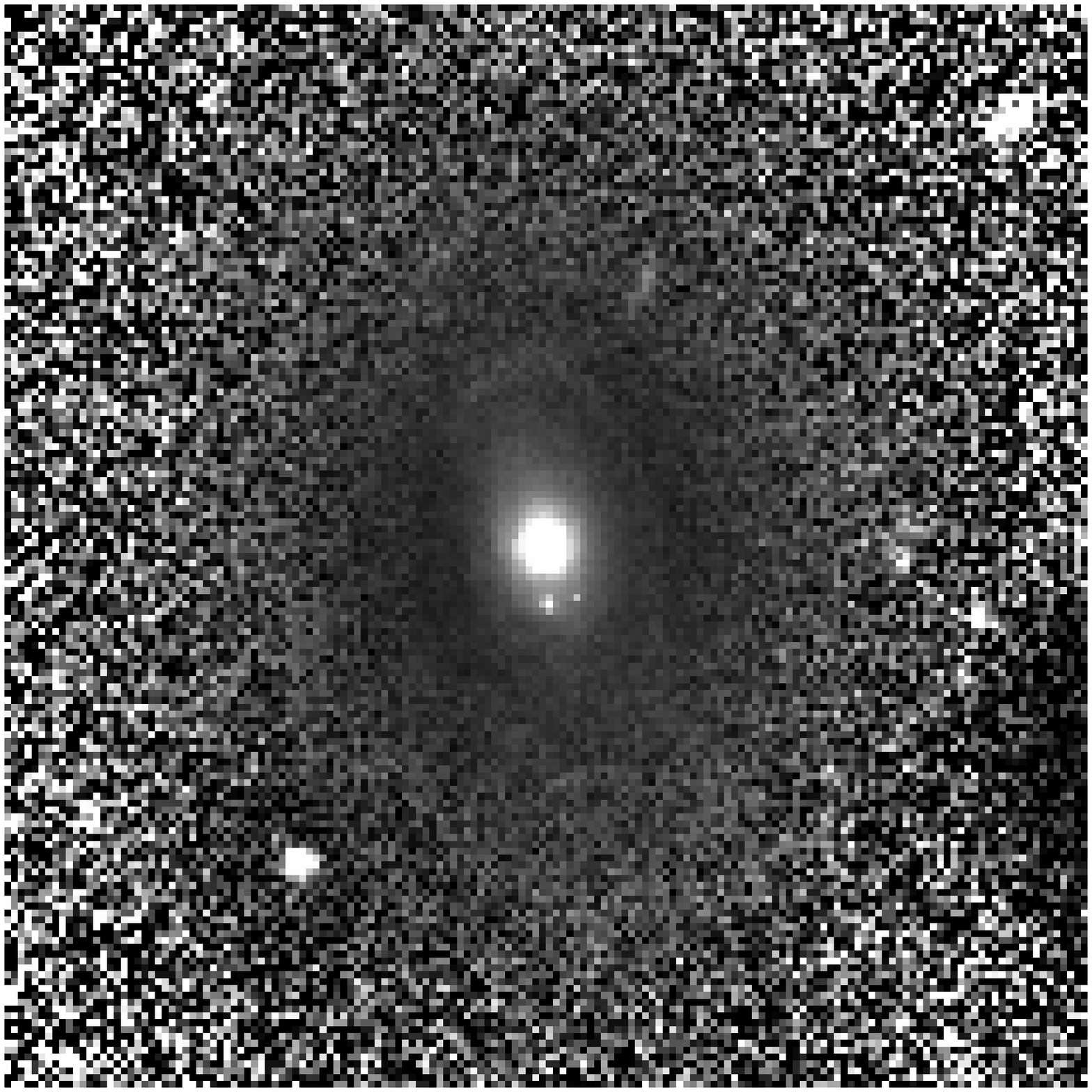}}\\    
\caption[SDSS $g$-band, residual images and unsharp masks for the HI-detected S0s]{SDSS $g$-band, residual images and unsharp masks (using circular profiles with kernel sizes of 20 pixels) for the S0s detected in HI, as well as the brighter dE VCC 611. Images are 1.7\arcmin across. The model from \textsc{iraf} does not quite extend to the outer edge of each galaxy, hence the appearance of a cutoff ring in the residual images. The type 2 VCC 94, which appears smooth in the $g$ band, is shown for comparison.}
\label{Residuals}
\end{center}
\end{figure*}

The second method is to create an unsharp mask. This is done by Gaussian smoothing the image with various kernel sizes (using the the \textit{gauss} task in \textsc{iraf}) in order to search for features of different sizes. Kernel sizes of 3, 5, 10 and 20 pixels are used. The original image is then divided by the smoothed image using \textit{imarith}. This has the advantage of being possible for all objects (but there is still no benefit in attempting this for VCC 758). Adopting the procedure in L06, both circular and elliptical unsharp mask images are created. This is because both can give rise to artificial structures and serve as a complimentary check on the other. Residual images and unsharp masks are shown in figures \ref{Unsharps} and \ref{Residuals}.

Most of the detections show little evidence of structure. As a comparison, the type 2 galaxy VCC 94 is also shown. With clear spiral structure evident in both the unsharp mask and the residual image, this demonstrates that the techniques used here are able to reveal extended features not visible in the SDSS RGB images. A caveat is the low luminosity of the dwarf ellipticals, but this is not an issue for the lenticulars.

VCC 180 does show a bar-like feature in its center, which is seen most clearly in the unsharp mask but is also visible in the residual image. Only VCC 180 and VCC 758, out of the 7 detected ETGs, seem to have any kind of unusual morphological feature, and that within VCC 180 is relatively small. There is certainly nothing approaching that seen in L06 figure 7, or VCC 94, which shows clear spiral structure extending over the whole diameter of the galaxy. However, 2 of the 4 detected S0s show some signs of structure, while it is very difficult to properly assess the dEs VCC 190 and VCC 1964 since they are very faint. 

Further differences between the detected ETGs become apparent when we consider the detection rates. While 25\% of the S0s listed in the VCC in this region are detected in HI, the fraction of detected dEs is less than 3\%. Arguably the latter fraction could be even lower, given the various problems presented by VCC 190, 1964 and 1394. In summary, while the numbers are small, the S0s are detected in HI relatively often and show some signs of structure, whereas the morphologically smooth dEs are almost never detected.

More generally, the VCC lists a total of 187 ETGs in this region, giving an HI-detected fraction of 3.7$\%$, broadly consistent with 2.3$\%$ reported by \citealt{aaETG} who used ALFALFA data. This latter figure corrects for the VCC brightness completeness limit, the contamination of background galaxies in those VCC members without redshift data, and those non-detections too close to the strong continuum source M87 (where they could not be detected). If the same corrections are applied to the VC1 region, the detected fraction rises to 8.8$\%$. This significantly higher detection rate is at least partly the result of the increased sensitivity of AGES relative to ALFALFA. It should be stressed that these are small number statistics, and the effects of cosmic variance are also important.

To attempt to improve our statistics we employ stacking. In this region there are 56 undetected early-type VCC galaxies whose spectra can be averaged to give an increase in sensitivity. The suitability of these objects rests on their having optical spectra (so that the spectra are all centered on the velocity of each galaxy, ensuring that the galaxies are averaged with each other when combined) and no HI-detected galaxies (including the Milky Way) or strong continuum sources nearby.

\begin{figure}
\begin{center}  
  \subfloat[All undetected ETGs. 56 objects, rms =  0.0876 mJy]{\includegraphics[width=65mm]{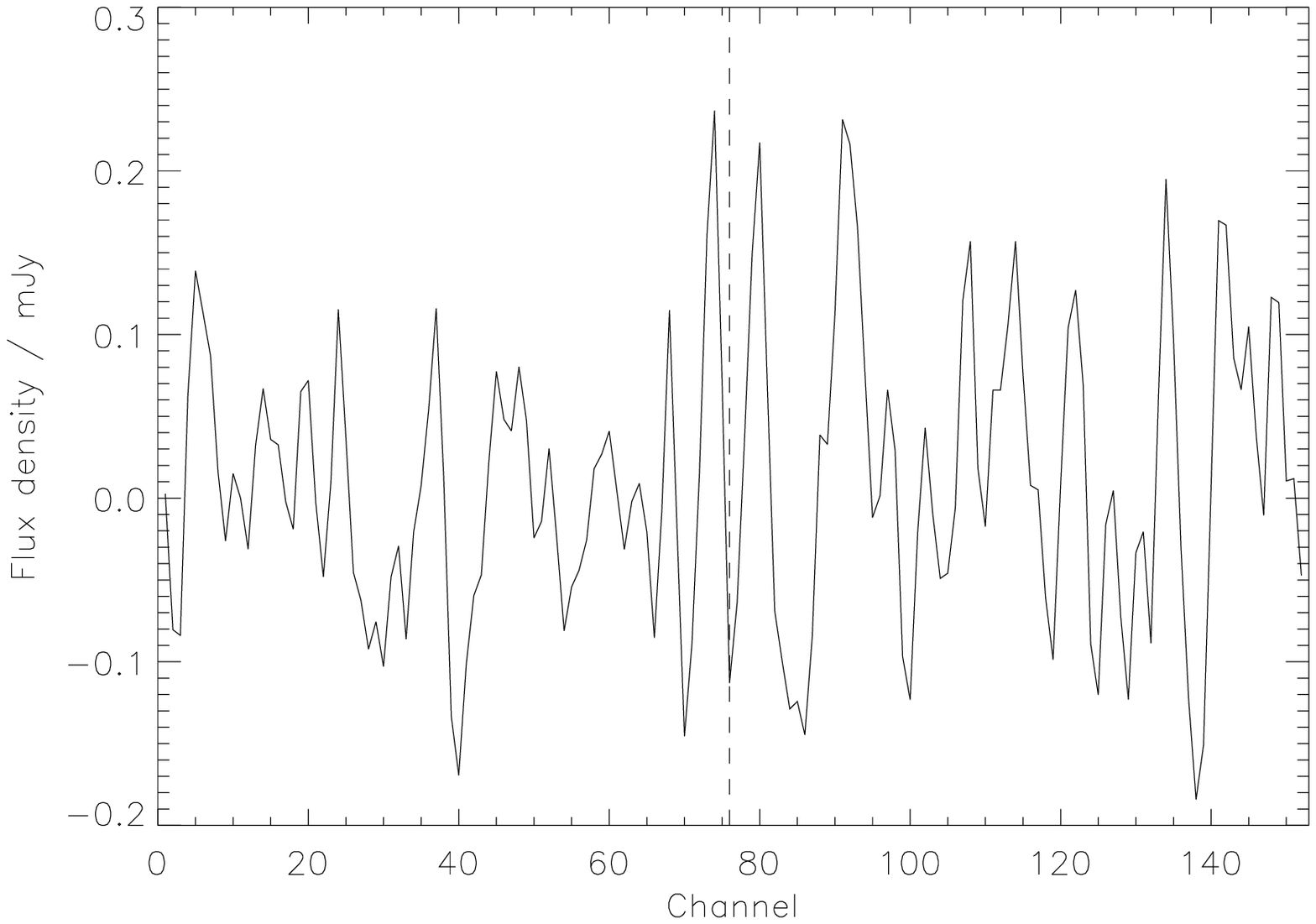}}\\
  \subfloat[All ETGs at 17 Mpc distance. 31 objects, rms = 0.1248 mJy]{\includegraphics[width=65mm]{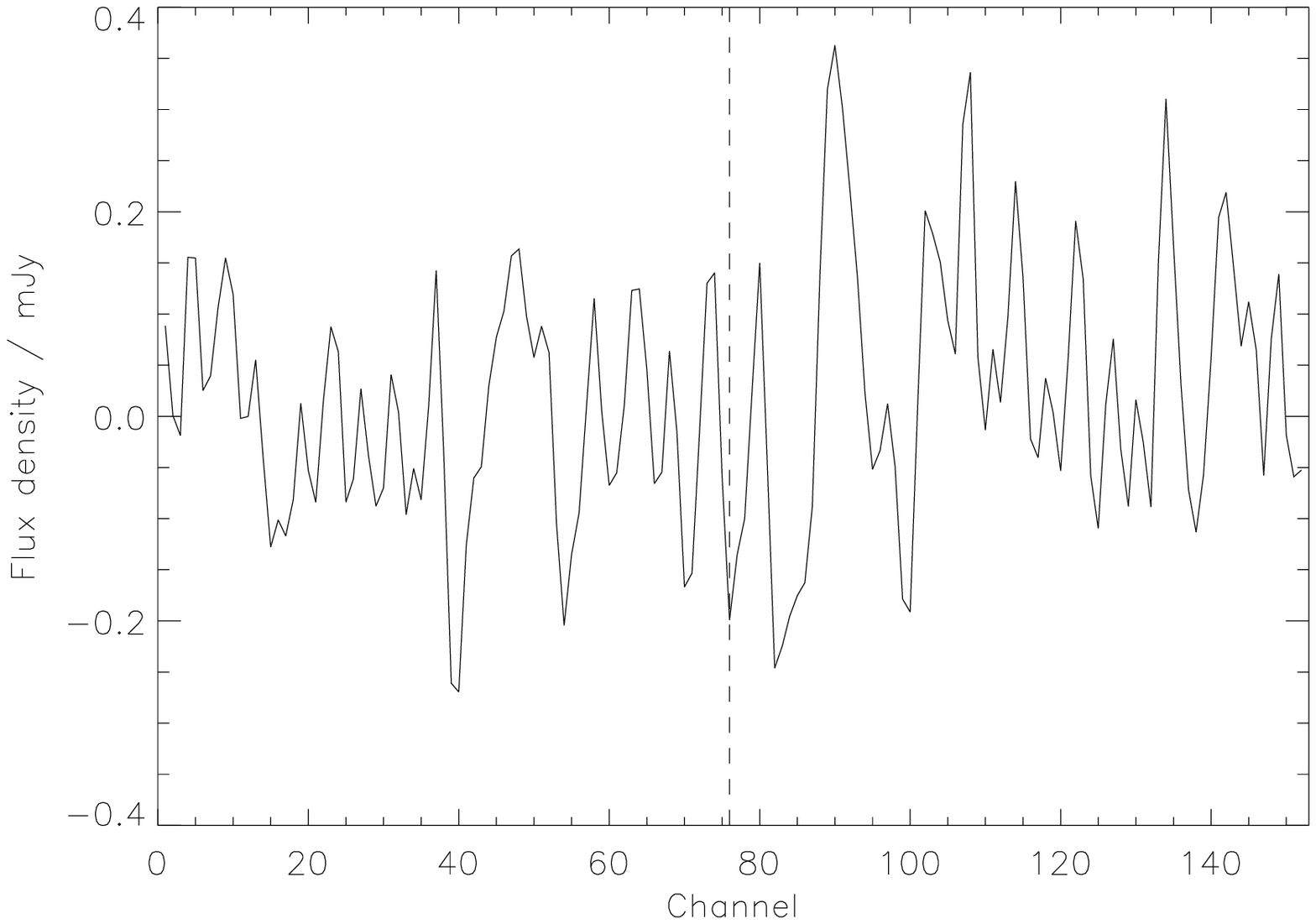}}\\ 
  \subfloat[S0s only. 15 objects, rms = 0.1877 mJy]{\includegraphics[width=65mm]{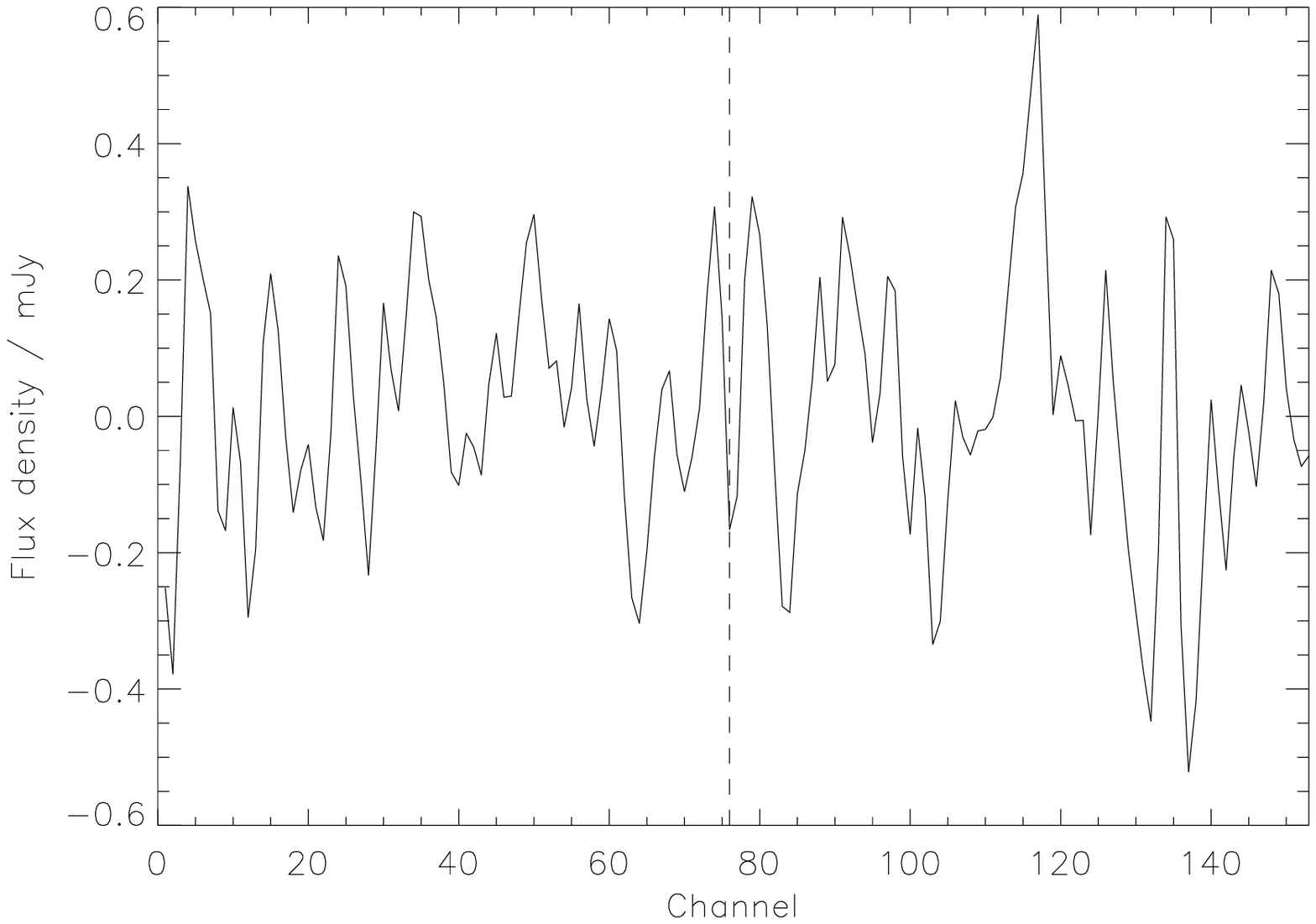}}\\
  \subfloat[dEs only. 41 objects, rms = 0.0997 mJy]{\includegraphics[width=65mm]{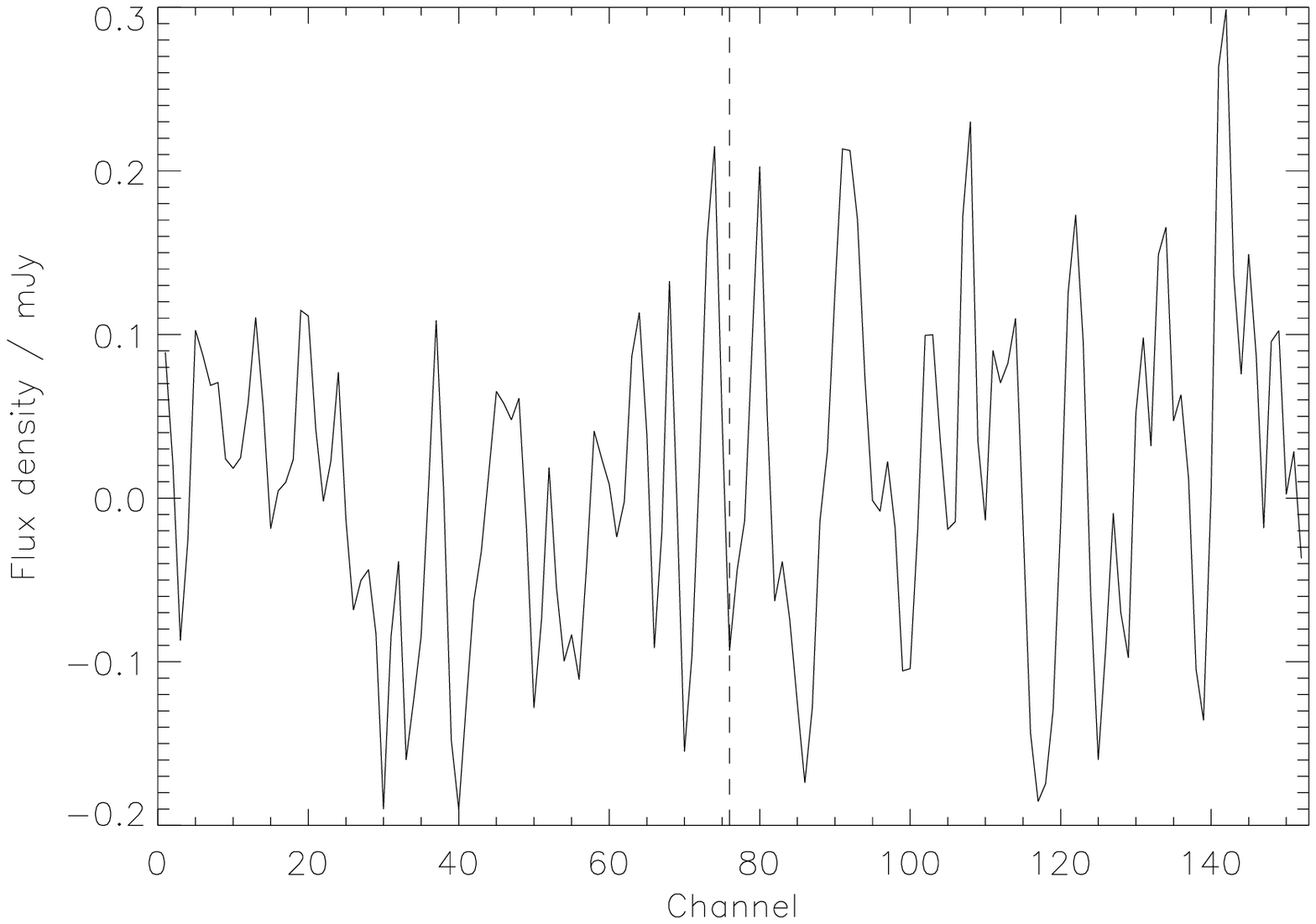}}
\caption[HI stacking of undetected ETGs]{Stacked spectra of HI-undetected ETGs. All spectra are weighted by rms and have been shifted so that their central channel, where a detection would be expected, is equivalent to 76, highlighted by the dashed line.}
\label{GalaxyStacking}
\end{center}
\end{figure}

Figure \ref{GalaxyStacking} shows the results of stacking the viable objects in various ways. The spectra are weighted by their rms, in order that the noisier spectra do not unduly raise the final rms level attained. We use the same weighting as \citealt{Fabello} :

\begin{equation}S_{stack} = \frac{\sum_{i=0}^N S_{i}.w_{i}}{\sum_{i=0}^N w_{i}}\end{equation}

Where the weighting $w_{i} = rms^{-2}$ and $S_{stack}$ is the resultant stacked spectrum. In figure \ref{GalaxyStacking} we separately combine our whole sample, only the lenticulars, only the dwarf ellipticals, and only the objects at 17 Mpc distance (since this increases the mass sensitivity achieved for a given rms). None of these procedures results in a detection. 

The lowest rms reached by stacking is 0.0876 mJy, equivalent to a mass sensitivity of 1.2$\times$10$^{6}$ M$_{\odot}$ at 17 Mpc (assuming a 4$\sigma$ top-hat detection with a width of 50 km/s). There is a slight complication that not all of the objects stacked in this sample are actually at 17 Mpc distance, but when we restrict the stacking to only the closest objects the mass sensitivity decreases only slightly to 1.7$\times$10$^{6}$ M$_{\odot}$. The lack of any detections by stacking to such high sensitivities implies that deep direct observations (of Virgo Cluster galaxies) will also fail to detect more ETGs in HI, which is to some extent borne out by the observations of \citealt{C03}.

It seems then that galaxies which are detected in HI are fundamentally different to those which are not. Our results, together with those of \citealt{C03}, imply that those galaxies which are not detected in HI are really devoid of gas. In contrast \citealt{Fabello} found that field galaxies not detected directly in HI nonetheless do possess some gas content, with stacking of ALFALFA data (approximately 4 times less sensitive than AGES) for 50 galaxies producing a detection. While field galaxies may lose gas, those in the cluster can become entirely depleted. We have already seen that the HI in the Virgo Cluster is also almost always associated with LTGs, while the ETGs are almost always absent from the HI detections. In the following sections we explore in more detail the properties of HI-detected and undetected galaxies.

\subsection{Colours}
\label{sec:colours}

\begin{figure*}
\begin{center}  
  \subfloat[Cluster members]{\includegraphics[width=85mm]{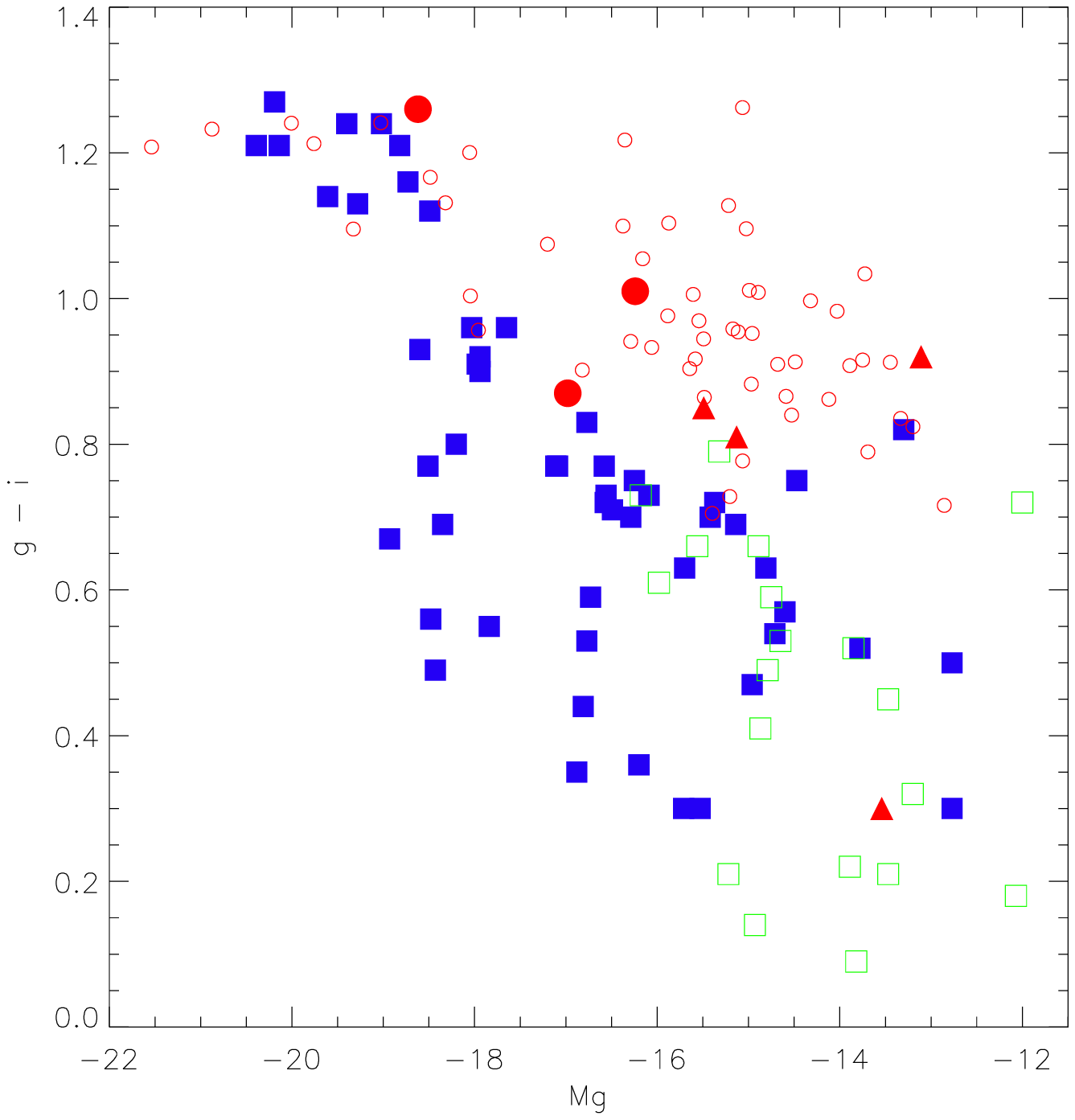}}
  \subfloat[Cluster members and background objects detected in HI.]{\includegraphics[width=85mm]{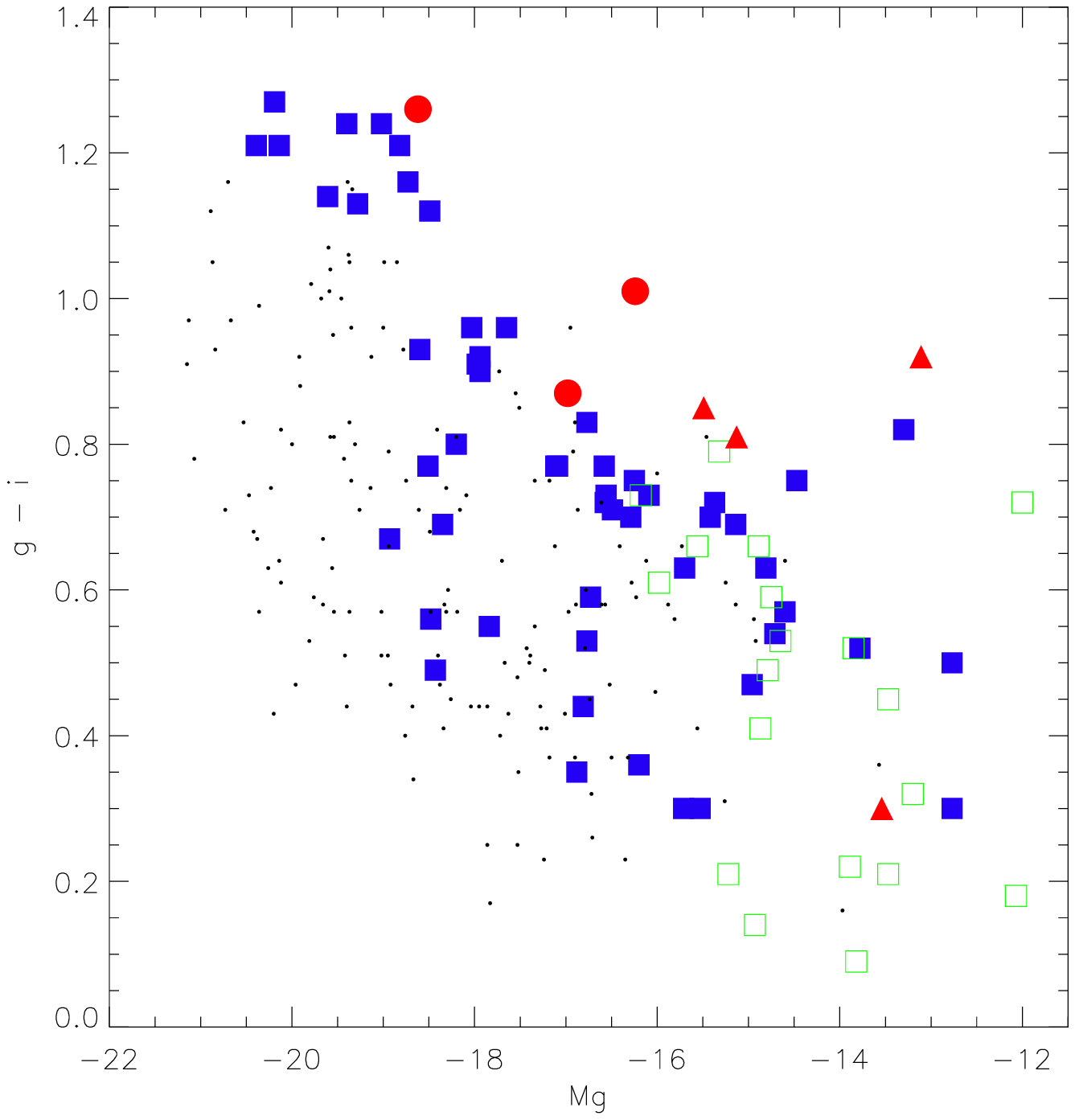}}
\end{center}
\caption[Optical colour-magnitude diagrams for cluster members and background objects in VC1]{Optical colour-magnitude diagrams for cluster members and background objects. Blue filed squares are late-type objects that are members of the VCC, open green squares show non-VCC detections. Small open red circles are VCC members not detected in HI. Large filled red circles are S0s detected in HI, while red triangles are dEs detected in HI. Black points are background HI detections. No correction was made for internal extinction.}
\label{CMDs}
\end{figure*}

Colour-magnitude diagrams for the HI detections are shown in figure \ref{CMDs}. As a comparison, photometry was also performed for the undetected galaxies with optical redshift measurements (76 objects) as well as the background HI-detected objects (154 objects). The requirement for an optical redshift ensures that the objects can be very reliably determined to be cluster members or background objects. In these plots only sure HI detections with a unique optical counterpart are shown. The majority of the undetected cluster members are, as we have seen, unambiguously early-type objects.

Objects in Virgo appear systematically redder than those in the background of the same luminosity, which could suggest an environmental influence of the cluster. It is difficult to disentangle this possibility from selection effects - objects in the background are selected by HI, and being more distant, must be richer in HI and therefore more likely to be bluer in comparison to \textit{any} nearby detections, not just those in the cluster (see next section).

As expected, the non-detections (being ETGs) define a red sequence, while the detections trace a blue sequence. The non-VCC galaxies detected in HI in the cluster are almost exclusively faint, blue objects - small, gas -rich objects (see section \ref{sec:gasratio}) are easier to detect from their HI content than in an optical survey.

Only one of the non-VCC detections appears to lie well on the red sequence. In this case the optical candidate counterpart is small and faint, and may be a misidentified background object (although there are no obvious alternative counterparts). Conversely, one of the detected dwarf ellipticals is seen to lie on the blue sequence. This very faint object (VCC 1964) appears to be irregular in the SDSS images, so this may be a morphological misclassification.

The other detected early-types are all clearly members of the red sequence or in the transition region between red and blue. One other HI-detection is also seen in the red sequence, VCC 867. This is classed as a dwarf irregular, but the assignation is questionable given its low optical luminosity and red colour.

\subsection{HI mass-to-light ratio}
\label{sec:gasratio}
\begin{figure}
\begin{center}y
\includegraphics[width=84mm]{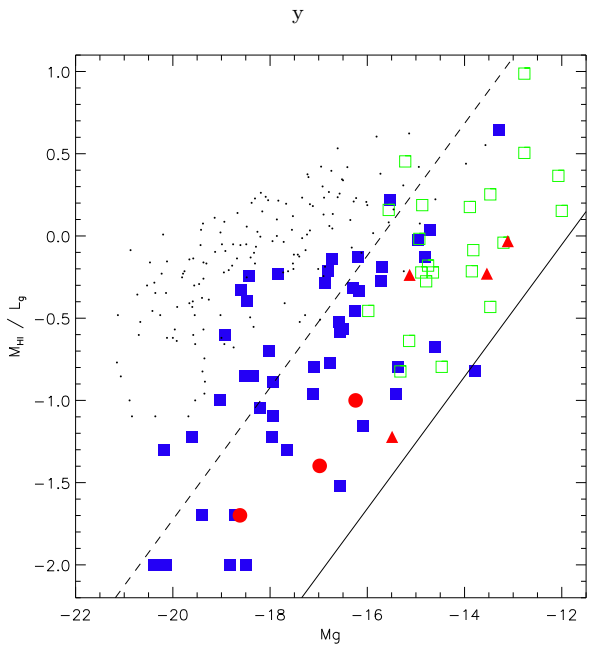}
\caption[HI mass-to-light ratio in VC1 as a function of M$_{g}$]{HI mass-to-light ratio. The lines indicate the sensitivity limits for a 4$\sigma$, 50 km/s top-hat profile at 0.5 mJy rms at  17Mpc (solid line) and 100 Mpc (dashed line). The y-axis is in logarithmic units. Blue squares are late-types, red are S0s and green are unclassified objects. Red triangles are dEs and black crosses are background detections.}
\label{VC1MHILG}
\end{center}
\end{figure}

The HI mass-to-light ratio varies as a function of absolute magnitude, as shown in figure \ref{VC1MHILG}. Fainter galaxies tend to be more gas-rich than brighter galaxies. This is at least partially a sensitivity effect - fainter galaxies require a higher gas fraction for the gas mass to be above the sensitivity limit. This does not explain why brighter galaxies do not have equally high mass-to-light ratios, which may perhaps relate to a ``saturation'' density of HI above which all gas is molecular (\citealt{leroy}).

Galaxies within the Virgo cluster generally have lower $M_{\rm{HI}}$/$L_{g}$ ratios than background galaxies of similar luminosities. This is an effect of both Malmquist bias and also related to environment, with cluster galaxies being subject to more gas loss than field objects (see section \ref{sec:def}).

It is apparent that with the exception of VCC 611, the detected dEs are relatively faint and gas-rich, of comparable $M_{\rm{HI}}$/$L_{g}$ ratio to the late-types of similar magnitude (i.e. dwarf irregulars). The detected S0s are brighter and also rather poorer in gas content compared to most other detections of similar magnitude (i.e. spirals).

\begin{figure}
\begin{center}
\includegraphics[width=84mm]{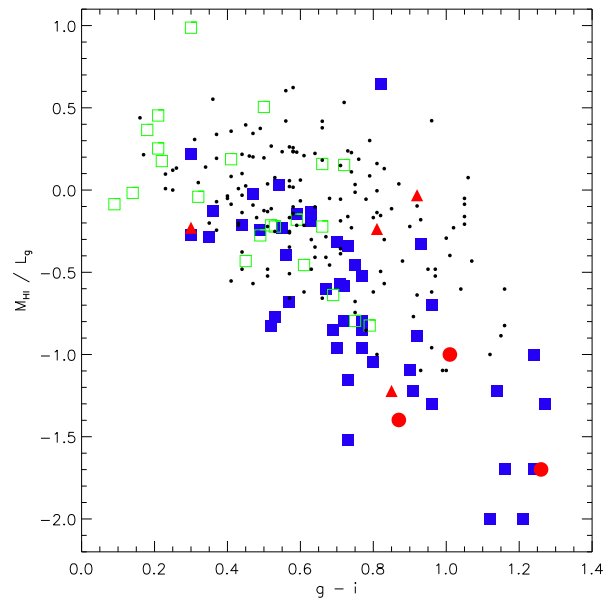}
\caption[HI mass-to-light ratio in VC1 as a function of $g - i$]{HI mass-to-light ratio (logarithmic scale) varying with $g$-$i$ colour. Blue squares are late-types, green are non-VCC galaxies and red are S0s. Red triangles are dEs, and black crosses are background galaxies.}
\label{VC1ColGasFrac}
\end{center}
\end{figure}

The effect of the variation in gas fraction at the same magnitude is perhaps reflected in the colours of the galaxies. Figure \ref{VC1ColGasFrac} plots the $M_{\rm{HI}}$/$L_{g}$ ratio against $g- i$ colour. Again there is much scatter but a clear trend - the most gas-rich galaxies are also the bluest.

It is difficult to assess whether there is any trend at all in terms of colour and gas fraction in terms of the ETGs. While the 3 detected S0s are among the reddest and most gas-poor objects, the detected dE VCC 611 is of comparable colour and gas fraction. The other 3 detected dEs are all of a very similar gas fraction to each other, but with a huge variation in colour - VCC 1964, is much bluer than the others, and as we saw in section \ref{sec:colours} it appears likely that this object is actually a dwarf irregular.

\subsection{HI deficiency}
\label{sec:def}
HI deficiency is a measure of how much gas a galaxy of given morphological type and optical diameter has lost in comparison to a similar field galaxy. Since this has been calibrated on a field sample (\citealt{haynesdef}, \citealt{giodef}) it is a more precise measure of actual gas loss than comparing relative  $M_{\rm{HI}}$/$L_{g}$ ratios. The HI deficiency is computed by the relation :
\begin{equation}D_{\rm{HI}}= \log(M_{{\rm{HI}}_{ref}}) - \log(M_{{\rm{HI}}_{obs}})\label{DefEqt}\end{equation}

Where $D_{\rm{HI}}$ is the HI deficiency, $M_{{HI}_{ref}}$ is the HI mass of the reference galaxy, and  $M_{{HI}_{obs}}$ is the HI mass of the observed galaxy. A galaxy with a deficiency of 1.0 therefore has 10\% of the gas of an equivalent field object, with a deficiency of 2.0 it would have 1\%, etc. The expected HI mass can be calculated by a linear equation, using the parameters of \citealt{b99} :
\begin{equation}\log(M_{{\rm{HI}}_{ref}}) = a + b\times \log(d)\label{HIref}\end{equation}
Where $a$ and $b$ depend upon the morphological type and $d$ is the optical diameter in kpc. The $a$ and $b$ parameters of field galaxies have only been calculated for spiral galaxies. We avoid calculating deficiencies for ETGs as they are rarely detected in HI so it makes little sense to speak of deficiency for these objects.

\begin{figure}
\begin{center}
\includegraphics[scale=0.4]{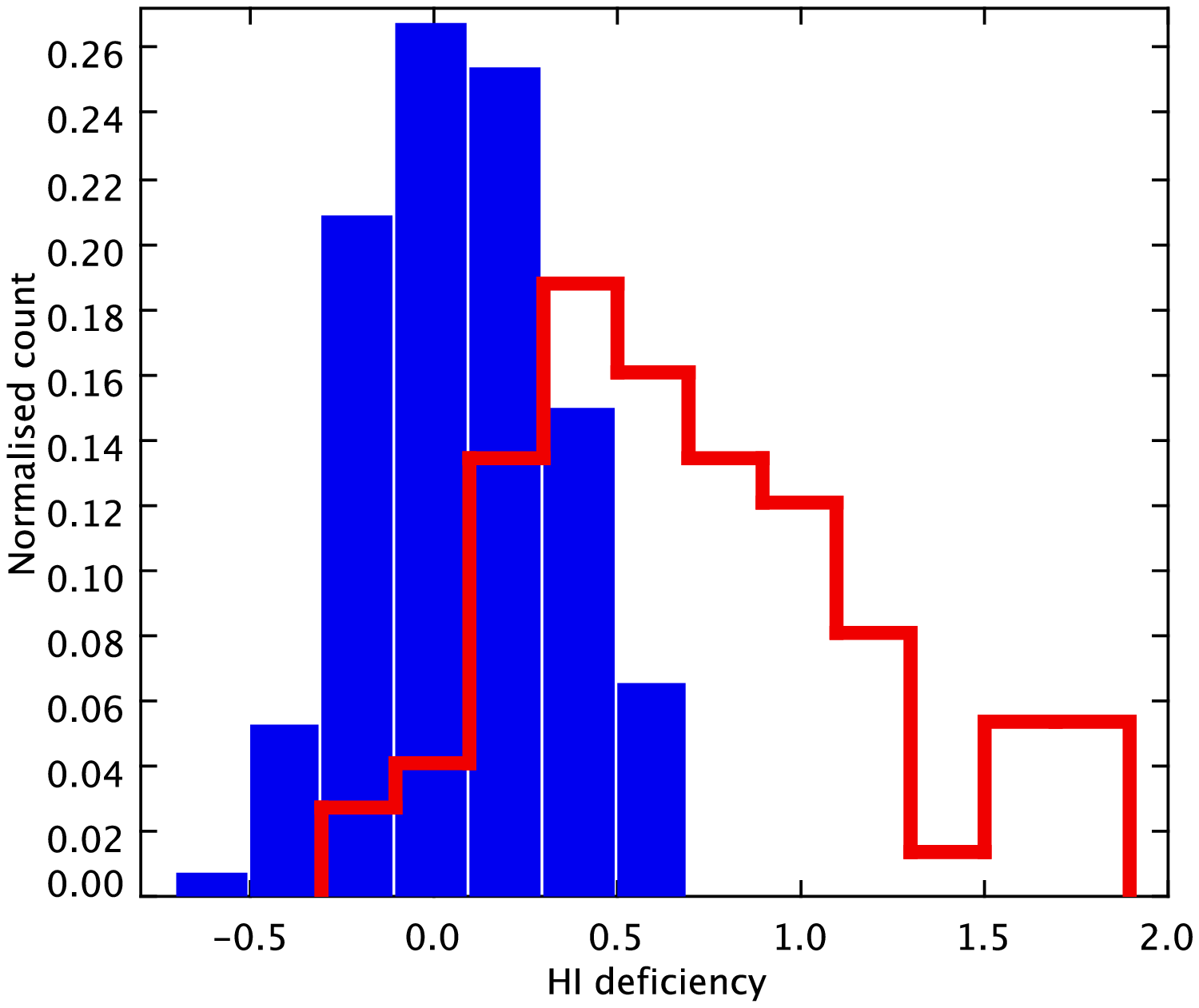}
\caption[HI deficiency distribution in VC1]{HI deficiency distribution for Virgo cluster objects (red) and background objects (blue), bin size 0.2.}
\label{VC1HIdef}
\end{center}
\end{figure}

The distribution of HI deficiency is shown in figure \ref{VC1HIdef}. The deficiency for an isolated object is typically -0.3 $<$ $D_{\rm{HI}}$ $<$ +0.3 (our background sample has more scatter than this, probably because most background objects are more distant and subject to greater measurement errors). The mean deficiency for this cluster sample is 0.7, with 76\% greater than 0.3. In comparison, the mean deficiency in the background objects is 0.1.

The deficiencies confirm that there is indeed an environmental influence of the cluster acting to remove gas. Cluster galaxies typically show greater deficiencies than background objects. Although some cluster galaxies show very low deficiencies, 35$\%$ of background objects have \textit{negative} deficiencies. 

\begin{figure}
\begin{center}
\includegraphics[scale=0.5]{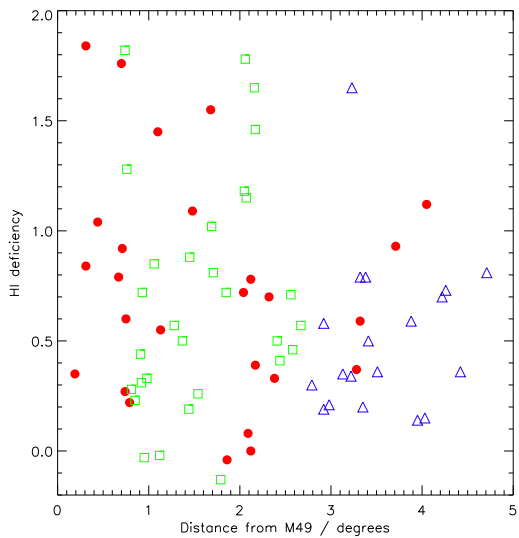}
\caption[Spatial variation of HI deficiency across VC1]{HI deficiency of objects with a morphological classification varying across the cube. Red circles are galaxies believed to be at 17 Mpc distance, green squares are at 23 Mpc and blue triangles are at 32 Mpc.}
\label{VC1DefVar}
\end{center}
\end{figure}

There is a hint of variation in deficiency with distance from M49, the dominant galaxy in this part of the cluster. Figure \ref{VC1DefVar} shows the deficiency as a function of projected distance from M49 and declination for comparison. Although the trend is weak, 81$\%$ of the 21 galaxies with deficiency $>$ 1.0 are within 2.5 degrees of M49. This bias in the distribution is not reproduced in the galaxies of lower deficiency, which are distributed evenly with distance from M49 - of the 57 objects width deficiency $<$ 1.0, 58\% are within 2.5 degrees of M49.

M49 lies in the centre of an X-ray emitting gas cloud, as shown in figure \ref{VMap}, and is believed to lie at 17 Mpc distance. As shown in figure \ref{VC1DefVar}, some of the strongly deficient objects within 2.5 degrees of M49 are purported to be at 23 Mpc. However, even if gas cloud associated with M49 is the cause of the high deficiencies, it does not necessarily imply that the cloud is 6 Mpc deep given the distance uncertainties discussed in section \ref{sec:distribution}. It is also worth briefly noting that deficiency is seen to remain high even as far as 4 degrees from M49 (as well as in the cloud at 32 Mpc distance). This indicates that the environmental influence of the cluster probably continues beyond the edge of this region.

\subsection{Optically undetected gas clouds and other exotica}
\label{sec:weird}

In this survey, 199 objects have been assigned sure optical counterparts. An additional 72 have only a single likely optical counterpart. Of the remainder, the majority are either unsure HI detections, have multiple possible optical counterparts, or a bright foreground star is present that could be obscuring the corresponding galaxy. But in a few instances the HI appears to be unassociated with any optical galaxies. We caution that determining which objects have merely an unusually faint optical counterpart, and which genuinely have none, is a highly subjective process. Within an area the size of the Arecibo beam there are invariably optical smudges that could potentially be associated with the HI. However, in the detections discussed below, the only possible optical counterparts are so extreme that their association with the HI must be questioned.

\begin{figure}
\begin{center}  
  \subfloat[Source 231]{\includegraphics[height=40mm]{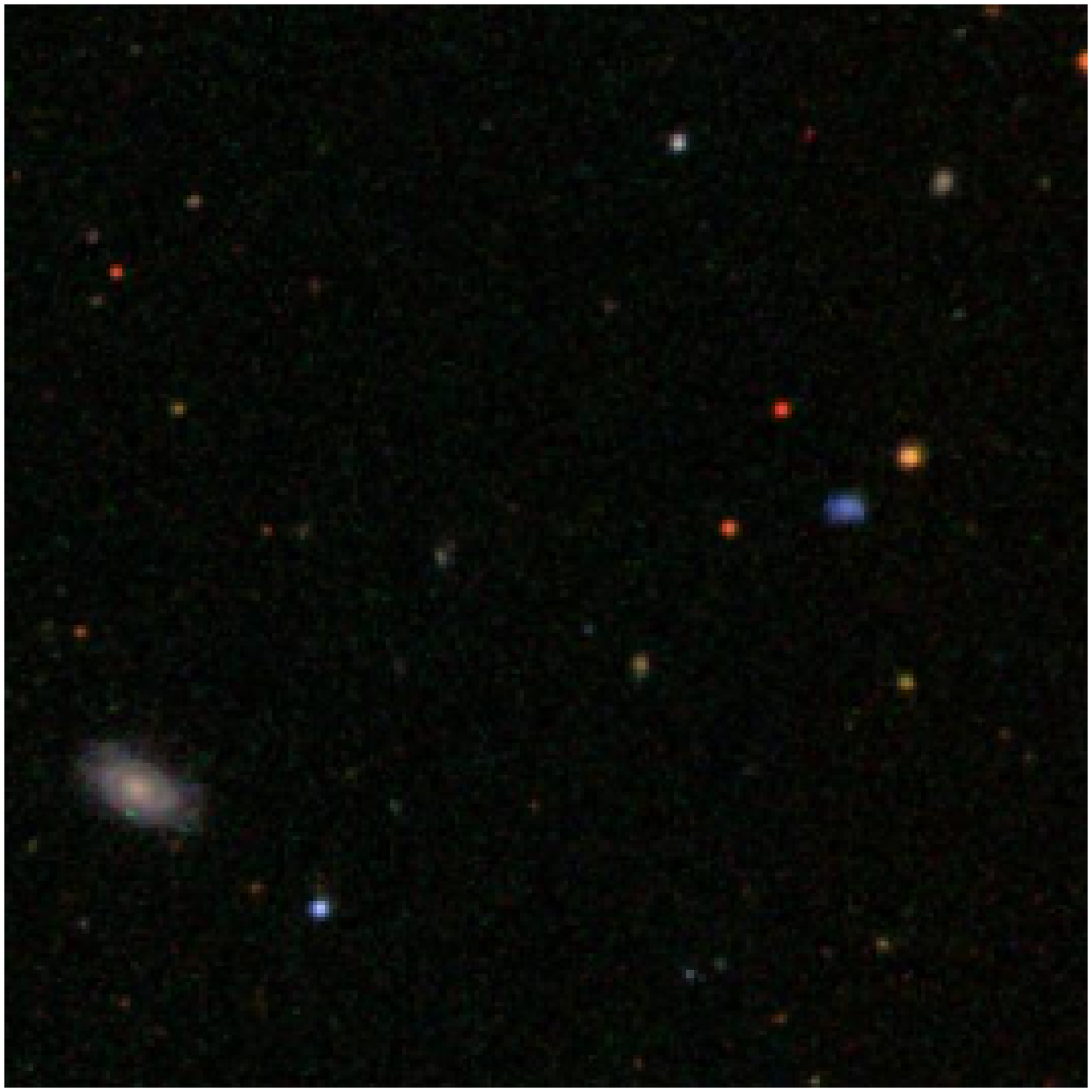}}
  \subfloat[Source 231]{\includegraphics[height=42mm]{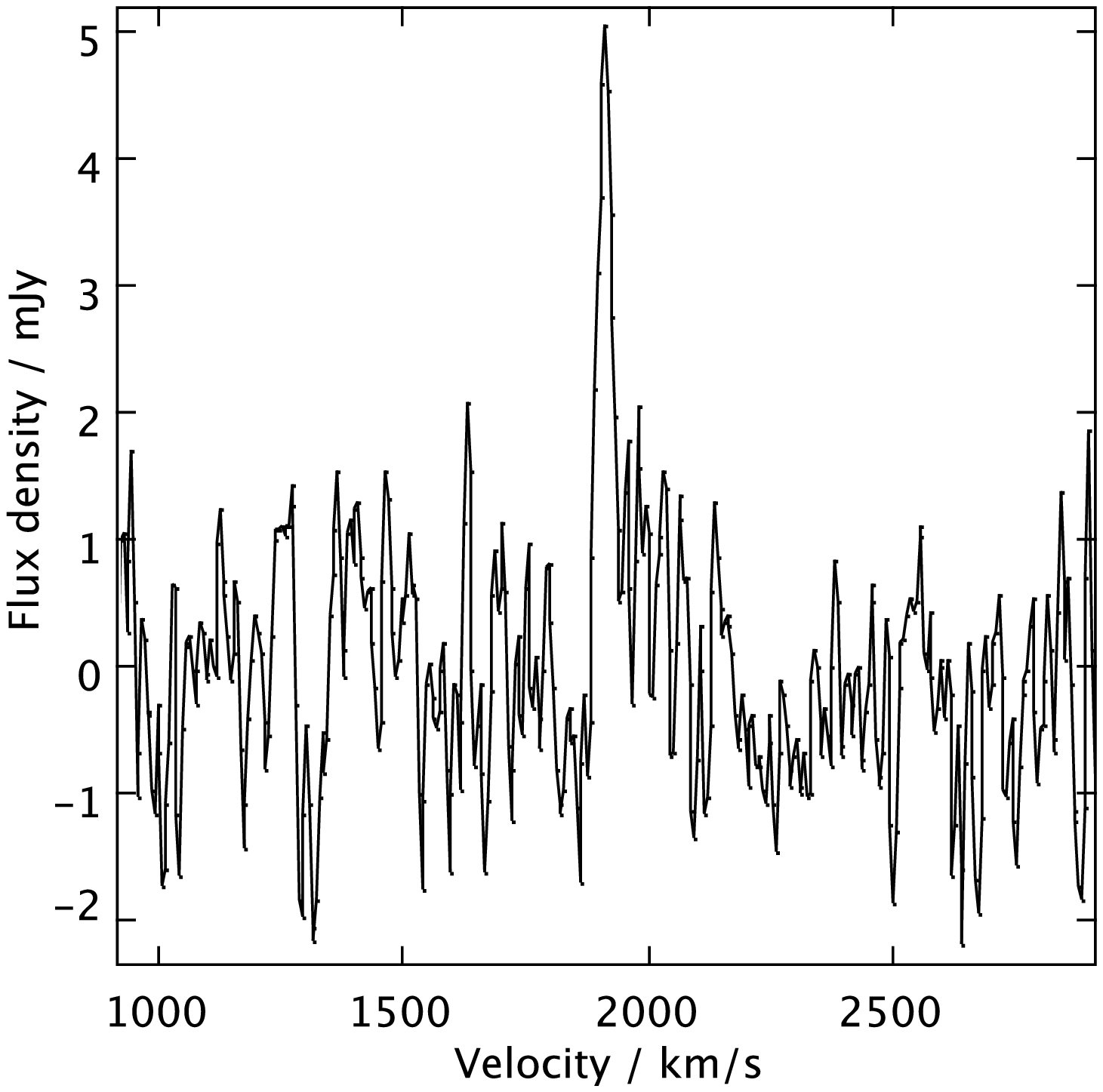}}\\
  \subfloat[Source 247]{\includegraphics[height=40mm]{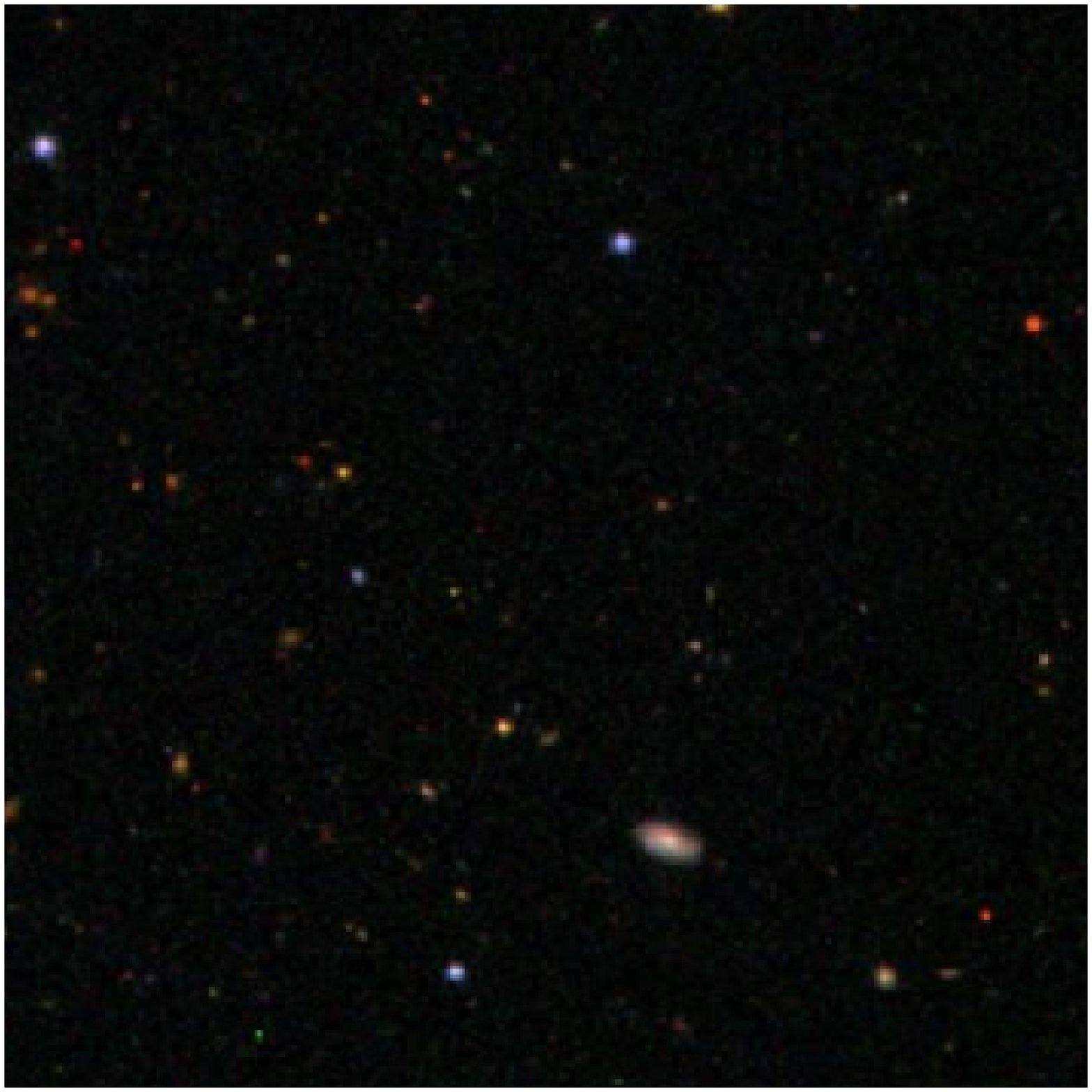}}
  \subfloat[Source 247]{\includegraphics[height=42mm]{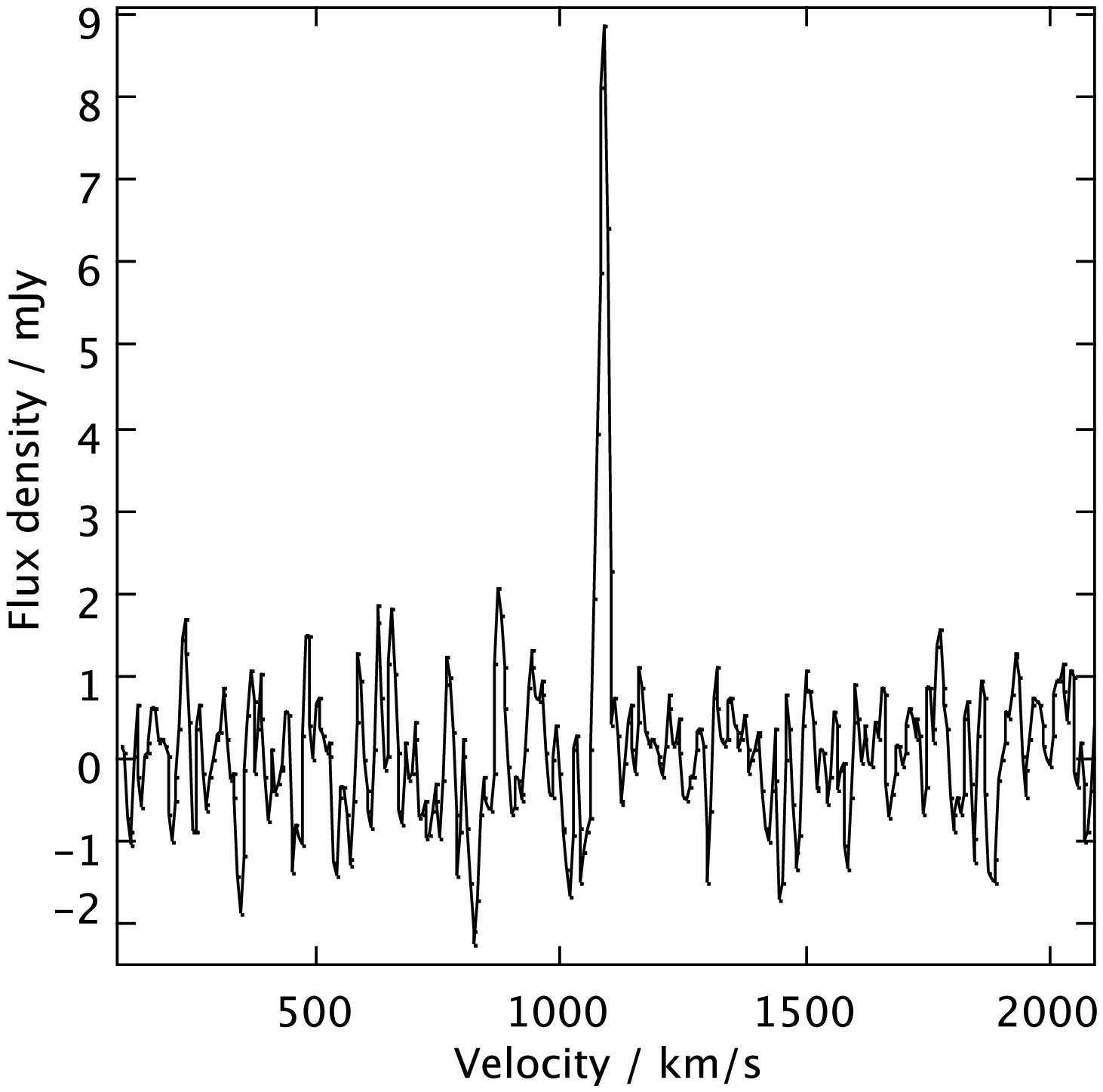}} \\  
  \subfloat[Source 257]{\includegraphics[height=40mm]{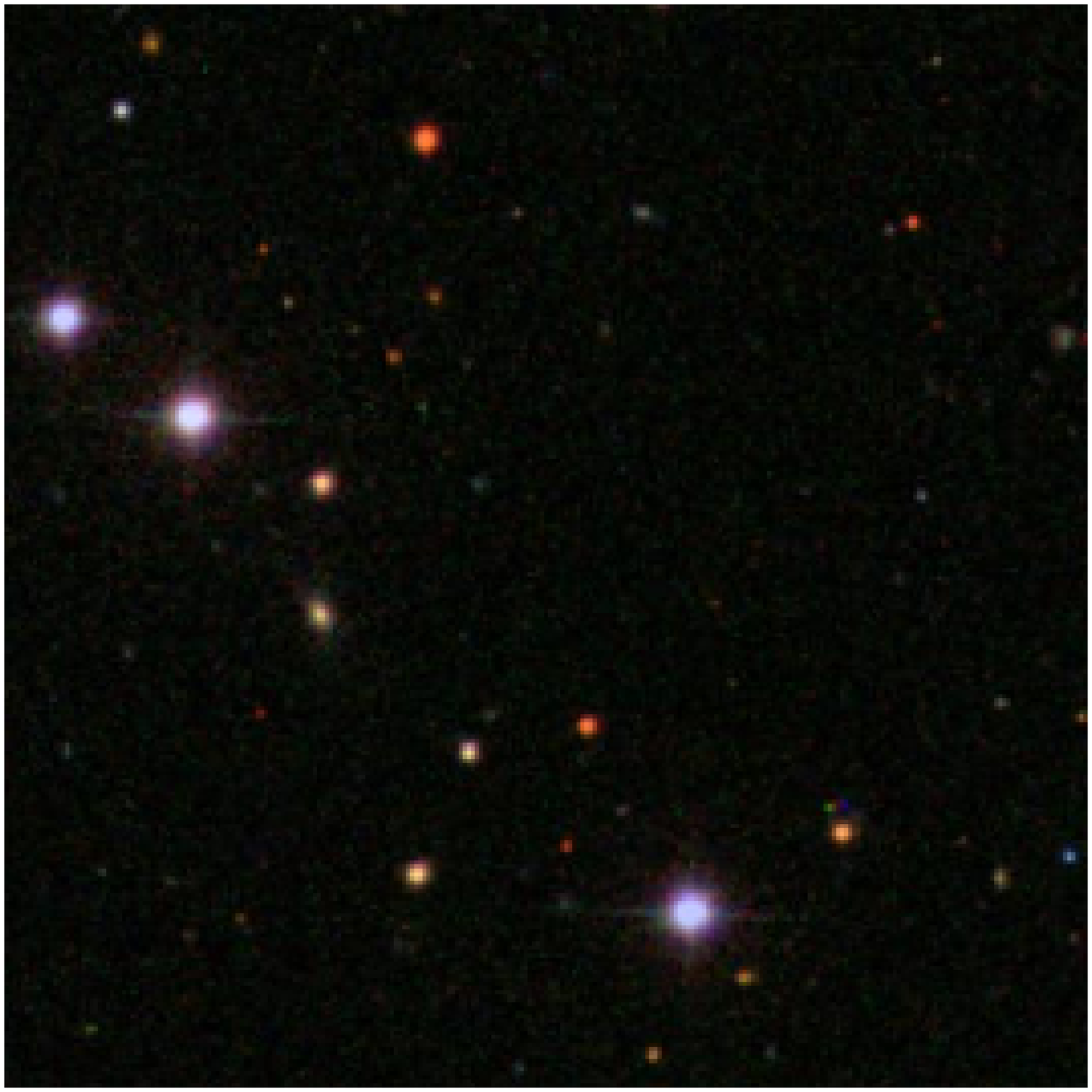}}
  \subfloat[Source 257]{\includegraphics[height=42mm]{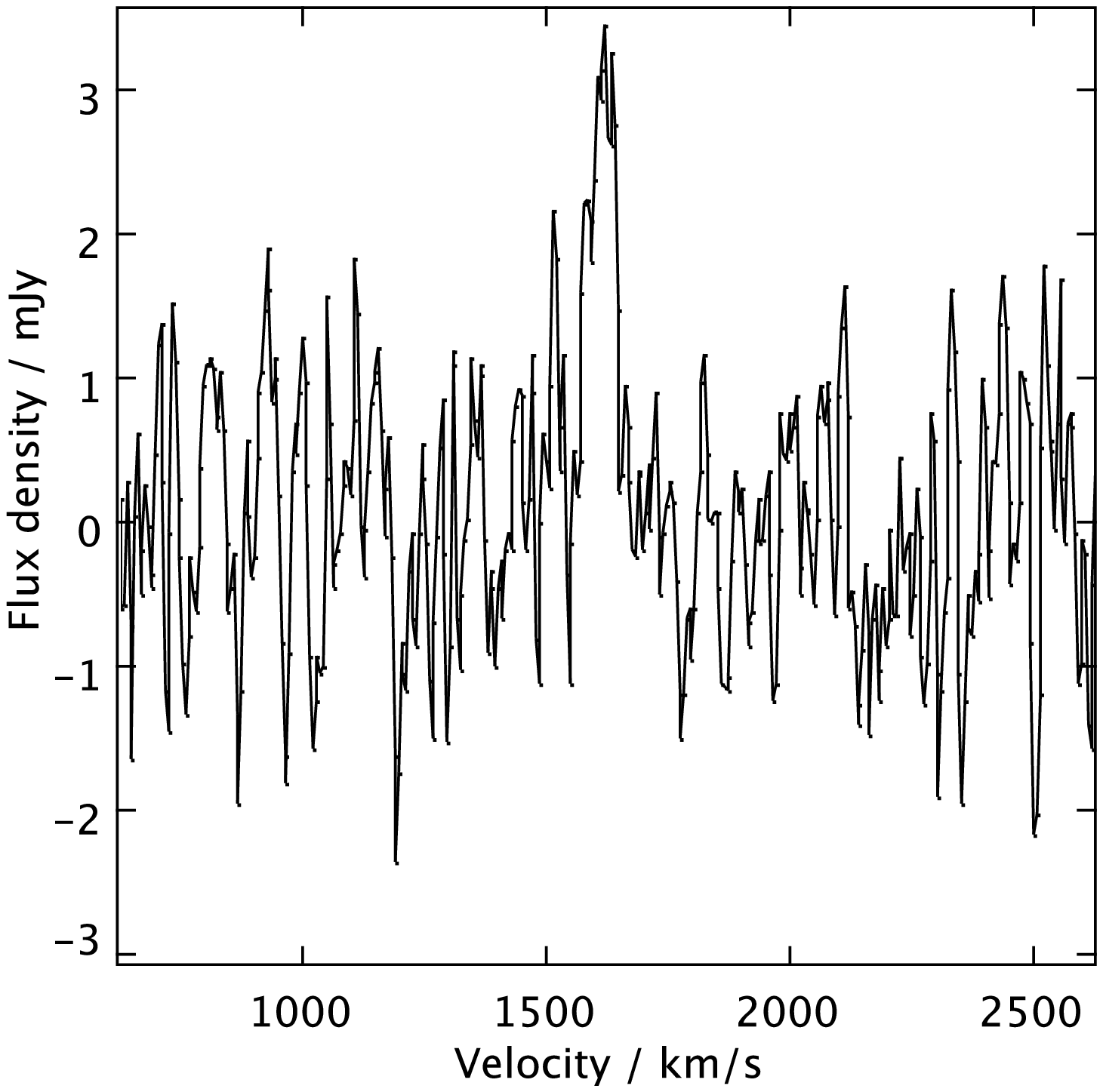}} 
\end{center}
\caption[SDSS images and HI spectra of dark galaxy candidates]{Dark galaxy candidates - SDSS RGB band images (3.3\arcmin across, north is up and east is to the left) and HI spectra (each spans 2,000 km/s of baseline).}
\label{DarkGalaxies1}
\end{figure}

\begin{figure}
\begin{center}     
  \subfloat[Source 258]{\includegraphics[height=40mm]{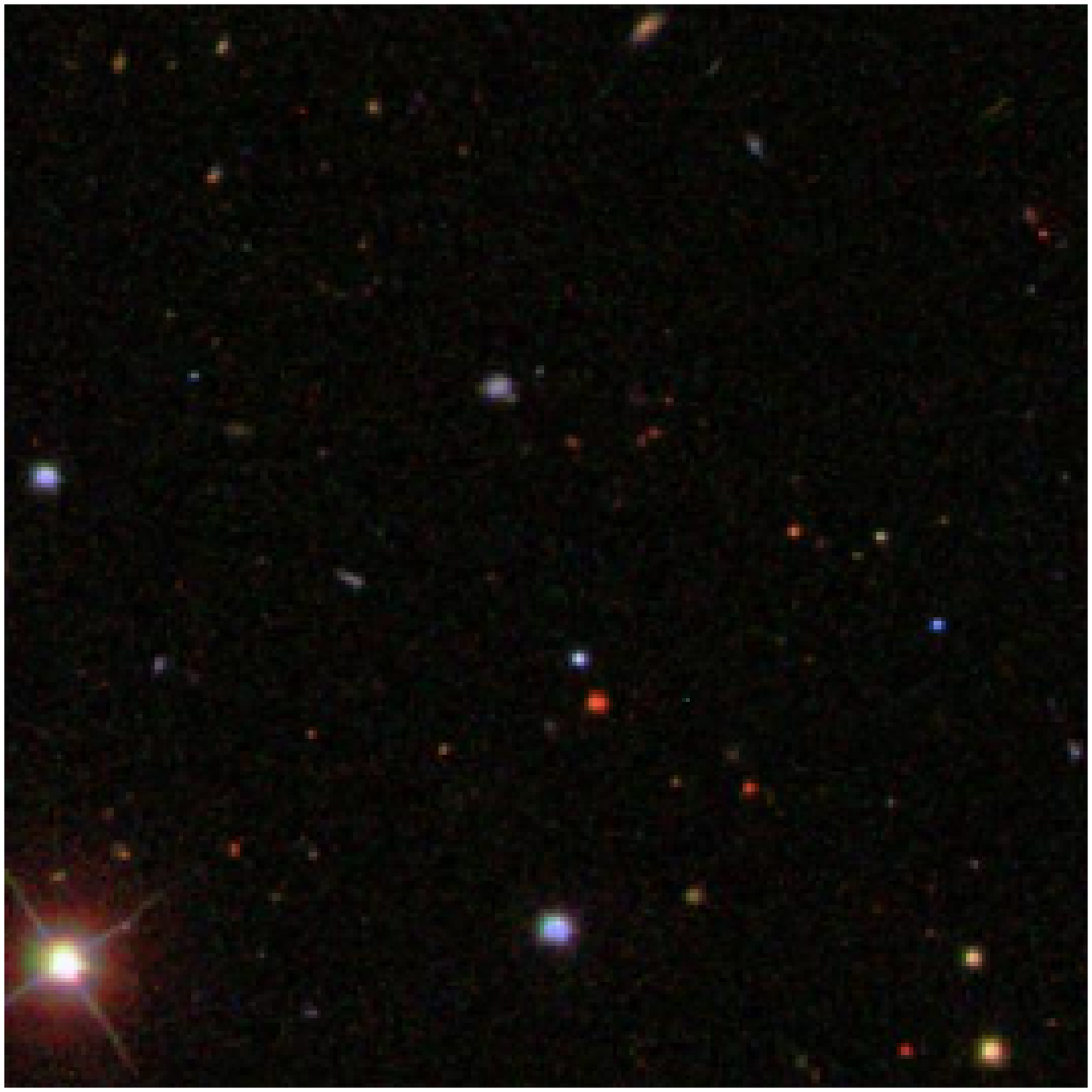}}
  \subfloat[Source 258]{\includegraphics[height=42mm]{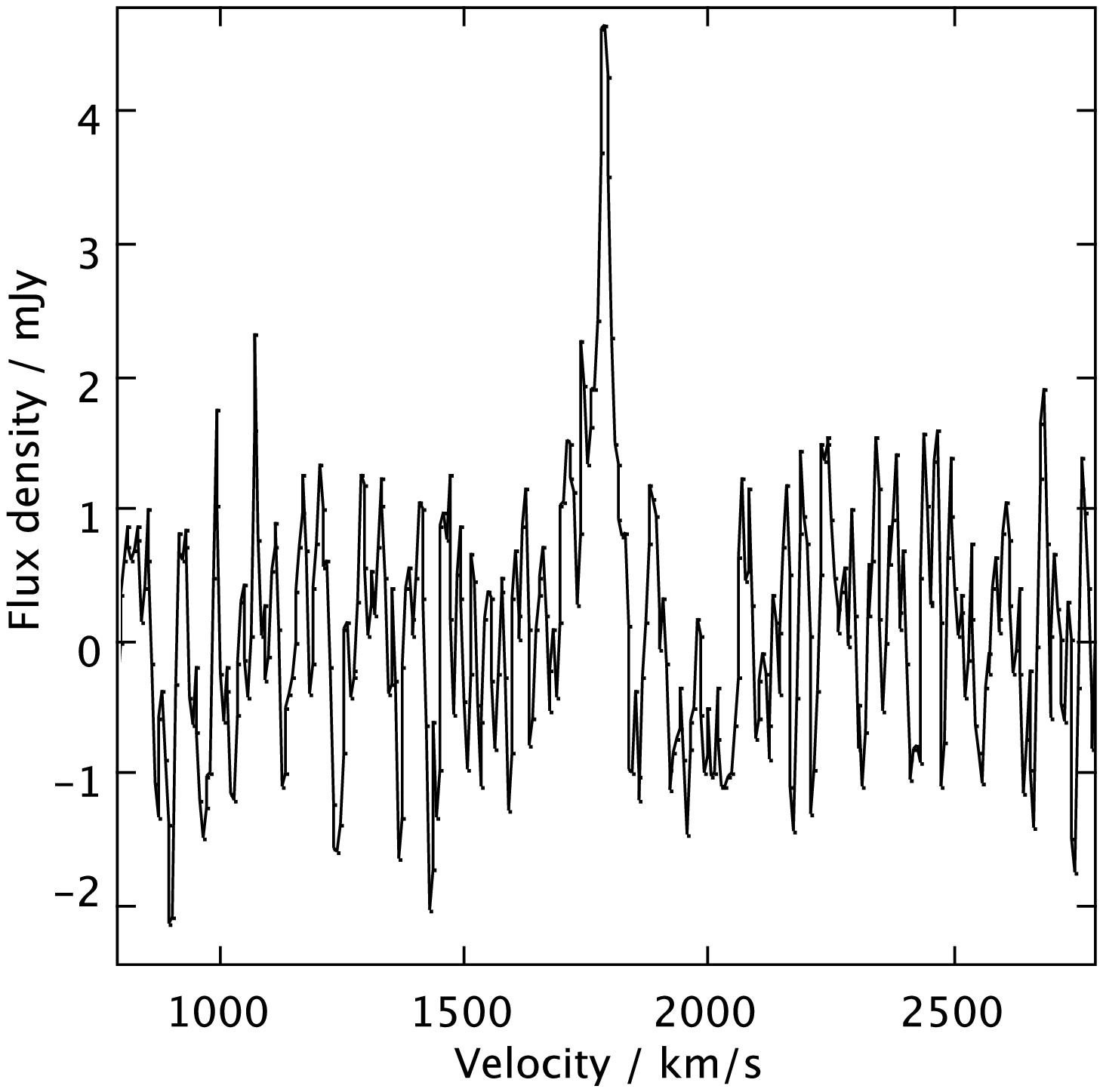}}\\
  \subfloat[Source 262]{\includegraphics[height=40mm]{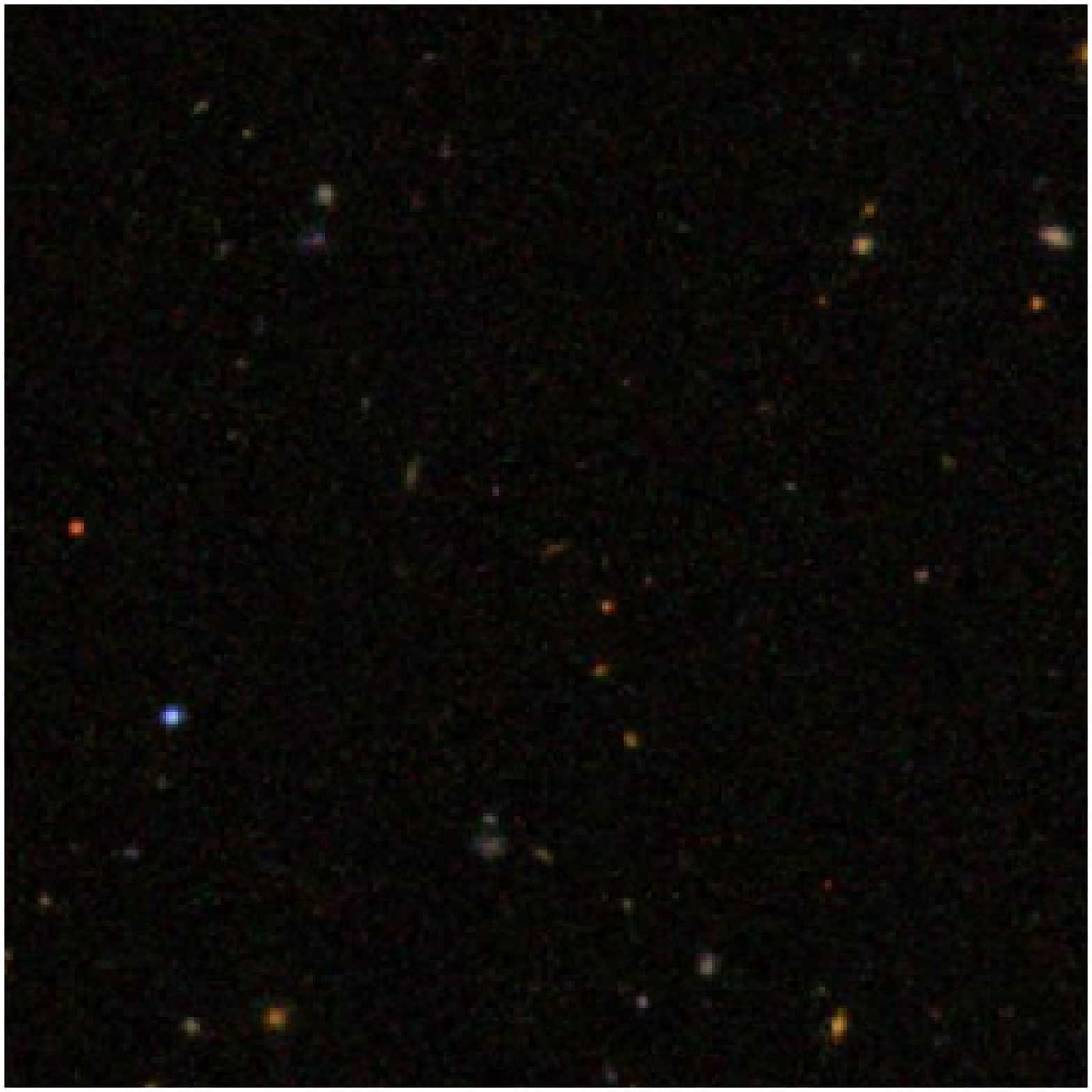}}
  \subfloat[Source 262]{\includegraphics[height=42mm]{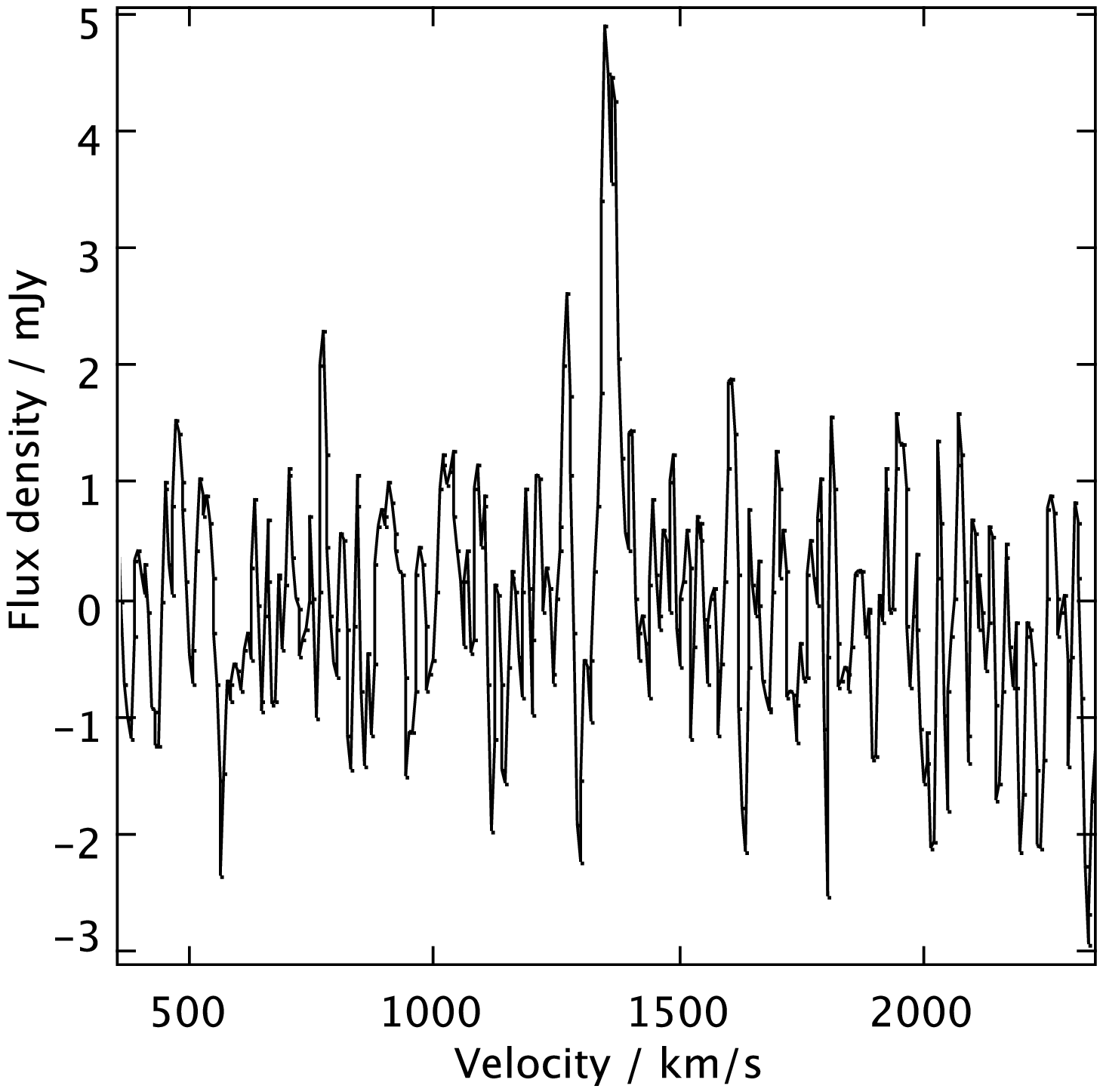}}\\  
  \subfloat[Source 266]{\includegraphics[height=40mm]{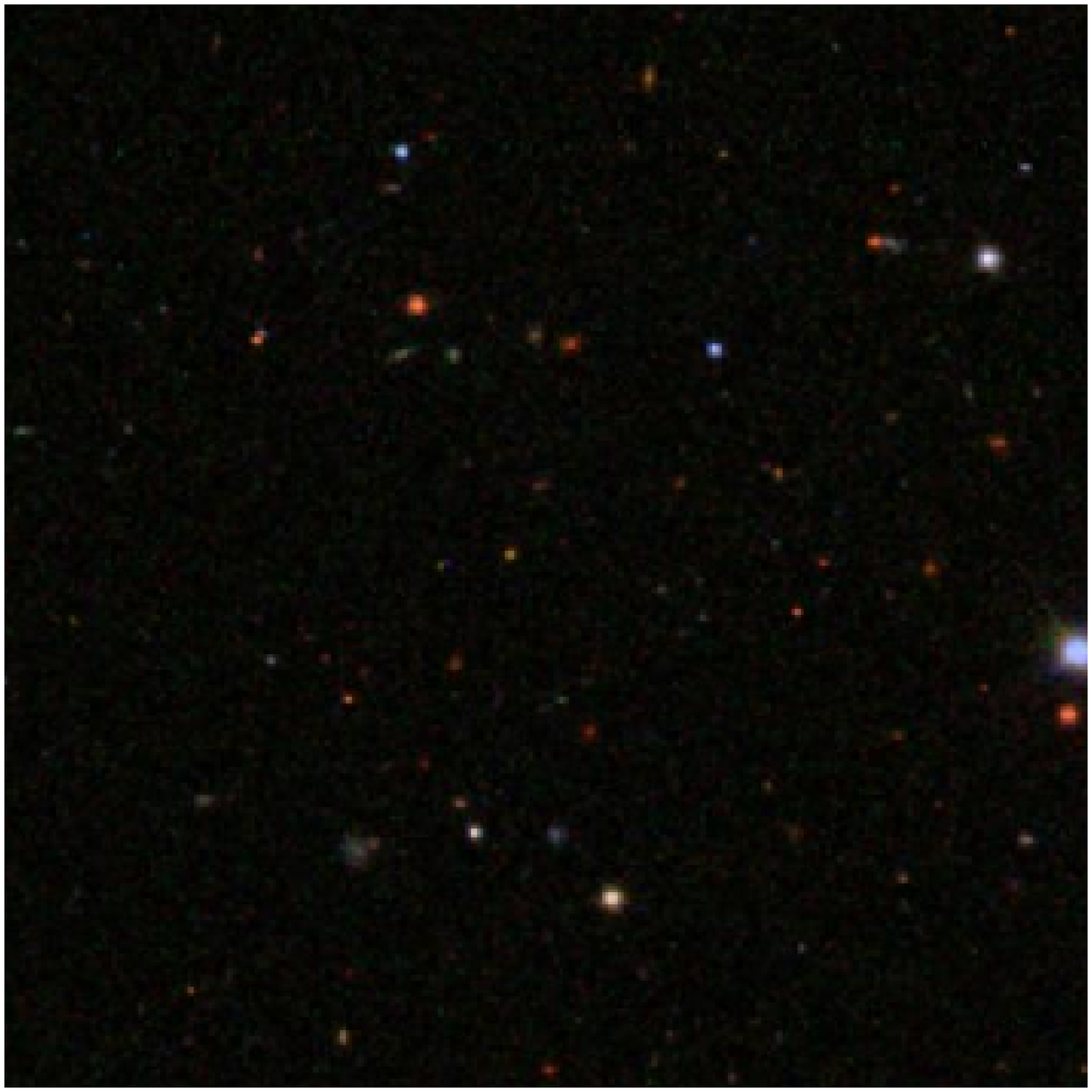}}
  \subfloat[Source 266]{\includegraphics[height=42mm]{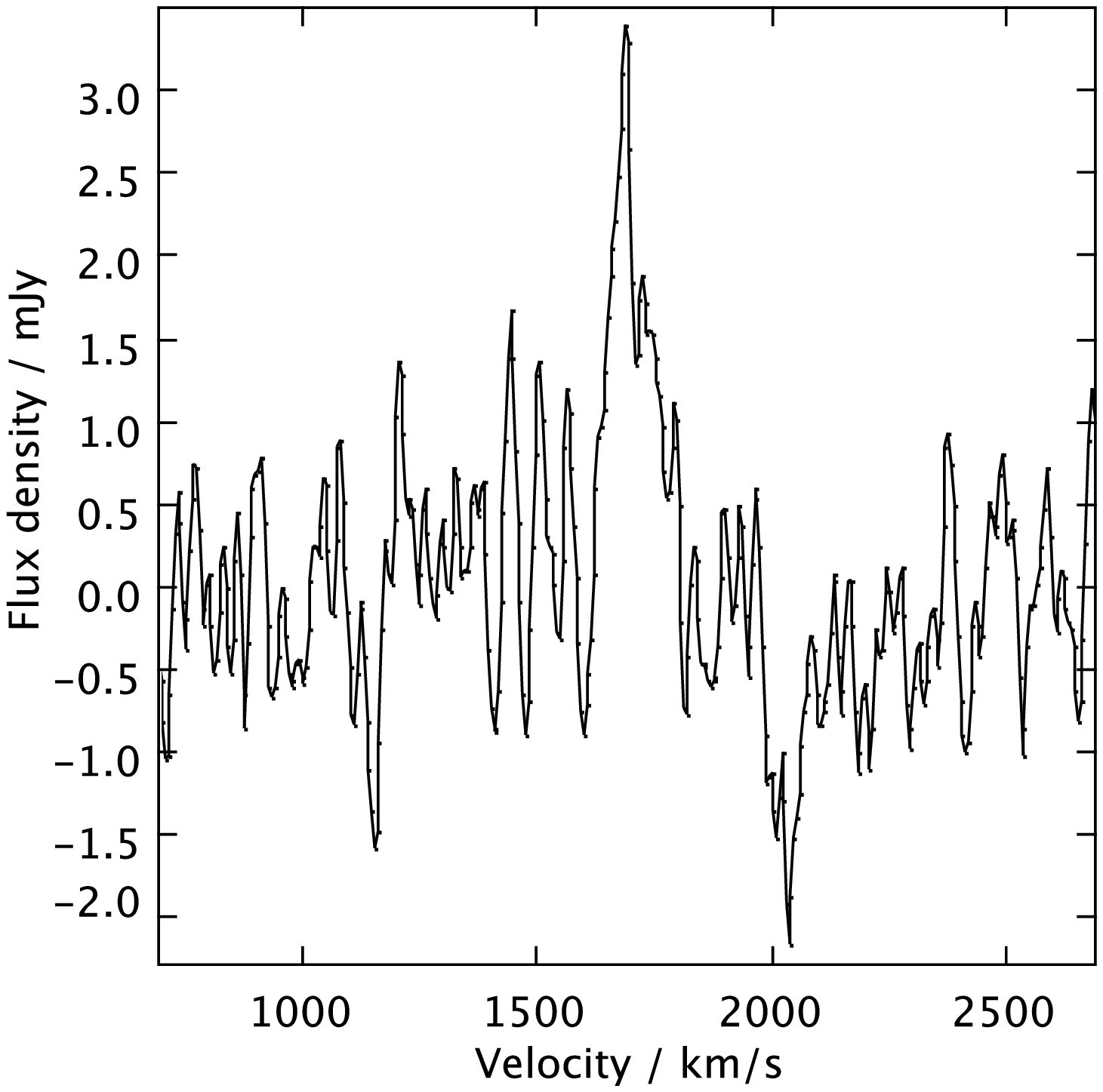}}\\ 
  \subfloat[Source 274]{\includegraphics[height=40mm]{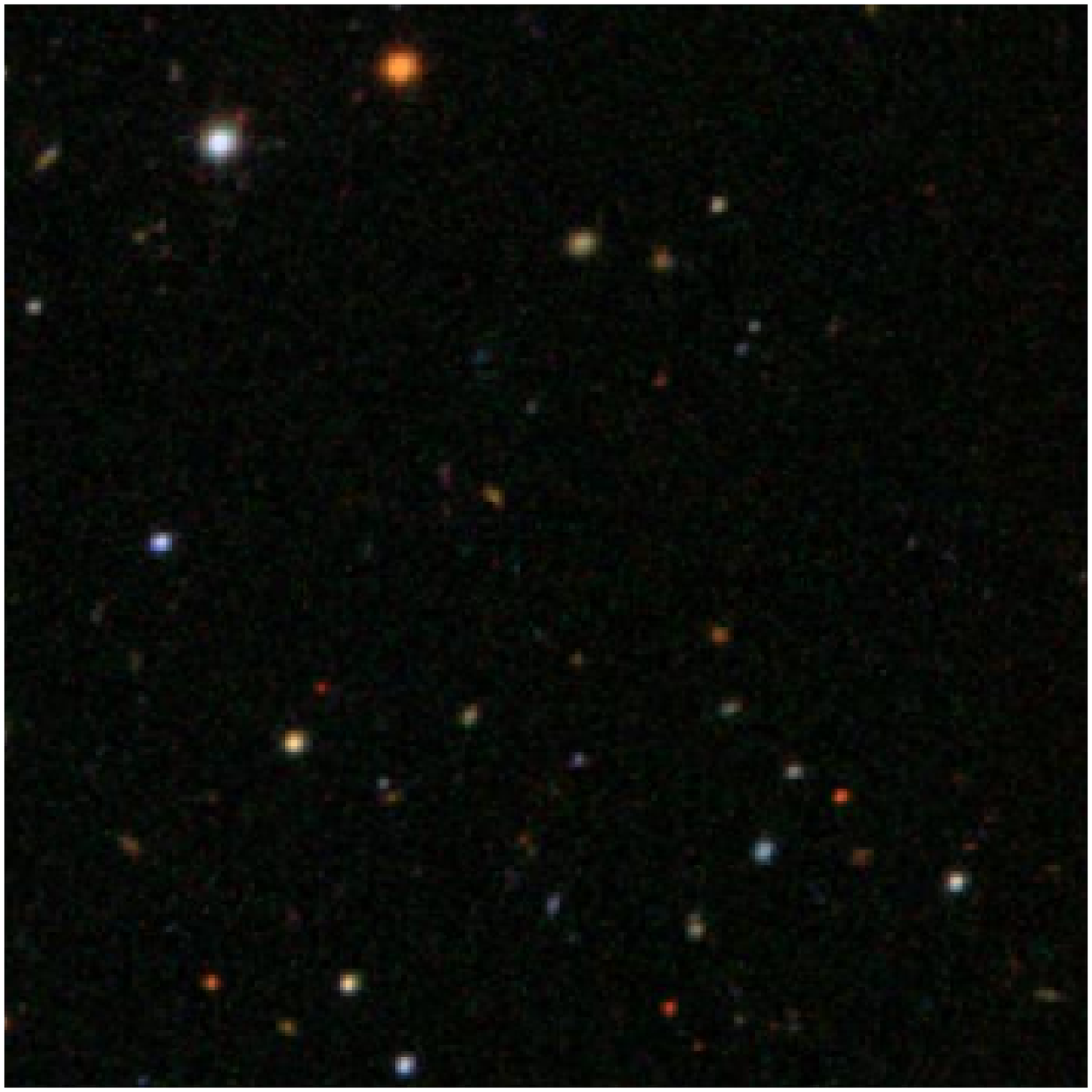}}
  \subfloat[Source 274]{\includegraphics[height=42mm]{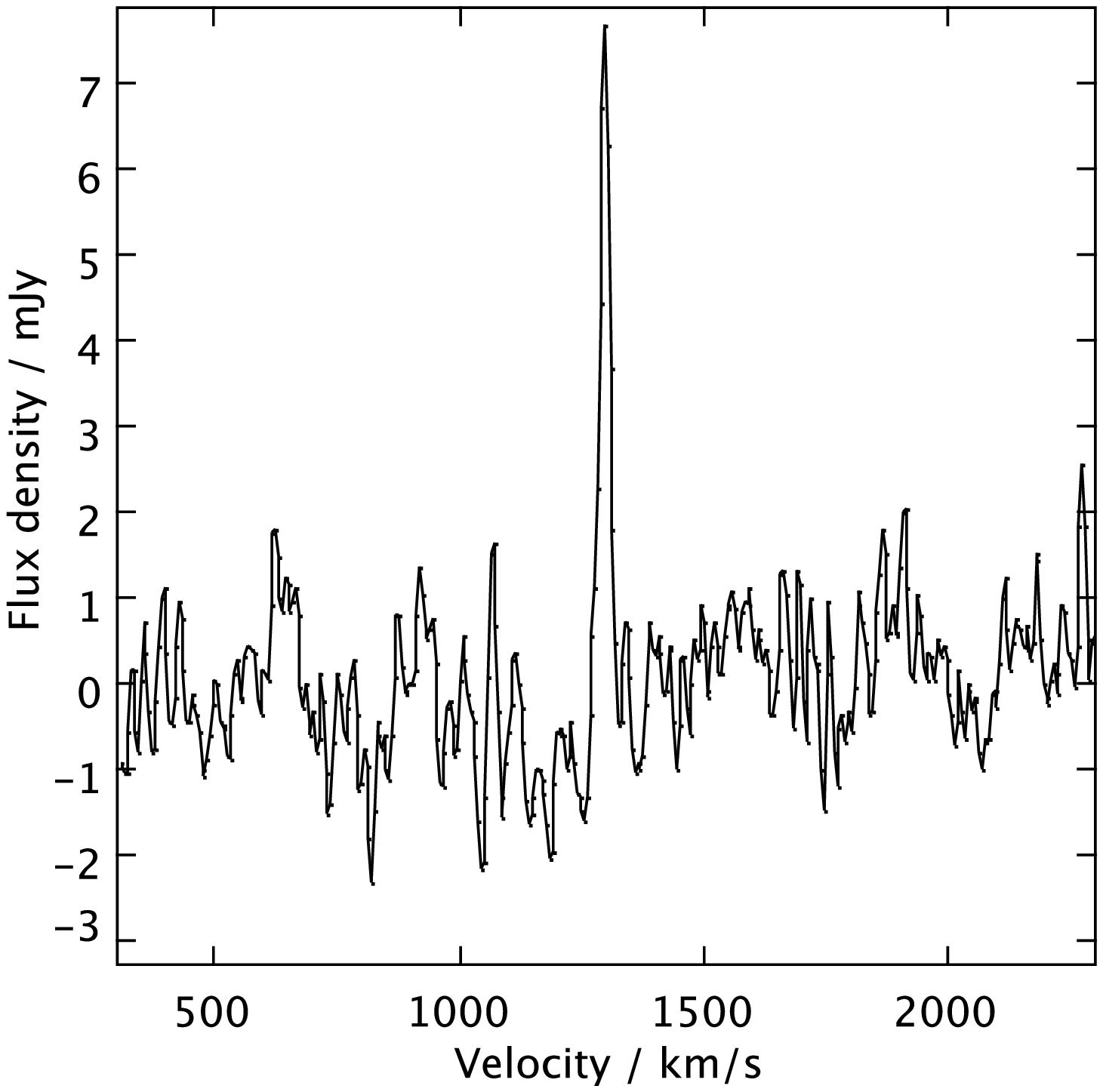}}
\end{center}
\caption[SDSS images and HI spectra of dark galaxy candidates]{Dark galaxy candidates - SDSS RGB band images (3.3\arcmin across, north is up and east is to the left) and HI spectra (each spans 2,000 km/s of baseline).}
\label{DarkGalaxies2}
\end{figure}

Only those cases where the HI detection is within the Virgo Cluster are considered here, since more distant galaxies (if of particularly high  $M_{\rm{HI}}$/$L$ ratios) might simply be too faint to detect optically. We show finding charts and spectra in figures \ref{DarkGalaxies1} and \ref{DarkGalaxies2}. The gas masses for these objects are amongst the lowest in the sample, with the highest at only 4$\times$10$^{7}$ M$_{\odot}$ (all have been confirmed by follow-up observations). They share few other similarities with each other - two detections have extremely low velocity widths (W20 $\sim$ 35 km/s) while the others are more typical of the rest of the sample (W20 $\sim$ 150 km/s).

The lack of normal optical counterparts for these objects suggests three possibilities : 1) The HI is associated with faint, optically unusual galaxies; 2) The HI was once associated with more typical optical galaxies but has since been removed; 3) The HI was never associated with any optical emission - these are so-called ``dark galaxies''. Of course it is possible that these detections are the result of all of these effects at work and there is no common explanation for all 7 of these objects.

We cannot entirely discount any of these possibilities. The first option requires optical redshifts to determine if any optical emission is genuinely associated with the HI. Without this, we can only say that the optical counterpart candidates would be extremely unusual. For instance the only remotely plausible candidate for AGESVC1\_231 would be the bluest object in the entire sample, and also very compact (optical diameter of 2.3 kpc) despite the relatively high HI velocity width (W20 of 152 km/s).

It is also difficult to explain these detections as tidal debris - however, the formation of gas clouds by gravitational interactions is not well understood. The purported dark galaxy VIRGOHI21 (\citealt{m07}) was found to be part of a much larger extended structure (\citealt{haynesV21}), greatly adding to the complexity of the situation. \citealt{duc} demonstrated that such features may, potentially, be formed from high-speed as well as low-speed interactions, so that the original interacting galaxies may become widely separated over time. While the above detections do not appear to be embedded in large-scale extended HI features, the survival of such streams in the intracluster medium depends upon many complex factors (see \citealt{kap}).

The third possibility, that these detections represent optically dark galaxies, presents its own difficulties. Ideally, of course, we would prefer a dark galaxy candidate to show all the features of more normal HI detections : a reasonably wide HI line with a clear double-horn, the signature of a rotating disc. While 5 of these detections have reasonable velocity widths, none are double-horned. On the other hand, many dwarf irregulars show both similar velocity widths and profile shapes. To prove both that these gas clouds do not have any optical counterparts \textit{and} did not form via tidal interactions is a formiddable challenge, requiring both optical redshift measurements as well as deeper (ideally higher-resolution) HI observations. Even then, accepting that the gas clouds are embedded in gravitationally-bound dark matter halos requires something of a leap of faith.

\begin{figure}
\begin{center}
\includegraphics[scale=0.5]{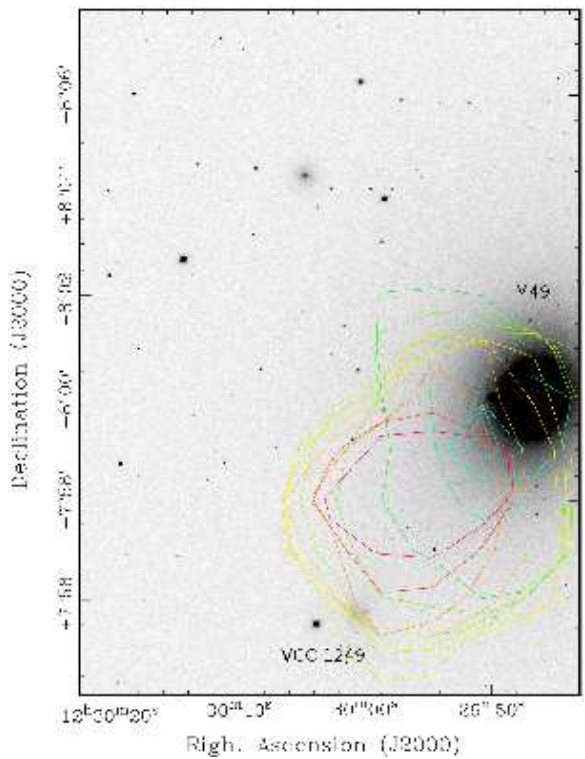}
\caption[Renzogram for AGES detection of HI associated with VCC 1249 and M49]{Renzogram for AGES detection of HI associated with VCC 1249 and M49. Contours are all at 5 mJy/beam with colour indicating velocity - red the lowest at 468 km/s, through orange, yellow, green and finally cyan at 533 km/s.}
\label{Renzo}
\end{center}
\end{figure}

VCC 1249 (AGESVC1\_281) is also worth mentioning, as it is the only clear example of a detection showing gas stripping actually occurring. We show a renzogram of this detection in figure \ref{Renzo} - contour maps of each channel are shown with a different colour for each velocity, overlaid on an SDSS $g$ band image. As the previous, higher-resolution observations of \citealt{1249} also demonstrated, the HI is separated from the late-type VCC 1249 and lies midway between that galaxy and M49 (the same is true in velocity : VCC 1249 is at  276 km/s, the HI is at 499 km/s and M49 at 997 km/s). Since this is the only example where we can claim confidently to have detected gas removed from a LTG (which is itself devoid of HI), it may be that in general the gas is rapidly ionized once removed from its host galaxy.

\section{Summary and discussion}
\label{sec:Conclusions}

AGES detects 89 objects within the Virgo Cluster in a 10 square degree area. 26 of these are new HI detections, while 21 of these are not members of the VCC. Of those 21 non-VCC galaxies detected in HI, 14 are optically brighter than the VCC completeness limit. The fact that AGES does not uncover many more new detections is likely due to its mass sensitivity, which is comparable to that of the VCC, and the fact that HI is almost always associated with an optical galaxy.

The basic structure of this region is of 3 clouds at 17, 23 and 32 Mpc distance. Both the HI and optical studies (i.e. the VCC) support this view, with all of the new HI detections being interspersed with the optical detections (both spatially and in position-velocity space). However, a greater fraction of the HI detections are found at higher velocities than the non-detections. This suggests that the gas removal mechanism(s) acting within the cluster proper are less significant within the infalling groups.

Although 76$\%$ of HI detections in this region are more deficient than 0.3, there are still 24$\%$ that are effectively non-deficient. This includes many of the new, non-VCC detections, which are generally small, blue, and gas-rich ($M_{g}$ fainter than -16, $g$-$i$ bluer than 0.7, and with  $M_{\rm{HI}}$/$L_{g}$ ratios greater than 0.3). Highly deficient (deficiency $>$ 1.0) giant galaxies apparently coexist with non-deficient dwarfs in the same population of objects. Simulations by \citealt{v01} and \citealt{rory} have shown that dwarf galaxies will lose all their gas through ram-pressure stripping on their first pass through the cluster center. Observationally some giant galaxies have lost more gas than some non-deficient dwarfs currently contain. Both of these statements imply that the non-deficient dwarfs cannot possibly have experienced the same environmental effects as the highly deficient giants. More likely the dwarfs are recent arrivals to the cluster, which have not yet had time to experience its full environmental effects.

Moreover there is a broad trend in HI deficiency. Most of the strongly deficient (greater than 1.0) galaxies are found close to M49, suggesting that the most efficient gas removal is only happening in this region. It also implies that the cluster is still assembling, otherwise galaxies that lose gas in the cluster center would have had time to move to the outer regions (so we would see highly deficient galaxies everywhere).

It is significant that we detect every (or very nearly) LTG in HI. This is a surprising result, as we should not be able to detect deficient late-type dwarfs. In fact we detect 20 galaxies which if they were more deficient than 1.5 would be undetectable. The question arises : where are all the undetected late-type objects ? It seems very unlikely that only the giant galaxies have been subjected to the full force of the gas removal mechanism acting within the cluster, so deficient giants imply the presence of deficient dwarfs. Yet apparently there are none. It seems that dwarf galaxies are either gas-rich late-types, or, as demonstrated by the lack of a detection through stacking, extremely gas poor early-types.

Unless all dwarf irregulars are recent arrivals to the cluster, this could suggest that galaxies which lose all of their gas content morphologically evolve into ETGs. This scenario has been previously described and modelled by \citealt{b08}, who found that ram pressure stripping can produce a rapid gas depletion and morphological transformation. The rapidity ($<$ 150 Myr) of the process would explain why no LTGs are undetected in HI. Thus any HI present in ETGs is either the last trace of their original gas, or has been replenished by dying stars returning some of their gas to the ISM. Certainly our results cannot rule out the \citealt{b08} scenario merely by detecting 3 ETGs in HI - rather, the fact that all late-types are detected (whereas so few dEs seem to have any HI at all) is consistent with their scenario.

\begin{figure*}
\begin{center}
\includegraphics[scale=0.35]{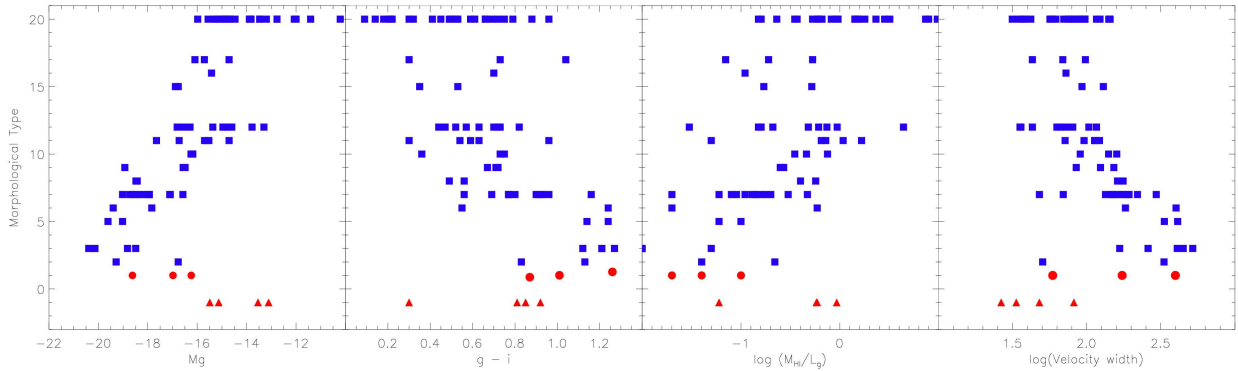}
\caption[Variation of properties with morphology]{Vartion of optical and HI properties with morphological type. From left to right : absolute $g$ magnitude, $g - i$ colour,  $M_{\rm{HI}}$/$L_{g}$ ratio, and logarithm of the average velocity width. S0s and dEs are highlighted by red circles and triangles respectively.}
\label{MorphStuff}
\end{center}
\end{figure*}

The S0s are, as we have seen, detected much more often than the dEs, have low $M_{\rm{HI}}$/$L_{g}$ ratios and 2/3 show signs of structure. Observationally, the detected S0s are broadly similar to spiral galaxies, but with particularly low gas fractions and a wide range of velocity widths (as can be seen directly in figure \ref{MorphStuff}. The natural implication is that these objects have evolved from spirals whose gas content has been depleted. Although \citealt{bg06} find that massive S0s are unlikely to be formed via gas removal from spirals, they leave open the possibility that this may explain the origin of low-mass S0s.

In complete contrast, the dEs are very rarely detected in HI, but those that are have high  $M_{\rm{HI}}$/$L_{g}$ ratios and show no obvious signs of structure. One of them appears to lie on the blue sequence, and the added complication of misidentification is more acute for these objects because of their low luminosities. Discounting this exception, the detected dEs are generally similar to dI's, except that they are much redder (as can be seen in figure \ref{MorphStuff}. Arguably this lends further credence to the model of \citealt{b08}. Since their gas fractions are comparable to the dI's, however, it is also possible that these particular detections have only recently entered the cluster. ALFALFA have found that HI is far more common in ETGs in the field, with around 44\% of morphologically early-type galaxies being detected in HI (as opposed to 9\% in the cluster) - see \citealt{aaETGField}. If the dEs detected here are indeed undergoing morphological evolution, then is raises the question of why they are red and structureless but with comparable gas fractions to many dIs.

In short, while it is perfectly possible that dEs in general have evolved from dIs within the cluster, the particular detections of this survey do not substantiate this view. In contrast the S0s we have detected appear much more likely candidates for examples of ongoing, gas-loss driven morphological evolution.

This region of the cluster is rich in HI-detected galaxies of all morphological types, many of which appear to be entering the cluster for the first time. In some ways it is an ideal location to search for dark galaxies. 7 candidates are found, about 7\% of the HI detections within the cluster in this region. This is well below earlier predictions of 23\% for AGES and there are a huge number of additional caveats. We cannot exclude the possibility that these detections do have optical counterparts, albeit ones with unusual properties. Nor can we disregard the alternative option that the HI detections were formed by gas removal from some unidentified ordinary galaxies. Even if neither of these scenarios are true, it would still be a substantial leap to say that these gas detections are really gravitationally bound in dark matter halos.

Yet it would also be premature to reject entirely the notion that these are indeed dark galaxies. Though the detection rate is below predictions, this may be due to the influence of the cluster environment. The alternative explanations for these objects also suffer from difficulties - the possible optical counterparts which are visible are too compact, and the unresolved nature of the HI detections makes it difficult to ascribe their formation to gas removal from any nearby galaxies. While the dark galaxy scenario remains a controversial hypothesis, current observations cannot determine if this explanation is any more or less valid than the others. 

\section*{Acknowledgments}

This work is based on observations collected at Arecibo Observatory. The Arecibo Observatory is operated by SRI International under a cooperative agreement with the National Science Foundation (AST-1100968), and in alliance with Ana G. Méndez-Universidad Metropolitana, and the Universities Space Research Association. 

This research has made use of the GOLDMine Database. 

This research has made use of the NASA/IPAC Extragalactic Database (NED) which is operated by the Jet Propulsion Laboratory, California Institute of Technology, under contract with the National Aeronautics and Space Administration. 

This work has made use of the SDSS. Funding for the SDSS and SDSS-II has been provided by the Alfred P. Sloan Foundation, the Participating Institutions, the National Science Foundation, the U.S. Department of Energy, the National Aeronautics and Space Administration, the Japanese Monbukagakusho, the Max Planck Society, and the Higher Education Funding Council for England. The SDSS Web Site is http://www.sdss.org/.

The SDSS is managed by the Astrophysical Research Consortium for the Participating Institutions. The Participating Institutions are the American Museum of Natural History, Astrophysical Institute Potsdam, University of Basel, University of Cambridge, Case Western Reserve University, University of Chicago, Drexel University, Fermilab, the Institute for Advanced Study, the Japan Participation Group, Johns Hopkins University, the Joint Institute for Nuclear Astrophysics, the Kavli Institute for Particle Astrophysics and Cosmology, the Korean Scientist Group, the Chinese Academy of Sciences (LAMOST), Los Alamos National Laboratory, the Max-Planck-Institute for Astronomy (MPIA), the Max-Planck-Institute for Astrophysics (MPA), New Mexico State University, Ohio State University, University of Pittsburgh, University of Portsmouth, Princeton University, the United States Naval Observatory, and the University of Washington.

{}

\begin{center}
\begin{table*}
\tiny
\caption[HI properties of AGES detections within the Virgo Cluster in VC1]{HI properties of objects detected within the Virgo Cluster. Bracketed values indicate errors as computed by \textsc{miriad}. Columns : (1) Source number in this catalogue (2) Right ascension J2000, error in seconds of time (3) Declination J2000, error in seconds of arc (4) Heliocentric velocity km/s (5) Maximum velocity width at 50$\%$ and (6) 20$\%$ of the peak flux (7) Total flux Jy (8) Distance Mpc (9) Estimated HI mass, $\log$ M$_{\odot}$ (10) Peak signal to noise (11) R.m.s. mJy}
\label{VirgoHIT}
\begin{tabular}{c c c c c c c c c c c}\\
\hline
  \multicolumn{1}{c}{(1) No.} & 
  \multicolumn{1}{c}{(2) R.A.} &
  \multicolumn{1}{c}{(3) Dec.} &
  \multicolumn{1}{c}{(4) Velocity} &
  \multicolumn{1}{c}{(5) W50} &
  \multicolumn{1}{c}{(6) W20} &
  \multicolumn{1}{c}{(7) Flux} &
  \multicolumn{1}{c}{(8) Dist} &
  \multicolumn{1}{c}{(9) $M_{\rm{HI}}$} &
  \multicolumn{1}{c}{(10) S/N} &
  \multicolumn{1}{c}{(11) rms} \\
\hline  AGESVC1\_199 & 12:20:27.60(1.2) & +06:53:13.0(10) & 2572(10) & 98(19) & 207(29) & 1.332(0.276) & 23.0 & 8.22 & 7.75 & 2.4\\
  AGESVC1\_200 & 12:21:01.40(0.7) & +07:04:14.0(10) & 2368(13) & 88(25) & 259(38) & 0.371(0.074) & 23.0 & 7.66 & 7.49 & 0.6\\
  AGESVC1\_201 & 12:18:17.10(0.7) & +07:07:12.0(11) & 2709(4) & 74(7) & 104(11) & 0.547(0.084) & 32.0 & 8.12 & 11.41 & 0.7\\
  AGESVC1\_202 & 12:13:50.20(0.7) & +07:11:56.0(10) & 2664(2) & 323(3) & 342(5) & 7.024(0.323) & 32.0 & 9.22 & 44.13 & 0.7\\
  AGESVC1\_203 & 12:21:36.40(0.8) & +07:14:34.0(13) & 2771(4) & 36(8) & 48(12) & 0.122(0.046) & 23.0 & 7.18 & 5.89 & 0.6\\
  AGESVC1\_204 & 12:17:27.30(0.7) & +07:19:19.0(10) & 2209(2) & 134(3) & 148(5) & 1.509(0.130) & 32.0 & 8.56 & 22.49 & 0.7\\
  AGESVC1\_205 & 12:13:00.20(0.7) & +07:17:37.0(10) & 2172(5) & 131(10) & 188(15) & 0.856(0.096) & 32.0 & 8.31 & 12.12 & 0.6\\
  AGESVC1\_207 & 12:16:33.50(0.7) & +07:27:30.0(10) & 2610(3) & 506(5) & 533(8) & 3.442(0.182) & 32.0 & 8.92 & 18.93 & 0.6\\
  AGESVC1\_210 & 12:20:07.30(0.7) & +07:41:18.0(10) & 2618(1) & 125(3) & 143(4) & 4.270(0.273) & 23.0 & 8.72 & 66.57 & 0.6\\
  AGESVC1\_211 & 12:16:24.30(0.7) & +07:47:21.0(12) & 2352(4) & 24(7) & 72(11) & 0.410(0.068) & 32.0 & 7.99 & 17.48 & 0.7\\
  AGESVC1\_212 & 12:29:37.70(0.7) & +07:49:08.0(10) & 2347(2) & 139(3) & 164(5) & 6.918(0.389) & 17.0 & 8.67 & 72.57 & 0.7\\
  AGESVC1\_213 & 12:19:21.90(0.7) & +07:52:13.0(10) & 2481(2) & 73(4) & 97(6) & 1.285(0.117) & 23.0 & 8.2 & 32.11 & 0.6\\
  AGESVC1\_214 & 12:10:45.00(0.7) & +07:51:39.0(12) & 2055(5) & 55(9) & 85(14) & 0.271(0.065) & 32.0 & 7.81 & 8.57 & 0.7\\
  AGESVC1\_215 & 12:16:11.00(0.9) & +07:55:16.0(14) & 2249(7) & 35(15) & 83(22) & 0.144(0.052) & 32.0 & 7.54 & 6.56 & 0.7\\
  AGESVC1\_216 & 12:16:46.20(0.7) & +08:03:06.0(11) & 2573(3) & 81(6) & 115(9) & 0.638(0.072) & 32.0 & 8.18 & 16.61 & 0.5\\
  AGESVC1\_217 & 12:21:27.40(0.7) & +08:09:02.0(11) & 2539(6) & 30(12) & 133(17) & 0.316(0.057) & 23.0 & 7.59 & 14.42 & 0.6\\
  AGESVC1\_218 & 12:18:00.10(0.7) & +08:10:00.0(10) & 2013(2) & 117(4) & 129(6) & 0.853(0.084) & 32.0 & 8.31 & 15.77 & 0.5\\
  AGESVC1\_219 & 12:15:38.40(0.7) & +08:17:05.0(10) & 2588(2) & 61(4) & 85(5) & 2.020(0.181) & 32.0 & 8.68 & 61.13 & 0.6\\
  AGESVC1\_220 & 12:19:02.70(0.7) & +08:51:21.0(10) & 2472(1) & 167(3) & 185(4) & 8.508(0.441) & 32.0 & 9.31 & 110.74 & 0.5\\
  AGESVC1\_222 & 12:42:44.60(0.7) & +07:20:20.0(10) & 2412(2) & 63(3) & 80(5) & 1.837(0.167) & 17.0 & 8.09 & 45.45 & 0.7\\
  AGESVC1\_224 & 12:33:41.00(3.5) & +09:06:30.0(10) & 2336(15) & 211(31) & 383(46) & 1.703(0.331) & 17.0 & 8.06 & 5.97 & 2.2\\
  AGESVC1\_225 & 12:39:23.40(0.7) & +07:57:36.0(10) & 2078(2) & 112(4) & 147(6) & 6.545(0.428) & 17.0 & 8.64 & 113.27 & 0.6\\
  AGESVC1\_226 & 12:13:03.00(0.7) & +07:01:52.0(10) & 2086(3) & 404(5) & 418(8) & 2.407(0.194) & 32.0 & 8.76 & 10.49 & 0.8\\
  AGESVC1\_227 & 12:23:37.20(0.7) & +06:57:01.0(10) & 1020(2) & 328(3) & 340(5) & 4.699(0.300) & 23.0 & 8.76 & 20.5 & 1.2\\
  AGESVC1\_228 & 12:24:04.80(0.7) & +07:02:06.0(10) & 1238(1) & 63(3) & 76(4) & 1.466(0.145) & 23.0 & 8.26 & 32.02 & 0.8\\
  AGESVC1\_229 & 12:23:47.20(0.7) & +07:11:04.0(10) & 1426(2) & 160(4) & 182(6) & 2.361(0.166) & 23.0 & 8.46 & 26.87 & 0.7\\
  AGESVC1\_230 & 12:16:24.00(0.8) & +07:12:29.0(12) & 1826(7) & 56(13) & 139(20) & 0.329(0.062) & 32.0 & 7.9 & 10.38 & 0.6\\
  AGESVC1\_231 & 12:18:17.90(1.3) & +07:21:40.0(11) & 1911(10) & 36(20) & 152(30) & 0.173(0.050) & 32.0 & 7.62 & 7.75 & 0.6\\
  AGESVC1\_232 & 12:23:58.20(0.7) & +07:26:55.0(10) & 1243(2) & 48(4) & 65(6) & 0.657(0.080) & 23.0 & 7.91 & 22.56 & 0.6\\
  AGESVC1\_233 & 12:25:53.60(0.7) & +07:33:04.0(10) & 1204(2) & 171(4) & 207(6) & 4.000(0.234) & 23.0 & 8.69 & 48.61 & 0.6\\
  AGESVC1\_234 & 12:27:11.30(0.7) & +07:38:27.0(10) & 1184(2) & 50(3) & 68(5) & 0.974(0.108) & 17.0 & 7.82 & 29.42 & 0.7\\
  AGESVC1\_235 & 12:20:06.20(0.7) & +07:33:51.0(11) & 1667(3) & 27(6) & 50(9) & 0.209(0.047) & 23.0 & 7.41 & 13.3 & 0.6\\
  AGESVC1\_236 & 12:27:15.40(0.7) & +07:40:09.0(10) & 1850(4) & 43(9) & 83(13) & 0.328(0.065) & 17.0 & 7.34 & 11.57 & 0.7\\
  AGESVC1\_237 & 12:14:09.00(0.7) & +07:46:20.0(10) & 1226(1) & 116(3) & 132(4) & 5.872(0.375) & 32.0 & 9.15 & 80.71 & 0.7\\
  AGESVC1\_238 & 12:23:59.70(0.7) & +07:47:06.0(10) & 1128(2) & 143(4) & 157(6) & 1.167(0.116) & 23.0 & 8.16 & 15.34 & 0.7\\
  AGESVC1\_239 & 12:26:04.80(1.0) & +07:53:59.0(20) & 1231(9) & 26(18) & 124(27) & 0.110(0.043) & 23.0 & 7.13 & 7.85 & 0.7\\
  AGESVC1\_240 & 12:26:46.30(0.7) & +07:55:05.0(10) & 1399(2) & 130(4) & 167(7) & 3.842(0.228) & 17.0 & 8.41 & 47.78 & 0.6\\
  AGESVC1\_241 & 12:26:01.20(0.7) & +08:10:07.0(10) & 1310(2) & 34(3) & 52(5) & 1.112(0.126) & 23.0 & 8.14 & 51.65 & 0.6\\
  AGESVC1\_242 & 12:24:04.50(0.7) & +08:17:40.0(10) & 1381(3) & 56(5) & 100(8) & 1.090(0.108) & 23.0 & 8.13 & 32.4 & 0.6\\
  AGESVC1\_243 & 12:26:46.90(0.7) & +08:15:45.0(10) & 1348(2) & 51(4) & 65(6) & 0.347(0.054) & 17.0 & 7.37 & 14.54 & 0.5\\
  AGESVC1\_244 & 12:22:38.00(0.7) & +08:17:47.0(10) & 1413(2) & 84(3) & 108(5) & 3.264(0.234) & 23.0 & 8.61 & 79.01 & 0.5\\
  AGESVC1\_245 & 12:23:04.10(0.9) & +08:20:12.0(13) & 1309(3) & 22(6) & 31(9) & 0.107(0.043) & 23.0 & 7.12 & 7.44 & 0.7\\
  AGESVC1\_246 & 12:26:18.40(0.7) & +08:21:00.0(10) & 1089(2) & 89(4) & 118(6) & 1.889(0.155) & 23.0 & 8.37 & 43.47 & 0.6\\
  AGESVC1\_247 & 12:24:59.20(1.0) & +08:22:38.0(11) & 1087(2) & 22(4) & 33(5) & 0.183(0.042) & 23.0 & 7.35 & 15.92 & 0.6\\
  AGESVC1\_248 & 12:24:13.90(0.8) & +08:31:52.0(11) & 1133(5) & 202(9) & 238(14) & 0.714(0.094) & 23.0 & 7.95 & 9.39 & 0.6\\
  AGESVC1\_249 & 12:28:18.60(0.7) & +08:43:36.0(10) & 1125(1) & 174(3) & 191(4) & 9.852(0.498) & 23.0 & 9.08 & 101.03 & 0.6\\
  AGESVC1\_250 & 12:26:37.90(0.8) & +08:52:33.0(10) & 1280(2) & 87(3) & 108(5) & 7.611(0.512) & 23.0 & 8.97 & 140.59 & 0.6\\
  AGESVC1\_251 & 12:28:55.30(0.7) & +08:49:20.0(10) & 1312(1) & 25(3) & 38(4) & 0.496(0.074) & 23.0 & 7.79 & 34.62 & 0.6\\
  AGESVC1\_252 & 12:16:38.40(0.7) & +08:49:42.0(10) & 1985(1) & 29(3) & 45(4) & 0.783(0.103) & 32.0 & 8.27 & 47.57 & 0.6\\
  AGESVC1\_253 & 12:14:13.00(0.7) & +08:54:31.0(10) & 1934(1) & 88(3) & 105(4) & 2.178(0.160) & 32.0 & 8.72 & 50.64 & 0.5\\
  AGESVC1\_255 & 12:22:04.80(0.7) & +09:03:04.0(11) & 1059(12) & 321(24) & 481(35) & 1.738(0.193) & 23.0 & 8.33 & 7.76 & 0.9\\
  AGESVC1\_256 & 12:37:02.70(0.7) & +06:52:43.0(26) & 1636(3) & 92(5) & 134(8) & 7.829(0.678) & 17.0 & 8.72 & 27.2 & 3.5\\
  AGESVC1\_257 & 12:36:55.10(0.8) & +07:25:48.0(12) & 1580(7) & 131(13) & 157(20) & 0.199(0.062) & 17.0 & 7.13 & 5.38 & 0.6\\
  AGESVC1\_258 & 12:38:07.20(1.0) & +07:30:45.0(14) & 1786(9) & 32(18) & 120(27) & 0.200(0.054) & 17.0 & 7.13 & 7.6 & 0.6\\
  AGESVC1\_259 & 12:43:06.70(0.7) & +07:39:00.0(10) & 1316(2) & 57(3) & 76(5) & 2.013(0.186) & 17.0 & 8.13 & 54.17 & 0.7\\
  AGESVC1\_260 & 12:32:44.80(0.8) & +07:48:13.0(12) & 1347(7) & 42(14) & 119(21) & 0.281(0.059) & 23.0 & 7.54 & 9.52 & 0.6\\
  AGESVC1\_261 & 12:31:49.00(0.9) & +07:50:33.0(12) & 1289(6) & 48(11) & 116(17) & 0.365(0.069) & 17.0 & 7.39 & 11.06 & 0.7\\
  AGESVC1\_262 & 12:32:27.20(0.7) & +07:51:52.0(10) & 1322(6) & 104(13) & 146(19) & 0.160(0.053) & 23.0 & 7.3 & 7.23 & 0.7\\
  AGESVC1\_263 & 12:38:20.40(0.7) & +07:53:16.0(10) & 1791(2) & 157(4) & 179(5) & 2.889(0.184) & 17.0 & 8.29 & 38.72 & 0.6\\
  AGESVC1\_265 & 12:36:33.00(0.7) & +08:03:09.0(12) & 1799(2) & 36(5) & 65(7) & 0.465(0.061) & 17.0 & 7.5 & 23.57 & 0.5\\
  AGESVC1\_266 & 12:36:06.50(0.9) & +08:00:07.0(13) & 1691(11) & 77(21) & 173(32) & 0.245(0.058) & 17.0 & 7.22 & 6.45 & 0.5\\
  AGESVC1\_267 & 12:30:59.60(0.7) & +08:04:38.0(12) & 1767(10) & 114(19) & 220(29) & 0.437(0.079) & 17.0 & 7.47 & 7.76 & 0.6\\
  AGESVC1\_268 & 12:44:45.50(0.7) & +08:05:58.0(11) & 1872(4) & 55(7) & 90(11) & 0.372(0.063) & 17.0 & 7.4 & 12.95 & 0.6\\
  AGESVC1\_269 & 12:34:19.60(0.7) & +08:11:35.0(11) & 1974(2) & 244(5) & 298(7) & 15.834(0.675) & 17.0 & 9.03 & 140.84 & 0.5\\
  AGESVC1\_270 & 12:48:22.60(0.7) & +08:29:18.0(10) & 1015(1) & 427(3) & 445(4) & 18.421(0.647) & 17.0 & 9.09 & 90.16 & 0.6\\
  AGESVC1\_271 & 12:46:01.30(0.7) & +08:27:43.0(35) & 1488(2) & 33(4) & 53(6) & 0.423(0.064) & 17.0 & 7.46 & 21.46 & 0.6\\
  AGESVC1\_272 & 12:37:41.60(0.7) & +08:33:27.0(10) & 1077(2) & 80(4) & 106(6) & 1.919(0.156) & 17.0 & 8.11 & 56.18 & 0.5\\
  AGESVC1\_273 & 12:33:27.90(0.7) & +08:40:00.0(11) & 1231(2) & 171(4) & 215(7) & 16.490(0.871) & 17.0 & 9.05 & 189.7 & 0.6\\
  AGESVC1\_274 & 12:30:25.60(0.8) & +08:38:05.0(12) & 1297(2) & 22(4) & 35(6) & 0.107(0.030) & 17.0 & 6.86 & 14.2 & 0.5\\
  AGESVC1\_275 & 12:43:17.00(0.7) & +08:54:53.0(30) & 1429(2) & 26(4) & 41(6) & 0.307(0.054) & 17.0 & 7.32 & 20.13 & 0.6\\
  AGESVC1\_276 & 12:25:42.70(0.7) & +07:12:27.0(10) & 995(1) & 285(3) & 304(4) & 15.822(0.682) & 23.0 & 9.29 & 101.03 & 0.7\\
  AGESVC1\_277 & 12:27:10.80(0.7) & +07:15:37.0(10) & 937(1) & 153(3) & 170(4) & 11.911(0.618) & 23.0 & 9.17 & 130.48 & 0.6\\
  AGESVC1\_278 & 12:27:29.40(0.7) & +07:38:35.0(10) & 873(2) & 132(3) & 149(5) & 2.428(0.171) & 17.0 & 8.21 & 29.58 & 0.7\\
  AGESVC1\_279 & 12:29:29.90(0.7) & +07:41:36.0(10) & 760(2) & 172(4) & 198(6) & 2.492(0.160) & 17.0 & 8.23 & 29.23 & 0.6\\
  AGESVC1\_280 & 12:24:45.60(0.8) & +07:55:34.0(12) & 801(2) & 21(4) & 35(6) & 0.201(0.049) & 23.0 & 7.39 & 14.51 & 0.7\\
  AGESVC1\_281 & 12:29:54.40(0.9) & +07:58:05.0(14) & 475(2) & 29(5) & 63(7) & 0.754(0.104) & 17.0 & 7.71 & 33.16 & 0.8\\
  AGESVC1\_282 & 12:25:24.10(0.9) & +08:16:54.0(48) & 943(7) & 69(13) & 164(20) & 0.351(0.057) & 23.0 & 7.64 & 11.42 & 0.5\\
  AGESVC1\_283 & 12:16:11.40(0.7) & +08:21:43.0(28) & 871(2) & 27(4) & 43(6) & 0.337(0.062) & 32.0 & 7.91 & 18.3 & 0.7\\
  AGESVC1\_284 & 12:24:38.80(0.7) & +08:30:13.0(10) & 880(2) & 114(3) & 130(5) & 1.516(0.123) & 23.0 & 8.27 & 26.88 & 0.6\\
  AGESVC1\_285 & 12:28:35.50(0.7) & +08:38:09.0(12) & 564(2) & 26(4) & 45(6) & 0.324(0.056) & 17.0 & 7.34 & 21.56 & 0.6\\
  AGESVC1\_286 & 12:21:57.60(0.7) & +08:40:23.0(10) & 856(1) & 41(3) & 55(4) & 0.956(0.108) & 23.0 & 8.07 & 40.95 & 0.6\\
  AGESVC1\_288 & 12:29:29.60(0.7) & +08:54:13.0(10) & 980(6) & 43(12) & 129(18) & 0.255(0.048) & 23.0 & 7.5 & 11.78 & 0.5\\
  AGESVC1\_289 & 12:24:54.80(0.9) & +07:25:42.0(26) & 784(15) & 303(31) & 490(46) & 0.711(0.118) & 23.0 & 7.94 & 6.22 & 0.7\\
  AGESVC1\_290 & 12:29:26.80(0.8) & +08:44:10.0(11) & 648(4) & 403(8) & 414(12) & 0.737(0.114) & 23.0 & 7.96 & 5.89 & 0.6\\
  AGESVC1\_291 & 12:37:45.90(0.7) & +07:06:19.0(10) & 65(2) & 68(5) & 113(7) & 4.981(0.386) & 17.0 & 8.53 & 121.29 & 0.6\\
  AGESVC1\_292 & 12:34:39.90(0.7) & +07:09:28.0(10) & 600(3) & 89(6) & 141(10) & 1.326(0.123) & 17.0 & 7.95 & 23.41 & 0.7\\
  AGESVC1\_293 & 12:29:59.10(0.7) & +08:26:01.0(12) & 609(3) & 82(6) & 102(9) & 0.439(0.063) & 17.0 & 7.47 & 12.34 & 0.5\\
  AGESVC1\_294 & 12:38:12.10(1.1) & +06:59:39.0(11) & -114(6) & 47(11) & 91(17) & 0.323(0.084) & 17.0 & 7.34 & 8.77 & 1.0\\
  GLADOS\_001 & 12:19:06.10(0.7) & +07:37:55.0(10) & 2675(2) & 53(4) & 71(6) & 0.523(0.065) & 23.0 & 7.81 & 18.91 & 0.5\\
  GLADOS\_003 & 12:12:24.70(1.2) & +06:58:37.0(15) & 2088(15) & 20(29) & 252(44) & 0.291(0.084) & 32.0 & 7.84 & 7.45 & 1.0\\
  GLADOS\_005 & 12:28:27.00(0.9) & +06:57:03.0(18) & 880(4) & 31(7) & 41(11) & 0.110(0.061) & 17.0 & 6.87 & 6.26 & 1.2\\
  GLADOS\_006 & 12:21:09.10(3.6) & +08:26:27.0(14) & 1155(9) & 27(19) & 97(28) & 0.103(0.042) & 23.0 & 7.1 & 6.26 & 0.6\\
  AGESVC1\_304 & 12:22:15.10(1.3) & +07:07:32.0(33) & 1082(14) & 227(27) & 292(41) & 0.303(0.093) & 23.0 & 7.57 & 4.08 & 0.7\\
  AGESVC1\_305 & 12:17:09.50(0.9) & +07:11:32.0(15) & 2306(12) & 408(24) & 487(36) & 0.520(0.102) & 32.0 & 8.09 & 5.12 & 0.6\\
  AGESVC1\_306 & 12:12:26.50(1.2) & +06:59:05.0(12) & 2089(13) & 19(26) & 214(40) & 0.388(0.092) & 32.0 & 7.97 & 7.57 & 0.9\\
\end{tabular}
\end{table*}
\end{center}

\begin{table*}
\tiny
\begin{center}
\caption[HI properties of AGES detections behind the Virgo Cluster in VC1]{HI properties of galaxies behind the Virgo Cluster. Columns are as for table \ref{VirgoHIT} (continued on next page).}
\label{BackHIT}
\begin{tabular}{c c c c c c c c c c c}\\
\hline
  \multicolumn{1}{c}{(1) No.} & 
  \multicolumn{1}{c}{(2) R.A.} &
  \multicolumn{1}{c}{(3) Dec.} &
  \multicolumn{1}{c}{(4) Velocity} &
  \multicolumn{1}{c}{(5) W50} &
  \multicolumn{1}{c}{(6) W20} &
  \multicolumn{1}{c}{(7) Flux} &
  \multicolumn{1}{c}{(8) Dist} &
  \multicolumn{1}{c}{(9) $M_{\rm{HI}}$} &
  \multicolumn{1}{c}{(10) S/N} &
  \multicolumn{1}{c}{(11) rms} \\
\hline
  AGESVC1\_004 & 12:42:53.50(0.9) & +08:38:00.0(15) & 18972(10) & 366(21) & 477(31) & 0.643(0.106) & 267.2 & 10.0 & 7.34 & 0.7\\
  AGESVC1\_005 & 12:25:15.10(2.1) & +09:01:45.0(20) & 17955(12) & 248(23) & 396(35) & 0.778(0.131) & 252.8 & 10.0 & 7.46 & 0.9\\
  AGESVC1\_009 & 12:43:34.70(0.8) & +08:12:30.0(12) & 16360(6) & 342(12) & 372(18) & 0.675(0.115) & 230.4 & 9.9 & 6.29 & 0.7\\
  AGESVC1\_010 & 12:35:21.80(1.1) & +08:16:25.0(28) & 16419(3) & 46(6) & 63(10) & 0.096(0.037) & 231.2 & 9.0 & 9.7 & 0.7\\
  AGESVC1\_011 & 12:38:39.40(0.9) & +08:32:22.0(13) & 16405(9) & 273(17) & 318(26) & 0.574(0.104) & 231.0 & 9.8 & 5.47 & 0.6\\
  AGESVC1\_012 & 12:46:05.50(0.9) & +08:37:15.0(12) & 15377(5) & 338(10) & 360(15) & 0.501(0.091) & 216.5 & 9.7 & 6.41 & 0.6\\
  AGESVC1\_013 & 12:24:35.60(0.7) & +07:32:42.0(11) & 14629(8) & 245(17) & 344(25) & 0.966(0.121) & 206.0 & 9.9 & 8.8 & 0.7\\
  AGESVC1\_014 & 12:23:00.60(1.1) & +08:11:40.0(12) & 14728(13) & 305(25) & 427(38) & 0.578(0.090) & 207.4 & 9.7 & 6.19 & 0.5\\
  AGESVC1\_015 & 12:22:10.00(0.7) & +08:24:57.0(13) & 14722(10) & 299(21) & 383(31) & 0.288(0.075) & 207.3 & 9.4 & 6.27 & 0.7\\
  AGESVC1\_016 & 12:17:50.70(0.7) & +08:25:52.0(11) & 14756(7) & 448(14) & 505(21) & 1.331(0.137) & 207.8 & 10.1 & 7.87 & 0.6\\
  AGESVC1\_017 & 12:20:47.20(0.7) & +08:42:04.0(10) & 14820(2) & 65(4) & 84(5) & 1.082(0.105) & 208.7 & 10.0 & 28.83 & 0.6\\
  AGESVC1\_018 & 12:20:30.00(0.7) & +08:49:18.0(11) & 14779(8) & 223(17) & 443(25) & 1.654(0.128) & 208.1 & 10.2 & 15.1 & 0.6\\
  AGESVC1\_019 & 12:20:20.90(1.1) & +08:53:46.0(30) & 14866(12) & 135(24) & 353(37) & 0.754(0.107) & 209.3 & 9.8 & 8.9 & 0.7\\
  AGESVC1\_020 & 12:10:04.70(0.7) & +07:09:01.0(12) & 14388(9) & 83(19) & 142(28) & 0.228(0.070) & 202.6 & 9.3 & 5.69 & 0.7\\
  AGESVC1\_022 & 12:23:49.70(0.9) & +07:47:09.0(14) & 14694(16) & 440(31) & 570(47) & 0.528(0.104) & 206.9 & 9.7 & 5.02 & 0.6\\
  AGESVC1\_023 & 12:38:13.90(0.8) & +07:09:06.0(12) & 14653(12) & 224(24) & 305(35) & 0.430(0.099) & 206.3 & 9.6 & 5.31 & 0.7\\
  AGESVC1\_024 & 12:41:02.40(0.7) & +08:04:13.0(10) & 14528(15) & 190(29) & 401(44) & 0.750(0.107) & 204.6 & 9.8 & 7.08 & 0.6\\
  AGESVC1\_025 & 12:40:19.00(2.7) & +08:10:01.0(13) & 14134(8) & 512(15) & 591(23) & 1.793(0.167) & 199.0 & 10.2 & 8.43 & 0.7\\
  AGESVC1\_026 & 12:39:43.20(0.8) & +08:25:36.0(13) & 14064(7) & 236(14) & 340(22) & 0.962(0.096) & 198.0 & 9.9 & 10.71 & 0.5\\
  AGESVC1\_027 & 12:48:12.20(1.3) & +08:25:24.0(24) & 14472(11) & 209(22) & 284(33) & 0.550(0.110) & 203.8 & 9.7 & 5.42 & 0.7\\
  AGESVC1\_028 & 12:39:54.60(1.7) & +08:30:49.0(13) & 14246(15) & 59(30) & 232(44) & 0.292(0.070) & 200.6 & 9.4 & 6.26 & 0.6\\
  AGESVC1\_029 & 12:35:09.70(0.7) & +08:41:07.0(11) & 14882(9) & 239(19) & 381(28) & 0.853(0.094) & 209.6 & 9.9 & 9.4 & 0.5\\
  AGESVC1\_030 & 12:39:12.70(0.7) & +08:56:18.0(10) & 14109(10) & 299(19) & 436(29) & 1.194(0.134) & 198.7 & 10.0 & 8.93 & 0.7\\
  AGESVC1\_031 & 12:40:49.00(0.7) & +08:58:24.0(10) & 14105(6) & 110(12) & 146(18) & 0.227(0.068) & 198.6 & 9.3 & 6.99 & 0.8\\
  AGESVC1\_032 & 12:33:17.00(1.1) & +07:07:46.0(10) & 14707(13) & 42(26) & 182(40) & 0.209(0.068) & 207.1 & 9.3 & 6.3 & 0.8\\
  AGESVC1\_033 & 12:38:33.50(0.9) & +08:21:04.0(11) & 14559(24) & 80(47) & 542(71) & 0.327(0.079) & 205.0 & 9.5 & 6.43 & 0.7\\
  AGESVC1\_035 & 12:11:27.20(0.8) & +07:56:34.0(14) & 13701(9) & 71(19) & 163(28) & 0.202(0.055) & 192.9 & 9.2 & 7.32 & 0.6\\
  AGESVC1\_037 & 12:11:00.60(0.8) & +08:16:50.0(11) & 13550(10) & 240(20) & 310(29) & 0.459(0.089) & 190.8 & 9.5 & 6.01 & 0.6\\
  AGESVC1\_038 & 12:14:04.00(0.7) & +08:48:57.0(10) & 13805(7) & 264(15) & 452(22) & 1.809(0.133) & 194.4 & 10.2 & 15.92 & 0.6\\
  AGESVC1\_039 & 12:48:04.10(0.7) & +06:58:47.0(10) & 13434(2) & 113(5) & 139(7) & 1.903(0.175) & 189.2 & 10.2 & 20.42 & 1.0\\
  AGESVC1\_040 & 12:43:35.60(0.7) & +07:26:10.0(11) & 14012(10) & 285(20) & 398(30) & 1.024(0.131) & 197.3 & 9.9 & 7.74 & 0.7\\
  AGESVC1\_041 & 12:37:45.20(0.8) & +07:42:51.0(11) & 13476(2) & 32(4) & 48(7) & 0.302(0.059) & 189.8 & 9.4 & 14.95 & 0.7\\
  AGESVC1\_042 & 12:45:31.70(0.7) & +07:41:41.0(14) & 13189(9) & 200(18) & 253(28) & 0.419(0.096) & 185.7 & 9.5 & 5.49 & 0.7\\
  AGESVC1\_043 & 12:43:12.00(0.7) & +07:59:13.0(10) & 13239(8) & 274(15) & 349(23) & 0.912(0.112) & 186.4 & 9.8 & 8.1 & 0.6\\
  AGESVC1\_044 & 12:43:06.20(1.1) & +08:05:28.0(12) & 13360(13) & 178(27) & 303(40) & 0.359(0.080) & 188.1 & 9.4 & 5.81 & 0.6\\
  AGESVC1\_045 & 12:40:19.80(0.7) & +08:12:13.0(10) & 13735(5) & 417(11) & 489(16) & 2.138(0.151) & 193.4 & 10.2 & 12.41 & 0.6\\
  AGESVC1\_046 & 12:42:48.00(0.7) & +08:14:26.0(12) & 13360(11) & 273(21) & 402(32) & 0.655(0.097) & 188.1 & 9.7 & 7.66 & 0.6\\
  AGESVC1\_047 & 12:41:19.00(0.7) & +08:19:46.0(11) & 13258(5) & 93(10) & 190(16) & 0.768(0.079) & 186.7 & 9.8 & 16.33 & 0.5\\
  AGESVC1\_048 & 12:41:34.50(0.7) & +08:21:53.0(13) & 13419(10) & 275(19) & 409(29) & 1.154(0.122) & 189.0 & 9.9 & 8.88 & 0.6\\
  AGESVC1\_049 & 12:38:31.70(0.8) & +08:35:29.0(11) & 13133(9) & 198(17) & 268(26) & 0.557(0.093) & 184.9 & 9.6 & 6.83 & 0.6\\
  AGESVC1\_051 & 12:38:10.30(0.7) & +08:50:17.0(10) & 13137(6) & 337(12) & 439(19) & 2.300(0.168) & 185.0 & 10.2 & 13.04 & 0.7\\
  AGESVC1\_052 & 12:43:58.70(0.8) & +08:54:44.0(18) & 13830(12) & 369(24) & 440(36) & 0.609(0.122) & 194.7 & 9.7 & 4.84 & 0.7\\
  AGESVC1\_053 & 12:47:24.80(0.9) & +08:52:20.0(12) & 13348(21) & 302(43) & 502(64) & 0.489(0.112) & 188.0 & 9.6 & 4.56 & 0.7\\
  AGESVC1\_056 & 12:17:28.60(1.5) & +07:04:38.0(14) & 12138(4) & 228(9) & 239(13) & 0.211(0.064) & 170.9 & 9.1 & 5.34 & 0.6\\
  AGESVC1\_057 & 12:48:21.90(0.8) & +07:06:25.0(11) & 12772(11) & 72(23) & 198(34) & 0.299(0.073) & 179.8 & 9.3 & 6.97 & 0.7\\
  AGESVC1\_058 & 12:33:54.50(1.4) & +08:56:57.0(15) & 12872(5) & 103(10) & 142(15) & 0.292(0.060) & 181.2 & 9.3 & 9.45 & 0.6\\
  AGESVC1\_059 & 12:34:43.80(0.7) & +09:00:27.0(11) & 12917(4) & 369(9) & 391(13) & 0.930(0.133) & 181.9 & 9.8 & 7.66 & 0.8\\
  AGESVC1\_063 & 12:14:46.00(0.7) & +07:21:28.0(11) & 11127(12) & 77(25) & 212(37) & 0.196(0.056) & 156.7 & 9.0 & 6.6 & 0.6\\
  AGESVC1\_064 & 12:22:19.10(0.7) & +07:37:23.0(10) & 11285(2) & 161(4) & 174(6) & 1.086(0.103) & 158.9 & 9.8 & 16.62 & 0.6\\
  AGESVC1\_065 & 12:25:31.20(0.9) & +07:38:54.0(11) & 11376(4) & 149(8) & 162(12) & 0.421(0.083) & 160.2 & 9.4 & 6.42 & 0.6\\
  AGESVC1\_067 & 12:22:23.20(0.7) & +07:53:00.0(11) & 11367(4) & 74(9) & 118(13) & 0.612(0.088) & 160.0 & 9.5 & 11.83 & 0.7\\
  AGESVC1\_068 & 12:13:22.60(1.0) & +07:51:08.0(12) & 11328(12) & 108(25) & 224(37) & 0.364(0.086) & 159.5 & 9.3 & 6.08 & 0.7\\
  AGESVC1\_069 & 12:26:08.20(0.8) & +07:57:51.0(10) & 11357(10) & 188(19) & 229(29) & 0.211(0.074) & 159.9 & 9.1 & 4.54 & 0.7\\
  AGESVC1\_070 & 12:12:37.00(1.2) & +08:00:45.0(56) & 11163(6) & 90(12) & 149(19) & 0.354(0.055) & 157.2 & 9.3 & 9.06 & 0.4\\
  AGESVC1\_071 & 12:14:13.30(0.7) & +08:06:28.0(11) & 11156(7) & 216(13) & 249(20) & 0.538(0.096) & 157.1 & 9.4 & 6.15 & 0.6\\
  AGESVC1\_072 & 12:30:03.90(0.7) & +08:13:08.0(10) & 11607(6) & 288(11) & 341(17) & 0.924(0.107) & 163.4 & 9.7 & 9.59 & 0.6\\
  AGESVC1\_073 & 12:27:47.60(1.1) & +08:13:22.0(13) & 11207(10) & 185(20) & 231(29) & 0.229(0.070) & 157.8 & 9.1 & 4.81 & 0.6\\
  AGESVC1\_074 & 12:16:54.10(0.7) & +08:22:06.0(10) & 11294(4) & 357(8) & 418(12) & 2.162(0.133) & 159.0 & 10.1 & 16.57 & 0.5\\
  AGESVC1\_075 & 12:27:48.30(0.7) & +08:52:41.0(10) & 11615(5) & 175(10) & 234(14) & 1.121(0.109) & 163.5 & 9.8 & 12.56 & 0.6\\
  AGESVC1\_076 & 12:21:03.20(0.7) & +09:06:20.0(10) & 11263(9) & 99(19) & 211(28) & 0.677(0.153) & 158.6 & 9.6 & 8.08 & 1.5\\
  AGESVC1\_077 & 12:29:59.00(0.8) & +08:05:43.0(11) & 11205(16) & 151(33) & 510(49) & 0.565(0.087) & 157.8 & 9.5 & 8.43 & 0.6\\
  AGESVC1\_078 & 12:15:41.70(1.4) & +08:18:52.0(12) & 11197(12) & 180(24) & 273(36) & 0.400(0.087) & 157.7 & 9.3 & 5.52 & 0.6\\
  AGESVC1\_079 & 12:44:45.20(0.7) & +07:21:53.0(11) & 11460(7) & 294(14) & 350(22) & 0.752(0.105) & 161.4 & 9.6 & 7.4 & 0.6\\
  AGESVC1\_080 & 12:38:15.70(0.7) & +07:48:57.0(10) & 11292(10) & 362(21) & 590(31) & 1.722(0.139) & 159.0 & 10.0 & 11.25 & 0.6\\
  AGESVC1\_081 & 12:42:43.00(0.8) & +07:51:21.0(40) & 11381(9) & 81(17) & 151(26) & 0.268(0.070) & 160.2 & 9.2 & 6.92 & 0.7\\
  AGESVC1\_082 & 12:33:26.60(0.8) & +07:50:50.0(12) & 11466(4) & 99(7) & 108(11) & 0.187(0.062) & 161.4 & 9.0 & 5.98 & 0.7\\
  AGESVC1\_083 & 12:31:38.20(0.7) & +07:51:08.0(11) & 11171(8) & 181(16) & 234(23) & 0.463(0.093) & 157.3 & 9.4 & 6.58 & 0.7\\
  AGESVC1\_084 & 12:36:37.30(0.7) & +07:58:43.0(11) & 11368(3) & 335(6) & 358(9) & 1.689(0.132) & 160.1 & 10.0 & 13.68 & 0.6\\
  AGESVC1\_085 & 12:33:51.10(0.7) & +08:01:35.0(11) & 11367(4) & 99(8) & 166(13) & 0.935(0.095) & 160.0 & 9.7 & 17.27 & 0.6\\
  AGESVC1\_086 & 12:32:01.90(2.2) & +08:56:07.0(15) & 11879(4) & 230(9) & 243(13) & 0.507(0.102) & 167.3 & 9.5 & 5.94 & 0.7\\
  AGESVC1\_088 & 12:19:02.00(0.7) & +07:48:05.0(10) & 10852(3) & 207(6) & 230(9) & 0.947(0.101) & 152.8 & 9.7 & 12.39 & 0.6\\
  AGESVC1\_089 & 12:18:42.40(0.7) & +07:55:29.0(11) & 10765(11) & 66(22) & 194(33) & 0.332(0.076) & 151.6 & 9.2 & 7.27 & 0.7\\
  AGESVC1\_090 & 12:15:40.90(0.7) & +08:07:57.0(10) & 11128(2) & 484(4) & 507(6) & 4.786(0.220) & 156.7 & 10.4 & 27.66 & 0.6\\
  AGESVC1\_091 & 12:35:16.10(0.7) & +08:09:21.0(10) & 10987(1) & 55(3) & 70(4) & 1.236(0.120) & 154.7 & 9.8 & 38.34 & 0.6\\
  AGESVC1\_092 & 12:10:23.60(0.8) & +08:09:18.0(16) & 10880(6) & 93(11) & 119(17) & 0.338(0.075) & 153.2 & 9.2 & 6.27 & 0.6\\
  AGESVC1\_093 & 12:17:04.40(0.8) & +08:30:48.0(12) & 10227(4) & 163(8) & 177(12) & 0.213(0.053) & 144.0 & 9.0 & 6.8 & 0.5\\
  AGESVC1\_094 & 12:15:30.50(0.9) & +08:29:50.0(13) & 10280(9) & 276(18) & 319(26) & 0.224(0.067) & 144.7 & 9.0 & 5.16 & 0.6\\
  AGESVC1\_095 & 12:12:22.90(0.7) & +08:46:17.0(10) & 10352(3) & 276(6) & 291(9) & 1.015(0.109) & 145.8 & 9.7 & 10.41 & 0.6\\
  AGESVC1\_096 & 12:27:55.00(0.7) & +08:05:09.0(10) & 10937(6) & 84(12) & 114(18) & 0.158(0.051) & 154.0 & 8.9 & 6.33 & 0.6\\
  AGESVC1\_098 & 12:45:14.20(0.7) & +08:24:14.0(12) & 10716(4) & 220(8) & 261(11) & 1.137(0.099) & 150.9 & 9.7 & 13.84 & 0.5\\
  AGESVC1\_099 & 12:34:30.70(0.7) & +08:23:25.0(18) & 10301(6) & 175(13) & 219(19) & 0.521(0.088) & 145.0 & 9.4 & 7.32 & 0.6\\
  AGESVC1\_100 & 12:48:26.40(0.8) & +08:25:04.0(11) & 10124(5) & 75(11) & 116(16) & 0.327(0.066) & 142.5 & 9.1 & 8.51 & 0.6\\
  AGESVC1\_101 & 12:36:00.10(0.7) & +08:35:54.0(10) & 10432(2) & 75(5) & 101(7) & 0.787(0.087) & 146.9 & 9.6 & 20.8 & 0.6\\
  AGESVC1\_102 & 12:43:04.60(0.7) & +08:38:03.0(10) & 10140(3) & 199(6) & 240(10) & 1.959(0.137) & 142.8 & 9.9 & 18.41 & 0.6\\
  AGESVC1\_103 & 12:35:28.80(0.7) & +08:41:33.0(10) & 10352(3) & 156(5) & 168(8) & 0.668(0.088) & 145.8 & 9.5 & 10.72 & 0.6\\
  AGESVC1\_104 & 12:32:39.50(1.0) & +08:39:29.0(11) & 10218(5) & 172(11) & 187(16) & 0.184(0.061) & 143.9 & 8.9 & 5.08 & 0.6\\
  AGESVC1\_105 & 12:33:17.10(0.8) & +08:45:27.0(14) & 10013(4) & 118(7) & 155(11) & 1.000(0.110) & 141.0 & 9.6 & 13.33 & 0.7\\
  AGESVC1\_106 & 12:32:50.60(1.0) & +08:45:45.0(34) & 10119(11) & 253(23) & 362(34) & 0.766(0.121) & 142.5 & 9.5 & 6.46 & 0.7\\
  AGESVC1\_107 & 12:29:15.70(0.9) & +08:15:26.0(14) & 10058(6) & 163(11) & 185(17) & 0.370(0.081) & 141.6 & 9.2 & 5.88 & 0.6\\
  AGESVC1\_108 & 12:29:52.10(0.7) & +08:19:49.0(10) & 9988(3) & 314(6) & 339(9) & 1.691(0.131) & 140.6 & 9.8 & 14.26 & 0.6\\
  AGESVC1\_109 & 12:25:10.50(0.8) & +08:21:44.0(11) & 9992(17) & 129(34) & 291(50) & 0.249(0.070) & 140.7 & 9.0 & 5.28 & 0.6\\
  AGESVC1\_110 & 12:27:05.00(0.8) & +08:40:43.0(11) & 9969(3) & 76(6) & 100(9) & 0.602(0.079) & 140.4 & 9.4 & 13.66 & 0.6\\
  AGESVC1\_111 & 12:15:06.50(0.7) & +08:58:26.0(10) & 9482(4) & 177(9) & 221(13) & 1.071(0.116) & 133.5 & 9.6 & 12.07 & 0.7\\
  AGESVC1\_112 & 12:27:52.30(0.8) & +09:03:36.0(11) & 9825(6) & 89(12) & 146(18) & 0.720(0.136) & 138.3 & 9.5 & 9.05 & 1.2\\
  AGESVC1\_113 & 12:15:16.80(0.7) & +09:04:46.0(13) & 9455(12) & 115(24) & 188(35) & 0.409(0.140) & 133.1 & 9.2 & 4.98 & 1.4\\
  AGESVC1\_114 & 12:34:05.40(0.7) & +08:00:29.0(10) & 9175(10) & 260(20) & 341(30) & 0.723(0.110) & 129.2 & 9.4 & 6.32 & 0.6\\
  AGESVC1\_115 & 12:32:59.60(0.7) & +08:09:34.0(15) & 9086(8) & 60(16) & 124(24) & 0.183(0.048) & 127.9 & 8.8 & 7.01 & 0.5\\
  AGESVC1\_116 & 12:46:37.80(0.8) & +08:54:11.0(16) & 9937(5) & 63(10) & 81(15) & 0.177(0.056) & 139.9 & 8.9 & 5.77 & 0.6\\
  AGESVC1\_117 & 12:36:05.90(0.7) & +07:42:47.0(10) & 8779(2) & 94(4) & 112(6) & 1.164(0.106) & 123.6 & 9.6 & 21.31 & 0.6\\
  AGESVC1\_118 & 12:39:30.10(0.7) & +08:08:11.0(10) & 8836(7) & 197(13) & 242(20) & 0.658(0.107) & 124.4 & 9.3 & 7.28 & 0.7\\
\end{tabular}
\end{center}
\end{table*}

\addtocounter{table}{-1}
\begin{table*}
\tiny
\begin{center}
\caption[HI properties of AGES detections behind the Virgo Cluster in VC1]{(continued from previous page) HI properties of galaxies behind the Virgo Cluster. Columns are as for table \ref{VirgoHIT}.}
\label{BackHIT}
\begin{tabular}{c c c c c c c c c c c}\\
\hline
  \multicolumn{1}{c}{(1) No.} & 
  \multicolumn{1}{c}{(2) R.A.} &
  \multicolumn{1}{c}{(3) Dec.} &
  \multicolumn{1}{c}{(4) Velocity} &
  \multicolumn{1}{c}{(5) W50} &
  \multicolumn{1}{c}{(6) W20} &
  \multicolumn{1}{c}{(7) Flux} &
  \multicolumn{1}{c}{(8) Dist} &
  \multicolumn{1}{c}{(9) $M_{\rm{HI}}$} &
  \multicolumn{1}{c}{(10) S/N} &
  \multicolumn{1}{c}{(11) rms} \\
\hline
  AGESVC1\_120 & 12:37:18.50(1.5) & +08:30:59.0(13) & 8454(10) & 76(21) & 152(31) & 0.217(0.056) & 119.0 & 8.8 & 5.92 & 0.5\\
  AGESVC1\_121 & 12:37:11.00(0.7) & +08:39:30.0(10) & 8480(3) & 356(6) & 370(9) & 1.223(0.137) & 119.4 & 9.6 & 8.76 & 0.7\\
  AGESVC1\_122 & 12:42:23.90(0.8) & +08:56:29.0(11) & 8854(11) & 124(23) & 225(34) & 0.394(0.088) & 124.7 & 9.1 & 6.2 & 0.7\\
  AGESVC1\_123 & 12:19:50.60(0.7) & +06:59:17.0(10) & 7204(2) & 57(3) & 77(5) & 4.547(0.401) & 101.4 & 10.0 & 74.93 & 1.1\\
  AGESVC1\_124 & 12:25:42.50(0.7) & +07:10:01.0(10) & 7162(3) & 281(6) & 306(10) & 1.366(0.121) & 100.8 & 9.5 & 12.48 & 0.6\\
  AGESVC1\_125 & 12:23:23.40(1.1) & +07:10:10.0(12) & 7258(8) & 116(17) & 185(25) & 0.253(0.067) & 102.2 & 8.7 & 7.02 & 0.7\\
  AGESVC1\_126 & 12:31:32.70(0.7) & +07:18:12.0(10) & 7319(4) & 188(8) & 240(12) & 1.303(0.114) & 103.0 & 9.5 & 15.49 & 0.6\\
  AGESVC1\_127 & 12:20:27.60(0.7) & +07:30:03.0(10) & 7325(2) & 57(3) & 77(5) & 1.287(0.120) & 103.1 & 9.5 & 35.85 & 0.6\\
  AGESVC1\_128 & 12:28:35.30(0.7) & +07:29:11.0(11) & 7537(8) & 183(15) & 260(23) & 0.779(0.102) & 106.1 & 9.3 & 8.44 & 0.6\\
  AGESVC1\_129 & 12:28:13.70(0.7) & +07:36:33.0(10) & 7429(3) & 117(6) & 142(9) & 0.961(0.107) & 104.6 & 9.3 & 13.94 & 0.7\\
  AGESVC1\_130 & 12:19:28.00(0.7) & +07:47:32.0(10) & 7276(3) & 154(6) & 195(9) & 1.580(0.123) & 102.4 & 9.5 & 18.91 & 0.6\\
  AGESVC1\_131 & 12:29:00.50(0.7) & +07:51:06.0(10) & 7457(4) & 329(8) & 342(12) & 0.669(0.106) & 105.0 & 9.2 & 6.22 & 0.6\\
  AGESVC1\_132 & 12:24:13.00(0.7) & +07:57:07.0(10) & 7221(2) & 151(4) & 187(7) & 3.184(0.196) & 101.7 & 9.8 & 40.67 & 0.6\\
  AGESVC1\_133 & 12:29:24.60(0.8) & +08:00:53.0(18) & 7355(9) & 212(18) & 325(28) & 0.598(0.082) & 103.5 & 9.1 & 8.42 & 0.5\\
  AGESVC1\_134 & 12:22:02.70(0.7) & +08:13:48.0(10) & 7292(4) & 140(7) & 169(11) & 0.916(0.101) & 102.7 & 9.3 & 11.34 & 0.6\\
  AGESVC1\_135 & 12:26:05.30(0.8) & +08:15:36.0(17) & 7372(6) & 246(12) & 274(18) & 0.445(0.086) & 103.8 & 9.0 & 6.43 & 0.6\\
  AGESVC1\_136 & 12:17:41.50(2.2) & +08:23:22.0(13) & 7958(3) & 134(6) & 150(9) & 0.893(0.119) & 112.0 & 9.4 & 10.24 & 0.8\\
  AGESVC1\_137 & 12:28:07.30(1.2) & +08:25:40.0(13) & 7393(8) & 116(17) & 213(25) & 0.530(0.084) & 104.1 & 9.1 & 8.55 & 0.6\\
  AGESVC1\_138 & 12:20:03.80(0.7) & +08:37:18.0(10) & 7377(2) & 293(4) & 333(7) & 7.754(0.340) & 103.9 & 10.2 & 68.81 & 0.5\\
  AGESVC1\_139 & 12:27:13.80(0.7) & +08:36:44.0(13) & 7176(6) & 139(13) & 197(19) & 0.388(0.071) & 101.0 & 8.9 & 8.82 & 0.6\\
  AGESVC1\_140 & 12:30:23.70(2.9) & +08:40:28.0(11) & 7302(9) & 149(19) & 238(28) & 0.335(0.071) & 102.8 & 8.9 & 7.23 & 0.6\\
  AGESVC1\_141 & 12:26:05.10(0.8) & +08:26:32.0(15) & 7615(7) & 320(14) & 356(20) & 0.537(0.096) & 107.2 & 9.1 & 6.18 & 0.6\\
  AGESVC1\_142 & 12:25:20.30(1.2) & +08:33:04.0(11) & 7563(9) & 105(17) & 149(26) & 0.300(0.076) & 106.5 & 8.9 & 5.33 & 0.6\\
  AGESVC1\_143 & 12:24:48.00(0.7) & +08:48:01.0(12) & 7197(5) & 148(11) & 191(16) & 0.391(0.071) & 101.3 & 8.9 & 9.1 & 0.6\\
  AGESVC1\_144 & 12:25:50.20(0.7) & +08:55:33.0(10) & 7573(5) & 269(11) & 352(16) & 1.611(0.129) & 106.6 & 9.6 & 13.45 & 0.6\\
  AGESVC1\_145 & 12:19:51.40(1.8) & +08:56:26.0(13) & 7413(5) & 155(9) & 166(14) & 0.340(0.090) & 104.4 & 8.9 & 4.98 & 0.7\\
  AGESVC1\_146 & 12:38:53.50(0.7) & +07:06:53.0(11) & 7185(9) & 321(17) & 424(26) & 1.062(0.118) & 101.1 & 9.4 & 8.65 & 0.6\\
  AGESVC1\_147 & 12:39:21.60(0.7) & +07:08:42.0(12) & 7219(5) & 122(11) & 152(16) & 0.427(0.086) & 101.6 & 9.0 & 7.41 & 0.7\\
  AGESVC1\_148 & 12:39:38.90(0.7) & +07:09:59.0(10) & 7268(3) & 270(7) & 323(10) & 2.988(0.183) & 102.3 & 9.8 & 21.2 & 0.7\\
  AGESVC1\_149 & 12:36:08.90(0.7) & +07:32:07.0(10) & 7214(2) & 301(5) & 322(7) & 1.937(0.149) & 101.6 & 9.6 & 16.05 & 0.7\\
  AGESVC1\_150 & 12:48:20.90(0.8) & +07:35:35.0(11) & 7101(7) & 174(14) & 235(20) & 0.546(0.093) & 100.0 & 9.1 & 8.35 & 0.7\\
  AGESVC1\_151 & 12:48:19.60(0.7) & +08:02:38.0(11) & 7388(3) & 120(7) & 142(10) & 0.529(0.078) & 104.0 & 9.1 & 10.99 & 0.6\\
  AGESVC1\_152 & 12:34:13.20(0.7) & +08:16:23.0(10) & 7351(2) & 259(4) & 283(6) & 3.136(0.179) & 103.5 & 9.8 & 28.74 & 0.6\\
  AGESVC1\_153 & 12:46:53.00(1.7) & +08:47:35.0(15) & 7323(6) & 56(12) & 86(18) & 0.226(0.067) & 103.1 & 8.7 & 6.19 & 0.7\\
  AGESVC1\_154 & 12:37:21.80(0.7) & +07:42:58.0(10) & 7268(10) & 57(19) & 98(29) & 0.143(0.061) & 102.3 & 8.5 & 4.52 & 0.7\\
  AGESVC1\_155 & 12:11:10.90(1.1) & +07:28:40.0(21) & 6233(8) & 90(17) & 142(25) & 0.226(0.063) & 87.7 & 8.6 & 5.98 & 0.6\\
  AGESVC1\_156 & 12:25:38.90(0.7) & +08:54:52.0(10) & 6890(2) & 228(5) & 245(7) & 1.553(0.136) & 97.0 & 9.5 & 13.8 & 0.7\\
  AGESVC1\_157 & 12:28:48.80(0.7) & +09:01:00.0(10) & 6470(2) & 250(4) & 266(6) & 2.763(0.202) & 91.1 & 9.7 & 16.29 & 0.9\\
  AGESVC1\_158 & 12:11:10.90(1.1) & +07:28:40.0(21) & 6233(8) & 90(17) & 142(25) & 0.226(0.063) & 87.7 & 8.6 & 5.98 & 0.6\\
  AGESVC1\_159 & 12:28:30.10(0.7) & +07:42:33.0(10) & 6398(9) & 65(18) & 125(27) & 0.142(0.053) & 90.1 & 8.4 & 6.14 & 0.7\\
  AGESVC1\_160 & 12:37:37.90(0.7) & +07:09:41.0(11) & 6522(14) & 127(28) & 291(42) & 0.499(0.098) & 91.8 & 8.9 & 6.41 & 0.7\\
  AGESVC1\_161 & 12:31:39.10(1.3) & +07:19:59.0(16) & 6423(9) & 51(17) & 117(26) & 0.196(0.061) & 90.4 & 8.5 & 6.64 & 0.7\\
  AGESVC1\_162 & 12:40:50.30(0.7) & +07:27:00.0(10) & 6502(4) & 57(7) & 92(11) & 0.330(0.064) & 91.5 & 8.8 & 12.75 & 0.7\\
  AGESVC1\_163 & 12:37:00.10(0.7) & +07:28:24.0(11) & 6452(9) & 69(19) & 221(28) & 0.363(0.061) & 90.8 & 8.8 & 9.84 & 0.5\\
  AGESVC1\_164 & 12:32:39.60(0.9) & +07:53:39.0(13) & 6521(5) & 122(11) & 165(16) & 0.450(0.076) & 91.8 & 8.9 & 8.82 & 0.6\\
  AGESVC1\_165 & 12:46:14.70(0.7) & +08:20:55.0(10) & 6459(2) & 455(4) & 475(5) & 3.702(0.178) & 90.9 & 9.8 & 29.29 & 0.5\\
  AGESVC1\_166 & 12:33:43.30(1.1) & +07:50:14.0(11) & 6563(11) & 88(22) & 164(33) & 0.272(0.078) & 92.4 & 8.7 & 5.41 & 0.7\\
  AGESVC1\_167 & 12:33:10.50(0.8) & +07:49:30.0(12) & 6066(11) & 369(21) & 429(32) & 0.606(0.120) & 85.4 & 9.0 & 5.02 & 0.7\\
  AGESVC1\_168 & 12:12:24.70(0.7) & +07:08:11.0(10) & 5652(5) & 163(9) & 214(14) & 0.990(0.104) & 79.6 & 9.1 & 11.7 & 0.6\\
  AGESVC1\_169 & 12:15:25.50(0.7) & +07:46:23.0(10) & 5952(2) & 191(5) & 210(7) & 1.308(0.113) & 83.8 & 9.3 & 16.69 & 0.6\\
  AGESVC1\_170 & 12:30:42.70(0.7) & +08:08:57.0(11) & 5690(11) & 164(22) & 260(33) & 0.523(0.094) & 80.1 & 8.8 & 6.25 & 0.6\\
  AGESVC1\_171 & 12:23:02.00(0.8) & +08:31:43.0(14) & 5854(3) & 54(6) & 70(9) & 0.243(0.054) & 82.4 & 8.5 & 10.0 & 0.6\\
  AGESVC1\_172 & 12:16:28.90(0.7) & +08:36:18.0(10) & 5734(2) & 123(4) & 137(5) & 1.250(0.110) & 80.7 & 9.2 & 19.58 & 0.6\\
  AGESVC1\_173 & 12:22:07.00(0.7) & +08:59:32.0(10) & 5842(2) & 59(3) & 82(5) & 3.187(0.271) & 82.2 & 9.7 & 65.89 & 0.8\\
  AGESVC1\_174 & 12:22:20.00(0.7) & +09:00:37.0(10) & 5820(5) & 102(10) & 144(15) & 0.669(0.104) & 81.9 & 9.0 & 9.78 & 0.8\\
  AGESVC1\_175 & 12:35:14.10(0.7) & +07:08:25.0(11) & 5640(2) & 172(4) & 183(7) & 0.738(0.091) & 79.4 & 9.0 & 11.41 & 0.6\\
  AGESVC1\_176 & 12:24:25.00(0.7) & +07:07:38.0(11) & 4218(2) & 79(4) & 91(5) & 0.700(0.082) & 59.4 & 8.7 & 17.12 & 0.6\\
  AGESVC1\_177 & 12:23:24.00(0.7) & +07:28:12.0(10) & 4253(2) & 270(4) & 295(6) & 4.012(0.207) & 59.9 & 9.5 & 32.15 & 0.6\\
  AGESVC1\_178 & 12:26:03.60(0.7) & +07:25:50.0(10) & 4559(3) & 46(7) & 80(10) & 0.316(0.052) & 64.2 & 8.4 & 13.57 & 0.5\\
  AGESVC1\_179 & 12:24:04.80(1.1) & +07:32:16.0(12) & 4203(3) & 26(6) & 44(9) & 0.141(0.040) & 59.1 & 8.0 & 10.55 & 0.6\\
  AGESVC1\_180 & 12:15:56.80(2.0) & +08:06:02.0(12) & 4007(8) & 56(16) & 120(24) & 0.192(0.054) & 56.4 & 8.1 & 7.11 & 0.6\\
  AGESVC1\_181 & 12:23:26.00(0.7) & +08:09:24.0(11) & 4661(9) & 215(18) & 268(26) & 0.329(0.077) & 65.6 & 8.5 & 5.78 & 0.6\\
  AGESVC1\_182 & 12:18:10.50(0.7) & +08:19:24.0(10) & 4319(3) & 67(5) & 97(8) & 0.815(0.096) & 60.8 & 8.8 & 19.24 & 0.7\\
  AGESVC1\_183 & 12:18:58.60(0.7) & +08:28:27.0(10) & 4064(5) & 92(10) & 124(14) & 0.405(0.073) & 57.2 & 8.4 & 8.58 & 0.6\\
  AGESVC1\_184 & 12:20:18.40(0.7) & +08:31:58.0(10) & 4430(2) & 83(4) & 101(6) & 1.085(0.103) & 62.3 & 8.9 & 24.66 & 0.6\\
  AGESVC1\_185 & 12:20:45.60(0.7) & +08:35:30.0(10) & 4419(5) & 78(10) & 197(15) & 1.015(0.099) & 62.2 & 8.9 & 20.07 & 0.6\\
  AGESVC1\_186 & 12:18:30.80(0.7) & +08:45:13.0(10) & 4150(4) & 81(9) & 148(13) & 0.749(0.085) & 58.4 & 8.7 & 16.98 & 0.6\\
  AGESVC1\_187 & 12:17:40.50(0.7) & +08:48:53.0(10) & 4300(4) & 105(8) & 204(13) & 1.522(0.123) & 60.5 & 9.1 & 25.26 & 0.6\\
  AGESVC1\_188 & 12:18:56.30(0.7) & +08:57:46.0(10) & 4433(4) & 336(8) & 363(12) & 1.378(0.141) & 62.4 & 9.1 & 9.63 & 0.7\\
  AGESVC1\_189 & 12:18:02.50(1.4) & +08:59:39.0(28) & 4541(10) & 39(20) & 131(31) & 0.182(0.062) & 63.9 & 8.2 & 6.62 & 0.8\\
  AGESVC1\_190 & 12:17:23.60(0.7) & +07:02:18.0(10) & 3738(4) & 149(8) & 201(11) & 1.525(0.132) & 52.6 & 8.9 & 16.56 & 0.7\\
  AGESVC1\_191 & 12:17:58.00(0.7) & +07:11:04.0(10) & 3727(2) & 343(3) & 364(5) & 18.905(0.763) & 52.4 & 10.0 & 106.22 & 0.7\\
  AGESVC1\_192 & 12:17:57.80(0.7) & +07:16:05.0(10) & 3916(3) & 259(6) & 321(9) & 5.452(0.259) & 55.1 & 9.5 & 41.52 & 0.6\\
  AGESVC1\_193 & 12:18:50.80(0.7) & +07:32:27.0(10) & 3915(6) & 70(12) & 146(18) & 0.311(0.048) & 55.1 & 8.3 & 11.47 & 0.4\\
  AGESVC1\_194 & 12:18:11.50(0.7) & +07:39:27.0(10) & 3960(2) & 134(3) & 158(5) & 5.588(0.329) & 55.7 & 9.6 & 75.58 & 0.6\\
  AGESVC1\_195 & 12:19:51.70(0.7) & +07:43:43.0(10) & 3785(2) & 122(4) & 147(6) & 1.813(0.148) & 53.3 & 9.0 & 29.32 & 0.7\\
  AGESVC1\_196 & 12:38:19.20(2.5) & +08:02:43.0(17) & 3966(11) & 98(23) & 170(34) & 0.227(0.067) & 55.8 & 8.2 & 5.18 & 0.6\\
  AGESVC1\_197 & 12:37:30.60(0.7) & +08:09:18.0(11) & 3943(7) & 126(14) & 170(21) & 0.390(0.079) & 55.5 & 8.4 & 6.63 & 0.6\\
  AGESVC1\_198 & 12:43:07.20(1.1) & +08:07:03.0(15) & 3909(13) & 72(27) & 221(40) & 0.225(0.055) & 55.0 & 8.2 & 6.49 & 0.5\\
  AGESVC1\_295 & 12:36:34.20(1.7) & +08:16:16.0(38) & 14844(12) & 73(24) & 167(36) & 0.231(0.065) & 209.0 & 9.3 & 5.67 & 0.6\\
  AGESVC1\_296 & 12:49:13.80(0.7) & +07:30:39.0(12) & 11544(5) & 100(11) & 128(16) & 0.587(0.134) & 162.5 & 9.5 & 7.03 & 1.2\\
  AGESVC1\_299 & 12:31:36.80(0.8) & +08:26:19.0(11) & 7423(10) & 45(20) & 164(30) & 0.193(0.052) & 104.5 & 8.6 & 7.88 & 0.6\\
  GLADOS\_004 & 12:24:52.40(0.7) & +07:54:03.0(22) & 6417(5) & 65(9) & 87(14) & 0.223(0.063) & 90.3 & 8.6 & 7.13 & 0.7\\
  GLADOS\_008 & 12:30:19.30(0.8) & +06:57:31.0(13) & 6401(17) & 95(33) & 331(50) & 0.502(0.104) & 90.1 & 8.9 & 6.54 & 0.8\\
  GLADOS\_009 & 12:47:03.30(1.6) & +08:16:51.0(11) & 6292(5) & 90(9) & 105(14) & 0.168(0.050) & 88.6 & 8.4 & 5.72 & 0.5\\
  GLADOS\_011 & 12:44:09.50(1.5) & +07:55:05.0(11) & 13349(13) & 140(25) & 242(38) & 0.243(0.067) & 188.0 & 9.3 & 5.62 & 0.6\\
  GLADOS\_012 & 12:28:03.00(1.0) & +08:11:19.0(24) & 11281(9) & 262(19) & 317(28) & 0.434(0.090) & 158.8 & 9.4 & 5.57 & 0.6\\
  GLADOS\_013 & 12:47:26.40(0.7) & +07:07:00.0(10) & 12817(15) & 234(30) & 324(45) & 0.267(0.078) & 180.5 & 9.3 & 4.4 & 0.6\\
  GLADOS\_015 & 12:09:40.00(1.3) & +08:42:47.0(12) & 12538(12) & 164(23) & 245(35) & 0.361(0.083) & 176.5 & 9.4 & 5.36 & 0.6\\
  GLADOS\_016 & 12:46:34.90(1.0) & +07:40:01.0(13) & 13236(6) & 230(12) & 255(18) & 0.311(0.075) & 186.4 & 9.4 & 5.73 & 0.6\\
  AGESVC1\_300 & 12:48:30.00(1.1) & +07:18:10.0(12) & 14784(5) & 207(10) & 222(15) & 0.414(0.096) & 208.2 & 9.6 & 5.37 & 0.7\\
  AGESVC1\_301 & 12:32:43.10(0.7) & +07:56:17.0(10) & 7244(3) & 143(7) & 164(10) & 0.714(0.093) & 102.0 & 9.2 & 9.87 & 0.6\\
  AGESVC1\_302 & 12:26:26.90(1.0) & +08:04:49.0(27) & 7450(4) & 263(9) & 271(13) & 0.389(0.101) & 104.9 & 9.0 & 4.5 & 0.7\\
  AGESVC1\_303 & 12:23:01.60(1.2) & +07:27:22.0(12) & 4196(8) & 276(16) & 319(24) & 0.408(0.087) & 58.9 & 8.5 & 5.64 & 0.6\\
\end{tabular}
\end{center}
\end{table*}

\begin{table*}
\tiny
\begin{center}
\caption[Optical properties of AGES detections within the Virgo Cluster in VC1]{\footnotesize Optical properties of HI detections in the Virgo Cluster. Columns : (1) Source number in this catalogue, (2) Flag for status of the optical counterpart used here - 0 indicates the object has a matching optical redshift, 1 indicates no optical redshift is available (3) Name in a major catalogue if present (4) Morphological type, using the system of the GOLDMine database (5) Absolute magnitude in the $g$ and (6) $i$ bands   (7) $g$-$i$ colour (8) HI mass to light ratio, $g$ band (9) Optical diameter in kpc  (10) Estimated HI deficiency  (11) Difference between optical and HI centres in arcminutes (12) Sky distance from M49 in degrees}
\label{OptVHIT}
\begin{tabular}{c c c c c c c c c c c c}\\
\hline
  \multicolumn{1}{c}{(1) No.} &	 
  \multicolumn{1}{c}{(2) Flag} & 	 	   
  \multicolumn{1}{c}{(3) Name} &	 
  \multicolumn{1}{c}{(4) MType} &   
  \multicolumn{1}{c}{(5) M$_{g}$} &
  \multicolumn{1}{c}{(6) M$_{i}$} &	 
  \multicolumn{1}{c}{(7) $g$-$i$} &	 
  \multicolumn{1}{c}{(8) $M_{\rm{HI}}$/$L_{g}$} &	 
  \multicolumn{1}{c}{(9) OptD(kpc)} &	  
  \multicolumn{1}{c}{(10) HIdef} &	  
  \multicolumn{1}{c}{(11) Sep.} &
  \multicolumn{1}{c}{(12) Dist.M49}\\    
\hline
  AGESVC1\_199 & 0 & VCC415 & 9 & -16.57 & -17.28 & 0.72 & 0.26 & 10.42 & 0.71 & 1.41 & 2.56\\
  AGESVC1\_200 & 0 & VCC450 & 1 & -16.24 & -17.25 & 1.01 & 0.1 & 12.42 &  & 0.75 & 2.36\\
  AGESVC1\_201 & 0 & None & 20 & -15.98 & -16.59 & 0.61 & 0.35 & 4.71 & 0.21 & 0.53 & 2.98\\
  AGESVC1\_202 & 0 & VCC94 & 2 & -19.28 & -20.41 & 1.13 & 0.22 & 23.6 & 0.15 & 0.12 & 4.03\\
  AGESVC1\_203 & 0 & VCC488 & 20 & -14.47 & -15.22 & 0.75 & 0.16 & 9.16 & 1.65 & 0.99 & 2.16\\
  AGESVC1\_204 & 1 & None & 20 & -15.56 & -16.22 & 0.66 & 1.44 & 10.11 & 0.35 & 0.31 & 3.13\\
  AGESVC1\_205 & 0 & FGC1384 & 10 & -16.18 & -16.91 & 0.73 & 0.46 & 11.62 & 0.7 & 0.32 & 4.22\\
  AGESVC1\_207 & 0 & VCC199 & 3 & -20.19 & -21.46 & 1.27 & 0.05 & 44.42 & 0.79 & 0.27 & 3.32\\
  AGESVC1\_210 & 0 & VCC393 & 7 & -18.51 & -19.28 & 0.77 & 0.14 & 15.36 & 0.5 & 0.24 & 2.41\\
  AGESVC1\_211 & 1 & VCC190 & -1 & -15.13 & -15.93 & 0.81 & 0.58 & 7.99 &  & 0.63 & 3.32\\
  AGESVC1\_212 & 0 & VCC1205 & 7 & -18.35 & -19.04 & 0.69 & 0.14 & 11.86 & 0.35 & 0.26 & 0.19\\
  AGESVC1\_213 & 0 & VCC343 & 9 & -16.49 & -17.2 & 0.71 & 0.27 & 7.31 & 0.46 & 0.08 & 2.58\\
  AGESVC1\_214 & 0 & None & 20 & -14.65 & -15.18 & 0.53 & 0.6 & 6.99 & 0.81 & 0.41 & 4.71\\
  AGESVC1\_215 & 0 & VCC180 & 1 & -16.98 & -17.85 & 0.87 & 0.04 & 13.04 &  & 0.68 & 3.37\\
  AGESVC1\_216 & 0 & VCC207 & 17 & -15.71 & -16.01 & 0.3 & 0.53 & 6.11 & 0.34 & 0.48 & 3.22\\
  AGESVC1\_217 & 0 & VCC479 & 20 & -15.14 & -15.83 & 0.69 & 0.23 & 8.18 & 1.15 & 0.41 & 2.07\\
  AGESVC1\_218 & 0 & AGC226127 & 20 & -14.87 & -15.28 & 0.41 & 1.54 & 9.87 & 0.58 & 0.16 & 2.92\\
  AGESVC1\_219 & 0 & VCC159 & 12 & -16.81 & -17.25 & 0.44 & 0.61 & 12.25 & 0.36 & 0.16 & 3.51\\
  AGESVC1\_220 & 0 & VCC318 & 8 & -18.43 & -18.92 & 0.49 & 0.57 & 25.76 & 0.3 & 0.09 & 2.79\\
  AGESVC1\_222 & 0 & VCC1933 & 11 & -14.71 & -15.25 & 0.54 & 1.08 & 5.66 & 0.37 & 0.0 & 3.28\\
  AGESVC1\_225 & 0 & VCC1791 & 15 & -16.88 & -17.23 & 0.35 & 0.52 & 11.07 & 0.33 & 0.49 & 2.38\\
  AGESVC1\_226 & 0 & VCC73 & 5 & -19.61 & -20.75 & 1.14 & 0.06 & 21.78 & 0.73 & 0.45 & 4.26\\
  AGESVC1\_227 & 0 & VCC656 & 5 & -19.02 & -20.26 & 1.24 & 0.1 & 21.65 & 0.72 & 0.43 & 1.85\\
  AGESVC1\_228 & 0 & VCC697 & 7 & -17.94 & -18.84 & 0.9 & 0.08 & 12.56 & 0.81 & 0.41 & 1.71\\
  AGESVC1\_229 & 0 & VCC667 & 7 & -17.94 & -18.86 & 0.92 & 0.13 & 21.76 & 1.02 & 0.33 & 1.69\\
  AGESVC1\_230 & 1 & None & 20 & -14.75 & -15.34 & 0.59 & 0.66 & 5.17 & 0.5 & 0.32 & 3.41\\
  AGESVC1\_231 & 1 & None & 20 & -13.82 & -13.91 & 0.09 & 0.82 & 2.36 & 0.19 & 0.94 & 2.92\\
  AGESVC1\_232 & 0 & None & 20 & -13.89 & -14.11 & 0.22 & 1.5 & 3.84 & 0.26 & 0.09 & 1.54\\
  AGESVC1\_233 & 0 & VCC851 & 7 & -18.03 & -18.99 & 0.96 & 0.2 & 23.67 & 0.85 & 0.27 & 1.06\\
  AGESVC1\_234 & 0 & None & 20 & -13.47 & -13.68 & 0.21 & 1.79 & 3.44 & 0.27 & 0.41 & 0.74\\
  AGESVC1\_235 & 1 & None & 20 & -13.2 & -13.52 & 0.32 & 0.91 & 2.41 & 0.41 & 0.93 & 2.44\\
  AGESVC1\_236 & 0 & VCC989 & 12 & -14.6 & -15.16 & 0.57 & 0.21 & 4.34 & 0.92 & 0.54 & 0.71\\
  AGESVC1\_237 & 0 & VCC105 & 9 & -18.93 & -19.6 & 0.67 & 0.25 & 30.49 & 0.59 & 0.23 & 3.88\\
  AGESVC1\_238 & 0 & VCC688 & 7 & -17.97 & -18.88 & 0.91 & 0.06 & 12.07 & 0.88 & 0.17 & 1.45\\
  AGESVC1\_239 & 1 & None & 20 & -13.47 & -13.92 & 0.45 & 0.37 & 1.46 & 0.31 & 0.08 & 0.92\\
  AGESVC1\_240 & 0 & VCC938 & 7 & -18.2 & -19.0 & 0.8 & 0.09 & 11.73 & 0.6 & 0.16 & 0.75\\
  AGESVC1\_241 & 1 & VCC867 & 12 & -13.3 & -14.12 & 0.82 & 4.41 & 3.51 & -0.03 & 1.44 & 0.95\\
  AGESVC1\_242 & 0 & AGC225022 & 20 & -14.93 & -15.07 & 0.14 & 0.96 & 4.67 & 0.19 & 0.15 & 1.44\\
  AGESVC1\_243 & 1 & None & 20 & -12.07 & -12.26 & 0.18 & 2.32 & 1.78 & 0.22 & 0.11 & 0.79\\
  AGESVC1\_244 & 0 & VCC566 & 11 & -15.53 & -15.82 & 0.3 & 1.66 & 5.7 & -0.13 & 0.0 & 1.79\\
  AGESVC1\_245 & 0 & VCC611 & -1 & -15.49 & -16.34 & 0.85 & 0.06 & 9.78 &  & 0.35 & 1.69\\
  AGESVC1\_246 & 0 & VCC888 & 12 & -16.29 & -16.99 & 0.7 & 0.48 & 12.86 & 0.72 & 0.0 & 0.93\\
  AGESVC1\_248 & 0 & VCC713 & 7 & -18.73 & -19.89 & 1.16 & 0.02 & 41.91 & 2.03 & 0.27 & 1.47\\
  AGESVC1\_249 & 0 & VCC1091 & 6 & -17.84 & -18.39 & 0.55 & 0.59 & 18.05 & 0.28 & 0.2 & 0.81\\
  AGESVC1\_251 & 1 & VCC1142 & 20 & -12.77 & -13.26 & 0.5 & 3.2 & 3.12 & 0.23 & 0.29 & 0.85\\
  AGESVC1\_252 & 1 & VCC203 & 20 & -12.77 & -13.08 & 0.3 & 9.71 & 5.7 & 0.2 & 0.3 & 3.35\\
  AGESVC1\_253 & 0 & AGC224244 & 20 & -15.22 & -15.43 & 0.21 & 2.84 & 9.49 & 0.14 & 0.17 & 3.95\\
  AGESVC1\_255 & 0 & VCC524 & 6 & -19.4 & -20.64 & 1.24 & 0.02 & 40.08 & 1.46 & 0.52 & 2.17\\
  AGESVC1\_256 & 0 & VCC1699 & 11 & -16.73 & -17.32 & 0.59 & 0.72 & 8.82 & 0.08 & 1.15 & 2.09\\
  AGESVC1\_258 & 1 & None & 20 & -12.0 & -12.73 & 0.72 & 1.42 & 0.96 & -0.0 & 0.49 & 2.12\\
  AGESVC1\_259 & 0 & VCC1952 & 12 & -14.96 & -15.43 & 0.47 & 0.94 & 7.96 & 0.59 & 0.32 & 3.32\\
  AGESVC1\_260 & 0 & VCC1455 & 12 & -15.37 & -16.09 & 0.72 & 0.16 & 9.01 & 1.28 & 0.56 & 0.76\\
  AGESVC1\_261 & 1 & VCC1394? & -1 & -13.11 & -14.04 & 0.92 & 0.93 & 5.7 &  & 2.22 & 0.53\\
  AGESVC1\_263 & 0 & VCC1758 & 7 & -16.58 & -17.36 & 0.77 & 0.3 & 12.59 & 0.78 & 0.3 & 2.12\\
  AGESVC1\_265 & 0 & VCC1675 & 2 & -16.77 & -17.59 & 0.83 & 0.04 & 12.38 & 1.55 & 0.48 & 1.68\\
  AGESVC1\_267 & 0 & VCC1330 & 3 & -18.82 & -20.03 & 1.21 & 0.01 & 20.81 & 1.84 & 0.05 & 0.31\\
  AGESVC1\_268 & 0 & VCC2007 & 16 & -15.42 & -16.12 & 0.7 & 0.11 & 4.74 & 0.93 & 0.65 & 3.71\\
  AGESVC1\_269 & 0 & VCC1555 & 7 & =20.91 & -21.78 & 0.87 & 0.03 & 55.72 & 0.57 & 0.36 & 1.15\\
  AGESVC1\_270 & 0 & VCC2070 & 3 & -20.10 & -21.31 & 1.21 & 0.08 & 50.54 & 0.66 & 0.10 & 4.63\\
  AGESVC1\_271 & 0 & VCC2033 & 17 & -16.09 & -16.82 & 0.73 & 0.07 & 6.54 & 1.12 & 1.17 & 4.05\\
  AGESVC1\_272 & 0 & VCC1725 & 15 & -16.77 & -17.3 & 0.53 & 0.17 & 9.22 & 0.72 & 0.1 & 2.04\\
  AGESVC1\_275 & 1 & VCC1964 & -1 & -13.54 & -13.85 & 0.3 & 0.59 & 6.37 &  & 0.94 & 3.47\\
  AGESVC1\_276 & 0 & VCC827 & 7 & -18.6 & -19.53 & 0.93 & 0.47 & 35.67 & 0.57 & 0.53 & 1.28\\
  AGESVC1\_277 & 0 & VCC975 & 8 & -18.48 & -19.05 & 0.56 & 0.4 & 22.23 & 0.33 & 0.19 & 0.98\\
  AGESVC1\_278 & 0 & VCC1011 & 10 & -16.25 & -17.0 & 0.75 & 0.35 & 11.6 & 0.79 & 0.09 & 0.67\\
  AGESVC1\_279 & 0 & VCC1193 & 7 & -17.09 & -17.87 & 0.77 & 0.16 & 12.48 & 0.84 & 0.26 & 0.31\\
  AGESVC1\_283 & 1 & AGC220261 & 20 & -14.89 & -15.55 & 0.66 & 0.6 & 7.65 & 0.79 & 0.7 & 3.38\\
  AGESVC1\_284 & 0 & VCC740 & 11 & -15.7 & -16.33 & 0.63 & 0.65 & 8.54 & 0.5 & 0.2 & 1.37\\
  AGESVC1\_285 & 0 & VCC1114 & 12 & -16.56 & -17.29 & 0.73 & 0.03 & 13.02 & 1.76 & 0.49 & 0.7\\
  AGESVC1\_286 & 0 & VCC514 & 7 & -17.11 & -17.88 & 0.77 & 0.11 & 16.07 & 1.18 & 0.08 & 2.05\\
  AGESVC1\_288 & 1 & None & 20 & -13.85 & -14.38 & 0.52 & 0.61 & 2.82 & 0.44 & 0.06 & 0.91\\
  AGESVC1\_289 & 0 & VCC758 & 1 & -18.62 & -19.88 & 1.26 & 0.02 & 15.77 &  & 0.95 & 1.33\\
  AGESVC1\_290 & 0 & VCC1190 & 3 & -20.14 & -21.36 & 1.21 & 0.01 & 51.46 & 1.82 & 0.91 & 0.74\\
  AGESVC1\_291 & 0 & VCC1726 & 10 & -16.2 & -16.55 & 0.36 & 0.75 & 10.35 & 0.39 & 0.17 & 2.17\\
  AGESVC1\_292 & 0 & VCC1575 & 11 & -17.65 & -18.61 & 0.96 & 0.05 & 12.11 & 1.09 & 0.18 & 1.48\\
  AGESVC1\_293 & 0 & None & 20 & -15.32 & -16.12 & 0.79 & 0.15 & 6.02 & 1.04 & 0.17 & 0.44\\
  AGESVC1\_304 & 0 & VCC534 & 3 & -18.49 & -19.62 & 1.12 & 0.01 & 22.43 & 1.78 & 1.29 & 2.06\\
  AGESVC1\_305 & 0 & VCC222 & 3 & -20.39 & -21.61 & 1.21 & 0.01 & 47.52 & 1.65 & 0.11 & 3.23\\
  AGESVC1\_306 & 0 & VCC52 & 12 & -14.81 & -15.44 & 0.63 & 0.74 & 4.71 & 0.36 & 0.34 & 4.42\\
  GLADOS\_001 & 0 & None & 20 & -14.79 & -15.28 & 0.49 & 0.53 & 5.08 & 0.57 & 0.37 & 2.67\\
  GLADOS\_005 & 0 & VCC1102 & 12 & -13.78 & -14.3 & 0.52 & 0.15 & 4.64 & 1.45 & 0.3 & 1.1\\
\end{tabular}
\end{center}
\end{table*}

\begin{table*}
\tiny
\begin{center}
\caption[Optical properties of AGES detections behind the Virgo Cluster in VC1]{\footnotesize Optical properties of HI detections behind the Virgo Cluster. Objects with multiple or unclear optical counterparts are omitted. Columns are as for table \ref{OptVHIT} except that the previous column (12), sky distance from M49, is omitted. (continued on next page)}
\label{OptBHIT}
\begin{tabular}{c c c c c c c c c c c}\\
\hline
  \multicolumn{1}{c}{(1) No.} &	 
  \multicolumn{1}{c}{(2) Flag} & 	 	 
  \multicolumn{1}{c}{(3) Name} &	 
  \multicolumn{1}{c}{(4) MType} &   
  \multicolumn{1}{c}{(5) M$_{g}$} &
  \multicolumn{1}{c}{(6) M$_{i}$} &	 
  \multicolumn{1}{c}{(7) $g$-$i$} &	 
  \multicolumn{1}{c}{(8) $M_{\rm{HI}}$/$L_{g}$} &	 
  \multicolumn{1}{c}{(9) OptD(kpc)} &	  
  \multicolumn{1}{c}{(10) HIdef} &	  
  \multicolumn{1}{c}{(11) Sep.} \\
\hline
  AGESVC1\_001 & 1 & None & 20 & -17.34 & -17.89 & 0.55 & 1.71 & 27.01 & 0.3 & 1.42\\
  AGESVC1\_005 & 0 & None & 20 & -20.12 & -20.93 & 0.82 & 0.7 & 43.15 & -0.06 & 0.41\\
  AGESVC1\_009 & 0 & None & 20 & -20.67 & -21.63 & 0.97 & 0.3 & 56.07 & 0.28 & 0.27\\
  AGESVC1\_010 & 1 & None & 20 & -17.18 & -17.93 & 0.75 & 1.08 & 38.03 & 0.83 & 0.92\\
  AGESVC1\_013 & 0 & None & 20 & -19.42 & -19.93 & 0.51 & 1.09 & 38.62 & -0.06 & 0.44\\
  AGESVC1\_014 & 0 & VCC602 & 20 & -19.37 & -20.42 & 1.05 & 0.69 & 36.7 & 0.11 & 0.23\\
  AGESVC1\_016 & 0 & VCC259 & 5 & -21.13 & -22.11 & 0.97 & 0.33 & 64.2 & -0.08 & 0.08\\
  AGESVC1\_017 & 0 & None & 20 & -20.38 & -21.05 & 0.67 & 0.52 & 61.99 & 0.23 & 0.28\\
  AGESVC1\_019 & 0 & None & 20 & -20.14 & -20.78 & 0.64 & 0.45 & 26.14 & -0.27 & 1.35\\
  AGESVC1\_020 & 0 & None & 20 & -19.54 & -20.11 & 0.57 & 0.22 & 28.14 & 0.34 & 0.47\\
  AGESVC1\_022 & 0 & None & 20 & -18.99 & -20.04 & 1.05 & 0.89 & 32.96 & 0.07 & 0.57\\
  AGESVC1\_023 & 1 & None & 20 & -18.76 & -19.16 & 0.4 & 0.89 & 29.03 & 0.07 & 0.27\\
  AGESVC1\_024 & 0 & None & 20 & -19.76 & -20.35 & 0.59 & 0.61 & 37.28 & 0.02 & 0.27\\
  AGESVC1\_027 & 1 & None & 20 & -18.31 & -19.06 & 0.74 & 1.69 & 32.67 & 0.06 & 0.36\\
  AGESVC1\_028 & 1 & None & 20 & -17.67 & -18.17 & 0.5 & 1.57 & 18.91 & -0.06 & 0.86\\
  AGESVC1\_029 & 0 & None & 20 & -19.14 & -19.88 & 0.74 & 1.29 & 47.5 & 0.13 & 0.46\\
  AGESVC1\_031 & 1 & None & 20 & -17.86 & -18.11 & 0.25 & 1.0 & 35.57 & 0.53 & 0.27\\
  AGESVC1\_032 & 1 & None & 20 & -17.95 & -18.38 & 0.44 & 0.93 & 18.92 & 0.06 & 0.34\\
  AGESVC1\_033 & 0 & None & 20 & -18.78 & -19.71 & 0.93 & 0.66 & 52.9 & 0.65 & 0.69\\
  AGESVC1\_035 & 1 & None & 20 & -16.92 & -17.71 & 0.79 & 2.0 & 18.59 & 0.12 & 1.24\\
  AGESVC1\_037 & 1 & None & 20 & -18.16 & -18.88 & 0.71 & 1.41 & 46.91 & 0.47 & 0.72\\
  AGESVC1\_039 & 0 & CGCG043-031 & 20 & -20.47 & -21.2 & 0.73 & 0.69 & 52.84 & -0.05 & 0.49\\
  AGESVC1\_041 & 0 & None & 20 & -18.33 & -18.91 & 0.58 & 0.79 & 32.28 & 0.38 & 0.28\\
  AGESVC1\_043 & 0 & None & 20 & -19.58 & -20.39 & 0.81 & 0.73 & 80.87 & 0.61 & 0.51\\
  AGESVC1\_046 & 0 & None & 20 & -19.31 & -20.11 & 0.8 & 0.68 & 44.0 & 0.28 & 0.67\\
  AGESVC1\_047 & 0 & AGC226167 & 20 & -20.26 & -20.89 & 0.63 & 0.33 & 50.97 & 0.33 & 0.42\\
  AGESVC1\_049 & 0 & None & 20 & -18.92 & -19.38 & 0.47 & 0.81 & 34.35 & 0.18 & 0.3\\
  AGESVC1\_050 & 0 & None & 20 & -18.94 & -19.6 & 0.66 & 0.22 & 33.6 & 0.72 & 0.7\\
  AGESVC1\_051 & 0 & None & 20 & -20.2 & -20.64 & 0.43 & 1.02 & 53.73 & -0.1 & 0.22\\
  AGESVC1\_052 & 0 & None & 20 & -19.13 & -20.05 & 0.92 & 0.8 & 45.79 & 0.31 & 0.76\\
  AGESVC1\_056 & 1 & None & 20 & -16.79 & -17.31 & 0.52 & 1.85 & 16.34 & 0.11 & 0.49\\
  AGESVC1\_057 & 0 & None & 20 & -19.37 & -20.2 & 0.83 & 0.27 & 57.21 & 0.86 & 0.38\\
  AGESVC1\_059 & 0 & VCC1579 & 20 & -21.15 & -22.07 & 0.91 & 0.17 & 64.28 & 0.44 & 0.2\\
  AGESVC1\_060 & 1 & None & 20 & -17.51 & -18.36 & 0.85 & 0.77 & 18.34 & 0.28 & 1.79\\
  AGESVC1\_062 & 1 & VCC1439 & 20 & -16.9 & -17.73 & 0.83 & 1.35 & 23.85 & 0.49 & 1.74\\
  AGESVC1\_064 & 0 & None & 20 & -19.02 & -19.6 & 0.57 & 1.05 & 25.91 & -0.19 & 0.05\\
  AGESVC1\_065 & 0 & None & 18 & -20.84 & -21.77 & 0.93 & 0.08 & 34.32 & 0.43 & 0.38\\
  AGESVC1\_067 & 0 & None & 20 & -18.61 & -19.32 & 0.71 & 0.88 & 23.86 & -0.01 & 0.17\\
  AGESVC1\_069 & 1 & None & 20 & -17.34 & -18.09 & 0.75 & 0.97 & 15.47 & 0.12 & 0.37\\
  AGESVC1\_070 & 1 & None & 20 & -17.53 & -17.27 & -0.26 & 1.33 & 22.51 & 0.2 & 1.86\\
  AGESVC1\_071 & 0 & None & 20 & -19.26 & -19.97 & 0.71 & 0.41 & 44.1 & 0.53 & 0.4\\
  AGESVC1\_074 & 0 & VCC210 & 7 & -20.73 & -21.44 & 0.71 & 0.44 & 44.68 & -0.08 & 0.69\\
  AGESVC1\_075 & 0 & AGC225023 & 20 & -19.81 & -20.34 & 0.53 & 0.56 & 38.62 & 0.07 & 0.0\\
  AGESVC1\_076 & 0 & AGC224529 & 20 & -20.12 & -20.73 & 0.61 & 0.24 & 56.46 & 0.6 & 1.36\\
  AGESVC1\_077 & 0 & None & 20 & -19.55 & -20.5 & 0.95 & 0.33 & 19.54 & -0.12 & 0.68\\
  AGESVC1\_081 & 1 & VCC1930 & 20 & -17.12 & -17.78 & 0.66 & 1.52 & 29.26 & 0.5 & 0.83\\
  AGESVC1\_082 & 1 & VCC1506 & 20 & -16.61 & -17.2 & 0.58 & 1.72 & 19.79 & 0.35 & 0.55\\
  AGESVC1\_083 & 0 & None & 20 & -18.29 & -18.89 & 0.6 & 0.86 & 36.04 & 0.44 & 0.53\\
  AGESVC1\_084 & 0 & AGC226140 & 20 & -19.38 & -20.44 & 1.06 & 1.19 & 76.98 & 0.43 & 0.15\\
  AGESVC1\_085 & 0 & VCC1525 & 6 & -21.07 & -21.85 & 0.78 & 0.14 & 37.73 & 0.01 & 0.14\\
  AGESVC1\_086 & 0 & None & 20 & -18.49 & -19.17 & 0.68 & 0.89 & 36.74 & 0.36 & 0.58\\
  AGESVC1\_088 & 0 & None & 20 & -18.19 & -18.76 & 0.57 & 1.83 & 28.95 & -0.01 & 0.2\\
  AGESVC1\_090 & 0 & VCC161 & 6 & -20.87 & -21.92 & 1.05 & 0.82 & 96.23 & -0.18 & 0.12\\
  AGESVC1\_091 & 0 & AGC226138 & 20 & -18.94 & -19.74 & 0.79 & 1.22 & 36.91 & 0.04 & 0.27\\
  AGESVC1\_092 & 1 & None & 20 & -16.74 & -17.19 & 0.45 & 2.49 & 16.51 & 0.01 & 0.58\\
  AGESVC1\_093 & 1 & None & 20 & -17.63 & -18.06 & 0.43 & 0.61 & 23.94 & 0.54 & 0.33\\
  AGESVC1\_094 & 0 & None & 20 & -18.41 & -19.22 & 0.82 & 0.32 & 20.07 & 0.38 & 0.3\\
  AGESVC1\_095 & 0 & VCC53 & 3 & -19.91 & -20.79 & 0.88 & 0.37 & 41.75 & -0.03 & 0.3\\
  AGESVC1\_098 & 0 & AGC226146 & 20 & -18.67 & -19.01 & 0.34 & 1.38 & 38.1 & 0.12 & 0.54\\
  AGESVC1\_099 & 0 & None & 20 & -18.4 & -18.91 & 0.51 & 0.74 & 35.41 & 0.44 & 1.09\\
  AGESVC1\_100 & 2 & None & 20 & -15.63 & -16.52 & 0.89 & 5.8 & 12.35 & -0.14 & 0.59\\
  AGESVC1\_101 & 0 & AGC226139 & 20 & -19.96 & -20.44 & 0.47 & 0.27 & 36.54 & 0.28 & 0.3\\
  AGESVC1\_102 & 0 & AGC226145 & 20 & -20.36 & -20.93 & 0.57 & 0.45 & 34.46 & -0.14 & 0.43\\
  AGESVC1\_103 & 1 & None & 20 & -17.83 & -18.0 & 0.17 & 1.64 & 27.8 & 0.15 & 0.58\\
  AGESVC1\_105 & 0 & AGC220773 & 20 & -19.37 & -19.94 & 0.57 & 0.55 & 43.94 & 0.35 & 0.57\\
  AGESVC1\_106 & 0 & None & 20 & -20.0 & -20.8 & 0.8 & 0.24 & 26.5 & 0.07 & 1.35\\
  AGESVC1\_108 & 0 & AGC226135 & 20 & -19.43 & -20.22 & 0.78 & 0.88 & 65.1 & 0.42 & 0.08\\
  AGESVC1\_109 & 1 & None & 20 & -16.57 & -17.15 & 0.58 & 1.81 & 19.72 & 0.35 & 0.5\\
  AGESVC1\_110 & 1 & None & 20 & -17.23 & -17.72 & 0.49 & 2.37 & 17.3 & -0.13 & 0.48\\
  AGESVC1\_111 & 0 & AGC224409 & 20 & -18.09 & -18.81 & 0.73 & 1.73 & 34.22 & 0.18 & 0.05\\
  AGESVC1\_112 & 0 & AGC224455 & 20 & -19.92 & -20.84 & 0.92 & 0.23 & 33.81 & 0.31 & 0.15\\
  AGESVC1\_113 & 0 & None & 20 & -17.72 & -18.12 & 0.4 & 0.92 & 23.76 & 0.32 & 0.73\\
  AGESVC1\_114 & 0 & None & 20 & -18.2 & -19.01 & 0.81 & 0.99 & 30.24 & 0.28 & 0.35\\
  AGESVC1\_115 & 1 & None & 20 & -15.73 & -16.39 & 0.66 & 2.39 & 10.58 & 0.09 & 0.92\\
  AGESVC1\_116 & 1 & None & 20 & -16.72 & -17.03 & 0.32 & 1.11 & 11.81 & 0.11 & 0.3\\
  AGESVC1\_118 & 1 & None & 20 & -16.95 & -17.91 & 0.96 & 2.64 & 24.05 & 0.18 & 0.09\\
  AGESVC1\_120 & 1 & None & 20 & -15.14 & -15.72 & 0.58 & 4.2 & 20.5 & 0.58 & 0.59\\
  AGESVC1\_121 & 0 & None & 20 & -19.59 & -20.59 & 1.01 & 0.4 & 39.2 & 0.32 & 0.13\\
  AGESVC1\_122 & 1 & None & 20 & -16.41 & -17.07 & 0.66 & 2.62 & 16.85 & 0.13 & 0.42\\
  AGESVC1\_123 & 0 & VCC377 & 6 & -20.23 & -20.98 & 0.74 & 0.59 & 47.2 & -0.16 & 0.31\\
  AGESVC1\_124 & 0 & AGC221659 & 20 & -18.85 & -19.91 & 1.05 & 0.62 & 24.49 & 0.06 & 0.19\\
  AGESVC1\_125 & 1 & None & 20 & -16.02 & -16.48 & 0.46 & 1.61 & 8.46 & -0.02 & 0.37\\
  AGESVC1\_126 & 0 & None & 20 & -18.68 & -19.12 & 0.44 & 0.73 & 34.56 & 0.32 & 0.13\\
  AGESVC1\_127 & 0 & None & 20 & -18.26 & -18.71 & 0.45 & 1.06 & 20.1 & -0.08 & 0.17\\
  AGESVC1\_128 & 0 & None & 20 & -18.48 & -19.04 & 0.57 & 0.56 & 21.89 & 0.18 & 0.22\\
  AGESVC1\_131 & 0 & VCC1152 & 5 & -20.36 & -21.35 & 0.99 & 0.08 & 43.44 & 0.63 & 0.14\\
  AGESVC1\_132 & 0 & VCC712 & 4 & -20.42 & -21.1 & 0.68 & 0.35 & 43.33 & -0.19 & 0.28\\
  AGESVC1\_133 & 0 & VCC1182 & 7 & -19.35 & -20.09 & 0.75 & 0.18 & 26.35 & 0.45 & 0.73\\
  AGESVC1\_134 & 1 & VCC520 & 16 & -16.61 & -17.33 & 0.72 & 3.41 & 39.34 & 0.58 & 0.37\\
  AGESVC1\_135 & 0 & None & 20 & -17.73 & -18.64 & 0.9 & 0.6 & 30.0 & 0.68 & 0.46\\
  AGESVC1\_136 & 1 & VCC247 & 20 & -17.21 & -17.62 & 0.41 & 2.28 & 20.01 & 0.0 & 0.31\\
  AGESVC1\_137 & 1 & None & 20 & -16.52 & -16.99 & 0.47 & 2.21 & 20.73 & 0.32 & 0.23\\
  AGESVC1\_139 & 0 & None & 20 & -18.34 & -18.75 & 0.41 & 0.28 & 22.11 & 0.53 & 0.54\\
  AGESVC1\_141 & 0 & None & 20 & -19.0 & -19.96 & 0.96 & 0.24 & 32.77 & 0.63 & 0.52\\
\end{tabular}
\end{center}
\end{table*}

\addtocounter{table}{-1}  
\begin{table*}
\tiny
\begin{center}
\caption[Optical properties of AGES detections behind the Virgo Cluster in VC1]{\footnotesize (continued from previous page) Optical properties of HI detections behind the Virgo Cluster. Objects with multiple or unclear optical counterparts are omitted. Columns are as for table \ref{OptVHIT} except that the previous column (12), sky distance from M49, is omitted.}
\label{OptBHIT}
\begin{tabular}{c c c c c c c c c c c}\\
\hline
  \multicolumn{1}{c}{(1) No.} &	 
  \multicolumn{1}{c}{(2) Flag} & 	 	 
  \multicolumn{1}{c}{(3) Name} &	 
  \multicolumn{1}{c}{(4) MType} &   
  \multicolumn{1}{c}{(5) M$_{g}$} &
  \multicolumn{1}{c}{(6) M$_{i}$} &	 
  \multicolumn{1}{c}{(7) $g$-$i$} &	 
  \multicolumn{1}{c}{(8) $M_{\rm{HI}}$/$L_{g}$} &	 
  \multicolumn{1}{c}{(9) OptD(kpc)} &	  
  \multicolumn{1}{c}{(10) HIdef} &	  
  \multicolumn{1}{c}{(11) Sep.} \\
\hline
  AGESVC1\_142 & 1 & None & 20 & -16.23 & -16.82 & 0.59 & 1.7 & 13.61 & 0.23 & 0.99\\
  AGESVC1\_143 & 0 & VCC749 & 4 & -19.68 & -20.68 & 1.0 & 0.08 & 32.81 & 0.58 & 0.29\\
  AGESVC1\_144 & 0 & VCC847 & 7 & -19.66 & -20.33 & 0.67 & 0.39 & 36.12 & 0.23 & 0.38\\
  AGESVC1\_145 & 0 & None & 20 & -19.35 & -20.32 & 0.96 & 0.11 & 24.44 & 0.63 & 0.44\\
  AGESVC1\_146 & 0 & VCC1774 & 18 & -19.58 & -20.62 & 1.04 & 0.25 & 27.27 & 0.25 & 0.2\\
  AGESVC1\_147 & 0 & None & 20 & -17.4 & -17.9 & 0.5 & 0.75 & 16.16 & 0.24 & 0.45\\
  AGESVC1\_148 & 0 & VCC1802 & 18 & -19.79 & -20.82 & 1.02 & 0.59 & 36.67 & 0.01 & 0.15\\
  AGESVC1\_149 & 0 & None & 20 & -19.46 & -20.46 & 1.0 & 0.51 & 31.54 & 0.09 & 0.12\\
  AGESVC1\_150 & 1 & None & 20 & -15.81 & -16.38 & 0.56 & 4.02 & 6.57 & -0.53 & 0.44\\
  AGESVC1\_152 & 0 & NGC4535A & 20 & -19.66 & -20.24 & 0.58 & 0.72 & 37.88 & 0.01 & 0.19\\
  AGESVC1\_154 & 1 & None & 20 & -15.88 & -16.47 & 0.58 & 1.04 & 19.06 & 0.84 & 0.09\\
  AGESVC1\_155 & 0 & None & 20 & -17.28 & -17.72 & 0.44 & 0.33 & 8.04 & 0.12 & 1.16\\
  AGESVC1\_156 & 0 & VCC822 & 7 & -18.95 & -19.46 & 0.51 & 0.6 & 30.38 & 0.2 & 0.2\\
  AGESVC1\_157 & 0 & AGC224506 & 20 & -18.38 & -18.85 & 0.47 & 1.59 & 29.71 & -0.01 & 0.13\\
  AGESVC1\_158 & 0 & None & 20 & -17.39 & -17.9 & 0.51 & 0.3 & 13.61 & 0.52 & 1.16\\
  AGESVC1\_160 & 1 & None & 20 & -16.71 & -16.97 & 0.26 & 1.36 & 10.28 & -0.08 & 0.33\\
  AGESVC1\_161 & 1 & None & 20 & -14.94 & -15.5 & 0.56 & 2.63 & 12.75 & 0.51 & 1.52\\
  AGESVC1\_162 & 1 & None & 20 & -16.35 & -16.58 & 0.23 & 1.25 & 13.75 & 0.33 & 0.12\\
  AGESVC1\_164 & 0 & None & 20 & -16.78 & -17.38 & 0.6 & 1.15 & 15.63 & 0.29 & 0.2\\
  AGESVC1\_165 & 0 & VCC2036 & 3 & -20.7 & -21.86 & 1.16 & 0.25 & 59.02 & -0.0 & 0.22\\
  AGESVC1\_166 & 0 & None & 20 & -16.87 & -17.58 & 0.71 & 0.65 & 16.25 & 0.53 & 0.64\\
  AGESVC1\_167 & 0 & VCC1480 & 5 & -19.34 & -20.49 & 1.15 & 0.13 & 27.56 & 0.6 & 0.5\\
  AGESVC1\_168 & 0 & None & 20 & -16.89 & -17.47 & 0.58 & 1.71 & 20.88 & 0.29 & 0.21\\
  AGESVC1\_169 & 0 & None & 20 & -18.31 & -18.88 & 0.57 & 0.68 & 23.11 & 0.2 & 0.16\\
  AGESVC1\_170 & 1 & None & 20 & -16.28 & -16.88 & 0.61 & 1.62 & 11.86 & 0.13 & 0.51\\
  AGESVC1\_171 & 1 & VCC609 & 20 & -15.26 & -15.57 & 0.31 & 2.03 & 10.08 & 0.32 & 0.55\\
  AGESVC1\_173 & 0 & VCC526 & 7 & -19.56 & -20.19 & 0.63 & 0.5 & 36.12 & 0.16 & 0.13\\
  AGESVC1\_174 & 0 & VCC544 & 11 & -17.53 & -18.02 & 0.48 & 0.68 & 13.63 & 0.11 & 1.44\\
  AGESVC1\_175 & 0 & None & 20 & -17.43 & -17.96 & 0.52 & 0.77 & 27.04 & 0.61 & 0.22\\
  AGESVC1\_176 & 1 & VCC724 & 20 & -16.0 & -16.75 & 0.76 & 1.54 & 16.28 & 0.5 & 0.27\\
  AGESVC1\_177 & 0 & VCC638 & 4 & -20.89 & -22.01 & 1.12 & 0.1 & 62.36 & 0.35 & 0.17\\
  AGESVC1\_178 & 1 & None & 20 & -15.56 & -15.97 & 0.41 & 1.21 & 7.46 & 0.19 & 0.25\\
  AGESVC1\_179 & 1 & VCC694 & 20 & -15.25 & -15.86 & 0.61 & 0.61 & 7.69 & 0.64 & 0.43\\
  AGESVC1\_180 & 1 & None & 20 & -13.57 & -13.93 & 0.36 & 3.57 & 7.3 & 0.5 & 0.38\\
  AGESVC1\_181 & 0 & None & 20 & -16.12 & -16.76 & 0.64 & 0.79 & 6.95 & 0.1 & 0.13\\
  AGESVC1\_182 & 0 & AGC220317 & 20 & -17.7 & -18.35 & 0.64 & 0.39 & 15.68 & 0.39 & 0.17\\
  AGESVC1\_184 & 0 & VCC406 & 6 & -18.75 & -19.5 & 0.75 & 0.21 & 19.04 & 0.4 & 0.1\\
  AGESVC1\_186 & 0 & AGC224416 & 20 & -17.01 & -17.44 & 0.43 & 0.63 & 12.11 & 0.26 & 0.35\\
  AGESVC1\_187 & 0 & VCC248 & 12 & -16.5 & -16.87 & 0.37 & 2.18 & 19.33 & 0.28 & 0.22\\
  AGESVC1\_188 & 0 & VCC313 & 3 & -19.39 & -20.55 & 1.16 & 0.15 & 26.23 & 0.34 & 0.11\\
  AGESVC1\_189 & 1 & None & 20 & -14.6 & -15.24 & 0.64 & 1.67 & 6.59 & 0.34 & 0.5\\
  AGESVC1\_190 & 0 & None & 20 & -17.27 & -17.68 & 0.41 & 0.82 & 10.52 & -0.06 & 0.23\\
  AGESVC1\_191 & 0 & VCC264 & 7 & -20.53 & -21.36 & 0.83 & 0.5 & 49.9 & 0.03 & 0.12\\
  AGESVC1\_192 & 0 & VCC265 & 3 & -19.6 & -20.68 & 1.07 & 0.37 & 23.92 & -0.2 & 0.34\\
  AGESVC1\_194 & 0 & VCC277 & 18 & -18.04 & -18.48 & 0.44 & 1.65 & 21.38 & -0.14 & 0.16\\
  AGESVC1\_195 & 0 & VCC380 & 17 & -17.24 & -17.47 & 0.23 & 1.02 & 19.85 & 0.33 & 0.25\\
  AGESVC1\_196 & 1 & None & 20 & -14.92 & -15.44 & 0.53 & 1.19 & 8.37 & 0.54 & 0.18\\
  AGESVC1\_197 & 0 & None & 20 & -16.97 & -17.53 & 0.57 & 0.31 & 11.7 & 0.57 & 0.05\\
  AGESVC1\_198 & 1 & None & 20 & -13.97 & -14.13 & 0.16 & 2.75 & 8.17 & 0.54 & 1.19\\
  AGESVC1\_296 & 0 & None & 20 & -19.4 & -19.83 & 0.44 & 0.42 & 23.88 & -0.01 & 0.56\\
  AGESVC1\_299 & 1 & None & 20 & -16.9 & -17.27 & 0.37 & 0.57 & 11.47 & 0.31 & 0.4\\
  AGESVC1\_300 & 1 & None & 20 & -18.48 & -19.05 & 0.57 & 1.14 & 39.24 & 0.31 & 0.55\\
  AGESVC1\_301 & 0 & None & 20 & -17.18 & -17.55 & 0.37 & 1.56 & 17.56 & 0.08 & 0.29\\
  AGESVC1\_302 & 0 & VCC908 & 7 & -19.54 & -20.35 & 0.81 & 0.1 & 25.69 & 0.61 & 1.46\\
  AGESVC1\_303 & 0 & None & 20 & -15.46 & -16.27 & 0.81 & 1.45 & 10.49 & 0.41 & 1.7\\
  GLADOS\_004 & 1 & None & 20 & -16.32 & -16.68 & 0.37 & 0.85 & 11.37 & 0.36 & 0.35\\
  GLADOS\_008 & 0 & None & 20 & -17.52 & -17.87 & 0.35 & 0.63 & 12.55 & 0.09 & 0.94\\
  GLADOS\_011 & 0 & None & 20 & -19.02 & -19.53 & 0.51 & 0.33 & 23.95 & 0.25 & 1.2\\
  GLADOS\_013 & 1 & None & 20 & -17.53 & -17.78 & 0.25 & 1.32 & 28.56 & 0.38 & 1.04\\
  GLADOS\_014 & 1 & None & 20 & -17.55 & -18.41 & 0.87 & 1.09 & 19.9 & 0.18 & 0.74\\
  GLADOS\_015 & 1 & None & 20 & -17.86 & -18.3 & 0.44 & 1.26 & 40.94 & 0.54 & 0.0 \\
\end{tabular}
\end{center}
\end{table*}

\bsp

\label{lastpage}

\end{document}